\documentclass[aps,prb,twocolumn,superscriptaddress,showpacs]{revtex4-2}
\usepackage{amsmath,amssymb,wasysym,graphicx}
\usepackage[utf8]{inputenc}
\usepackage[T1]{fontenc}
\usepackage{xcolor}

\IfFileExists{newtxtext.sty}
{\usepackage{newtxtext,newtxmath}}
{\IfFileExists{stix.sty}
	{\usepackage{stix}}
	{\IfFileExists{mathptmx.sty}
		{\usepackage{mathptmx}}{} } }

\usepackage{textcomp}

\usepackage{bm}

\IfFileExists{siunitx.sty}{\usepackage{booktabs,siunitx}}{}

\usepackage{color}
\definecolor{LinkColor}{rgb}{0.256,0.439,0.588}
\usepackage{hyperref}
\hypersetup{
	pdfauthor={good guys},
	pdftitle={good title},
	colorlinks=true,
	citecolor=LinkColor,
	linkcolor=LinkColor,
	urlcolor=LinkColor
}

\newcommand{\beq} {\begin{equation}}
\newcommand{\eeq} {\end{equation}}
\newcommand{\bea} {\begin{eqnarray}}
\newcommand{\eea} {\end{eqnarray}}
\newcommand{\be} {\begin{equation}}
\newcommand{\ee} {\end{equation}}

\newcommand{\ket}[1]{\left|#1\right>}

\begin{document}
\title{Emergent gapless spiral phases and conformal Lifshitz criticality in the cluster Ising model with off-diagonal interactions}

\author{Wei-Lin Li}
\altaffiliation{The first two authors contributed equally.}
\affiliation {Key Laboratory of Atomic and Subatomic Structure and Quantum Control (Ministry of Education), Guangdong Basic Research Center of Excellence for Structure and Fundamental Interactions of Matter, School of Physics, South China Normal University, Guangzhou 510006, China}
\affiliation {Guangdong Provincial Key Laboratory of Quantum Engineering and Quantum Materials, Guangdong-Hong Kong Joint Laboratory of Quantum Matter, Frontier Research Institute for Physics, South China Normal University, Guangzhou 510006, China}

\author{Dan-Dan Liang}
\altaffiliation{The first two authors contributed equally.}
\affiliation {Key Laboratory of Atomic and Subatomic Structure and Quantum Control (Ministry of Education), Guangdong Basic Research Center of Excellence for Structure and Fundamental Interactions of Matter, School of Physics, South China Normal University, Guangzhou 510006, China}
\affiliation {Guangdong Provincial Key Laboratory of Quantum Engineering and Quantum Materials, Guangdong-Hong Kong Joint Laboratory of Quantum Matter, Frontier Research Institute for Physics, South China Normal University, Guangzhou 510006, China}

\author{Zhi Li}
\email{lizphys@m.scnu.edu.cn}
\affiliation {Key Laboratory of Atomic and Subatomic Structure and Quantum Control (Ministry of Education), Guangdong Basic Research Center of Excellence for Structure and Fundamental Interactions of Matter, School of Physics, South China Normal University, Guangzhou 510006, China}
\affiliation {Guangdong Provincial Key Laboratory of Quantum Engineering and Quantum Materials, Guangdong-Hong Kong Joint Laboratory of Quantum Matter, Frontier Research Institute for Physics, South China Normal University, Guangzhou 510006, China}

\author{Xue-Jia Yu}
\email{xuejiayu@fzu.edu.cn}
\affiliation{Department of Physics, Fuzhou University, Fuzhou 350116, Fujian, China}
\affiliation{Fujian Key Laboratory of Quantum Information and Quantum Optics,
College of Physics and Information Engineering,
Fuzhou University, Fuzhou, Fujian 350108, China}

\date{\today}

\begin{abstract}
We perform a comprehensive analytical study of the exotic quantum phases and phase transitions emerging from the cluster-Ising model with off-diagonal Gamma interactions. Specifically, we map out the ground-state phase diagram by analyzing both local and nonlocal order parameters, together with the energy spectra. The results reveal two pairs of gapped phases—namely and antiferromagnetic (AFM) long-range ordered phases, symmetry-protected topological (SPT) phases—as well as two distinct gapless spiral phases induced by the off-diagonal interactions, which are related by a duality transformation and are numerically confirmed through the long-distance behavior of various order parameters. Remarkably, four distinct phase transition lines emerge in the phase diagram. Two of them, which separate the distinct gapped or gapless phases, are described by the Ising and three-copy Ising conformal field theories, respectively. In contrast, the remaining two transition lines—between the gapless spiral and gapped phases—belong to a nonconformal Lifshitz criticality with dynamical critical exponent $z = 2$. More importantly, the intersection of these four transition lines gives rise to a new Lifshitz multicritical point exhibiting emergent conformal symmetry, marking a fundamental departure from all previously known nonconformal Lifshitz points. This work provides a valuable reference for future investigations of exotic gapless phases and their transitions in exactly solvable many-body systems.


\end{abstract}

\maketitle

\section{INTRODUCTION}
\label{sec:introduction}
Exploring exotic quantum phases and phase transitions is a fundamental issue in modern condensed matter and statistical physics~\cite{landau2013statistical,cardy1996scaling,sachdev1999quantum,sachdev2023quantum}. Traditionally, Landau theory has provided the standard paradigm for classifying them based on distinct patterns of symmetry-breaking~\cite{landau2013statistical}. However, the discovery of topological phases has challenged this framework, demonstrating that phases with the same symmetries can still be distinguished by topological characteristics~\cite{Hasan2010RMP,Qi2011RMP,senthil2015symmetry,Wen2017RMP}. A prototypical class of such phases is characterized by a bulk energy gap and nontrivial topology protected by global symmetries—now known as symmetry-protected topological (SPT) phases~\cite{senthil2015symmetry}. These include free-fermion systems with nontrivial topological band structures~\cite{Hasan2010RMP,Qi2011RMP,Kane2005PRL,Kane2005PRL_b,Bernevig2006PRL,Bernevig2006Science,Fu2007PRL} and bosonic SPT phases in strongly interacting many-body systems~\cite{Gu2009PRB,Chen2010PRB,Pollmann2012PRB,Chen2013PRB}. Notably, symmetry-protected topology has been recently extended to certain gapless quantum critical systems, leading to the concept of gapless SPT~\cite{Keselman2015PRB,Scaffidi2017PRX,Verresen2018PRL,Verresen2021PRX,Yu2022PRL,Yu2024PRL}. These phases challenge the traditional belief of condensed matter and statistical physics and have attracted considerable attention in recent years~\cite{Parker2018PRB,JIANG2018753,Jones2019JSP,verresen2020topologyedgestatessurvive,Thorngren2021PRB,Duque2021PRB,Jones2023PRL,Wen2023PRB,huang2025topologicalholographyquantumcriticality,Mondal2023PRB,Yu2024PRB,Prembabu2024PRB,Su2024PRB,Li2024SciPost,Zhong2024PRA,ando2024gauginglatticegappedgaplesstopological,Zhang2024PRA,Yang2025CP,Zhou2025CP,Cardoso2025PRB,Flores2025PRL,Wen2025PRB,Li2025SciPost,yu2025gaplesssymmetryprotectedtopologicalstates,tan2025exploringnontrivialtopologyquantum,zhong2025quantumentanglementfermionicgapless,yang2025deconfinedcriticalityintrinsicallygapless,li2024noninvertiblesymmetryenrichedquantumcritical,wen2025topologicalholography21dgapped,wen2025stringcondensationtopologicalholography}. A representative lattice model that exhibits both gapped and gapless SPT ground states is the one-dimensional cluster-Ising model, which incorporates two-spin Ising and three-spin cluster interactions~\cite{Son_2011,Smacchia2011PRA,Montes2012PRE,Lahtinen2015PRL,Giampaolo2015PRA,Ohta2016PRB,Verresen2017PRB,Nie2017PRE,Ding2019PRE,Jones2021PRR,Guo2022PRA,Kuno_2022,Yu2024PRB,Verga2023PRB,Alcaraz2024PRE,Li2025PRA}. The competition between these interactions gives rise to antiferromagnetic long-range ordered and gapped SPT phases. The critical points between them host nontrivial topological edge modes and are thus classified as gapless SPTs~\cite{Verresen2021PRX,Yu2022PRL}. This model has garnered widespread interest for two main reasons: (i) it is exactly solvable and serves as a valuable toy model for exploring SPT physics; and (ii) it can be implemented in state-of-the-art experimental platforms, such as ultracold atoms in triangular optical lattices~\cite{Becker_2010,Petiziol2021PRL} and superconducting quantum processor~\cite{Jones2021PRR,Smith2022PRR,shen2025robustsimulationsmanybodysymmetryprotected,tan2025exploringnontrivialtopologyquantum}.

On a different front, recent progress~\cite{Sorensen2021PRX,Yang2020PRL,Liu2020PRE,Luo2021PRB,Zhao2022PRA,LIU2021126122,yang2025emergentsu21conformalsymmetry,Kheiri2024PRB,Abbasi2025SciPostCore,Saito_2024,Mahdavifar2024} has drawn enormous attention to off-diagonal Gamma exchange interactions in quantum spin chains. These interactions refer to a bond-dependent coupling involving the product of two different spin components on neighboring sites, originally proposed in the Kitaev honeycomb model~\cite{Jackeli2009PRL,hermanns2018physics}. Off-diagonal Gamma interactions have been shown to induce a variety of exotic quantum phases and phase transitions, including the stabilization of quantum spin liquid phases~\cite{Jackeli2009PRL,Luo2021npj,Takikawa2019PRB,WangPRB2021}, magnetic long-range order~\cite{Rossler2006Nature,Heinze2011NP,WangPRB2021}, gapless chiral phases~\cite{Furukawa2012PRB,Luo2022PRB}, and Lifshitz critical points that typically lack conformal symmetry~\cite{ginsparg1988applied,francesco2012conformal,hornreich1980lifshitz,Liu2020PRE,Chepiga2021PRR,Wang2022SciPost,Yu2024PRB}. Importantly, in one-dimensional quantum spin models, Gamma interaction terms can be mapped onto fermionic bilinear terms via the Jordan-Wigner transformation, rendering the model exactly solvable in the language of free fermions. Moreover, these interactions are experimentally accessible through quantum simulators employing trapped atoms~\cite{Zhao2022PRA,Kunimi2024PRA}, photons~\cite{Pitsios2017NC,Douglas2015,GT2015NP}, or even in realistic solid-state materials~\cite{Kuepferling2023RMP,Yang2023NRP,Dmitrienko2014NP,BD1996PRL,RC2010Sci,KM2002PRB,BO2013PRL}. Therefore, off-diagonal Gamma interactions provide an ideal platform for exploring novel quantum phases and phase transitions from both analytical and experimental perspectives. Given these advantages, a natural and compelling question arises: can the interplay between symmetry-protected topology and off-diagonal interactions lead to novel quantum phases and phase transitions?

To make progress in answering the above questions, in this work, we conduct a comprehensive investigation of the ground-state properties of a one-dimensional cluster-Ising model with off-diagonal Gamma interactions. Specifically, by applying the Jordan-Wigner transformation, we map the model onto a free-fermion system and analytically determine its ground-state phase diagram. The resulting phase diagram features antiferromagnetic (AFM) long-range ordered phases, gapped SPT phases, and two distinct gapless spiral phases related by a duality transformation, as further confirmed numerically through the scaling behavior of order parameters and the energy spectra. Moreover, the phase diagram reveals four distinct phase transition lines. Two of them separate the AFM and SPT phases, as well as the two gapless spiral phases, and are described by the Ising conformal field theory (CFT) with central charge $c=1/2$ and the three-copy Ising CFT with $c=3/2$, respectively. In contrast, the remaining two transition lines, which separate the gapless spiral phases from the gapped ones, correspond to a nonconformal Lifshitz criticality characterized by a dynamical critical exponent $z=2$. Strikingly, we unambiguously demonstrate that the intersection point of these four transition lines corresponds to a novel Lifshitz multicritical point with emergent conformal symmetry—fundamentally distinct from the conventional nonconformal Lifshitz points previously studied.

The remainder of this paper is organized as follows. In Sec.\ref{sec:model}, we introduce the model and outline the analytical treatment of the Hamiltonian. To provide a quick overview of the key physics, we summarize the ground-state phase diagram in Sec.~\ref{sec:phase}. In Sec.\ref{sec:order}, we identify the distinct gapped and gapless phases by examining various physical quantities. The phase transitions and critical behaviors between these phases are explored in Sec.\ref{sec:variation}. Finally, a summary and discussion are provided in Sec.~\ref{sec:summary}. Detailed analytical and numerical calculations are provided in the Appendix.

\section{MODEL AND METHOD}
\label{sec:model}
\begin{figure}[h]
    \centering
\includegraphics[width=1.0\linewidth]{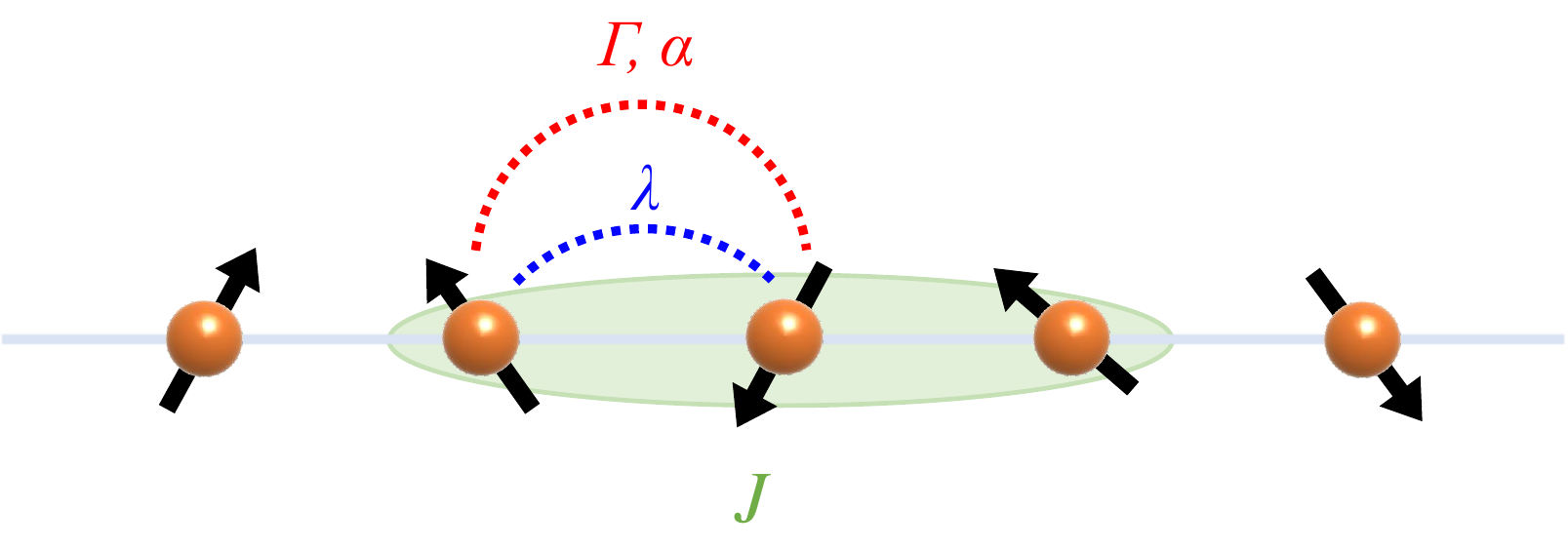}
    \caption{(Color Online) A schematic plot of the one-dimensional cluster-Ising model with off-diagonal Gamma interactions. The green-filled ellipsoids represent the three-spin cluster interaction $J$, the blue dotted lines denote the two-spin Ising interaction $\lambda$, and the red dashed lines indicate the off-diagonal Gamma interactions, including their strength $\Gamma$ and relative weight $\alpha$. In this work, we set $J = 1$ as the energy unit and fix $\Gamma = 0.5$.}
    \label{CImodel}
\end{figure}

To begin with, we introduce a cluster-Ising spin-1/2 chain with an off-diagonal Gamma interaction. The Hamiltonian of the model is given by (a schematic illustration of the model is shown in Fig.~\ref{CImodel}):
\begin{equation}\label{Hami}
\begin{aligned}
H=& -J\sum_{l=1}^N \sigma_{l-1}^x \sigma_l^z \sigma_{l+1}^x+\lambda \sum_{l=1}^N \sigma_l^y \sigma_{l+1}^y\\
+& \Gamma \sum_{l=1}^N (\sigma_{l}^{x}\sigma_{l+1}^{y}+\alpha\sigma_{l}^{y} \sigma_{l+1}^x),
\end{aligned}
\end{equation}
Here, $\sigma^{a}_{l}$ ($a = x, y, z$) denotes the Pauli matrix corresponding to the $a$-component of the $l$th spin. The parameters $J$, $\lambda$, $\Gamma$, and $\alpha$ respectively control the strengths of the three-spin cluster interaction, the two-spin Ising interaction, the Gamma interaction, and the relative weight of the off-diagonal exchange terms. When $\Gamma = \alpha = 0$, the model reduces to the one-dimensional cluster-Ising model, which is exactly solvable via a free-fermion mapping and exhibits rich SPT physics. Specifically, the system exhibits AFM long-range order when the Ising interaction dominates, and a gapped SPT phase when the cluster interaction is dominant. The competition between the cluster and Ising terms gives rise to nontrivial quantum critical points (QCPs) that host topologically protected edge states, corresponding to a special class of gSPT phases~\cite{Scaffidi2017PRX,Verresen2021PRX,Yu2022PRL}. For $\alpha = -1$ and $1$, the off-diagonal Gamma interaction reduces to the antisymmetric Dzyaloshinskii–Moriya (DM) interaction and its symmetric counterpart, both of which have been extensively studied over the past few decades~\cite{Kuepferling2023RMP,Yang2023NRP,Jafari2008PRB,Kargarian2009PRA,Ma2011PRA}. In this work, we focus on the interplay between SPT physics and off-diagonal exchange interactions and conduct a comprehensive study of the resulting exotic phenomena using exact solvable methods. Throughout the paper, we set $J = 1$ as the energy unit and fix $\Gamma = 0.5$ in the following numerical simulations.

To solve the interacting spin model in one dimension, we apply the Jordan–Wigner transformation to map Eq.~\eqref{Hami} into a fermionic representation. The transformation is defined as:
\begin{equation}\label{jw}
    \sigma_{l}^{z}=1-2c_{l}^{\dagger}c_{l},
\end{equation}
\begin{equation}
    \sigma_{l}^{+}=\prod_{j<l}(1-2c_{j}^{\dagger}c_{j})c_{l},
\end{equation}
where $c_l^{\dagger}$($c_l$) is the creation (annihilation) operator at site $l$. We then perform a Fourier transform:
\begin{equation}
c_l=\frac{1}{\sqrt{N}} \sum_{k} e^{-i k l} c_k.
\end{equation}
Substituting this into the Hamiltonian, we obtain its Bogoliubov–de Gennes (BdG) form:
\begin{equation}
\label{E1}
\begin{aligned}
\hat H_{k}&=\sum_{k}
\begin{pmatrix}
c_{k}^{\dagger} & c_{-k}\\
\end{pmatrix}
\begin{pmatrix}
A_{k} & B_{k}\\
B_{k}^{*} & -A_{-k}\\
\end{pmatrix}
\begin{pmatrix}
c_{k}\\
c_{-k}^{\dagger}
\end{pmatrix},
\end{aligned}
\end{equation}
where $A_{k}=\lambda\cos k-J\cos(2k)+\Gamma(\alpha-1)\sin k$ and $B_{k}=\Gamma(\alpha+1)\sin k+i[\lambda\sin k+J\sin(2k)]$. To diagonalize Eq.~\eqref{E1}, we perform a Bogoliubov transformation:
\begin{equation}
    \gamma_{k}=u_{k}c_{k}+v_{k}e^{i\varphi_{k}}c_{-k}^{\dagger},~\gamma_{k}^{\dagger}=u_{k}c_{k}^{\dagger}+v_{k}e^{-i\varphi_{k}}c_{-k},
\end{equation}
with the phase angle defined as $\varphi_{k}=\frac{\lambda\sin k+J\sin(2k)}{\Gamma(\alpha+1)\sin k}$. This yields the diagonalized Hamiltonian:
\begin{equation}
    H=\sum_{k} \varepsilon_{k}({\gamma}^{\dagger}_{k}\gamma_{k}-\frac{1}{2}),
\end{equation}
where the quasiparticle energy spectrum is given by:
\begin{equation}\label{lambdak}
\begin{aligned}
\varepsilon_{k}&=2\sqrt{x_{k}^{2}+y_{k}^{2}+z_{k}^{2}}+2\Gamma(\alpha-1)\sin k,
\end{aligned}
\end{equation}
with $x_{k}=\Gamma(\alpha+1)\sin k$, $y_{k}=\lambda\sin k+J\sin (2k)$, $z_{k}=\lambda\cos k-J\cos (2k)$. Based on these analytical results, explicit expressions for physical observables relevant to diagnosing different phases and phase transitions—such as order parameters, energy gaps, and entanglement entropy—are provided in Appendix~\ref{appA}.

\section{QUANTUM PHASE DIAGRAM}\label{sec:phase}
Before delving into the details of our analytical results, we summarize the main findings and present the global quantum phase diagram of the model defined in Eq.\eqref{Hami}. A schematic illustration of the phase diagram is shown in Fig.~\ref{fig2}. The tuning parameters $(\alpha,\lambda)$ drive the system into distinct gapped and gapless phases, which are summarized below (see also Table~\ref{tab1} for the key properties of each phase):

\begin{figure}[h]
    \centering
\includegraphics[width=1.0\linewidth]{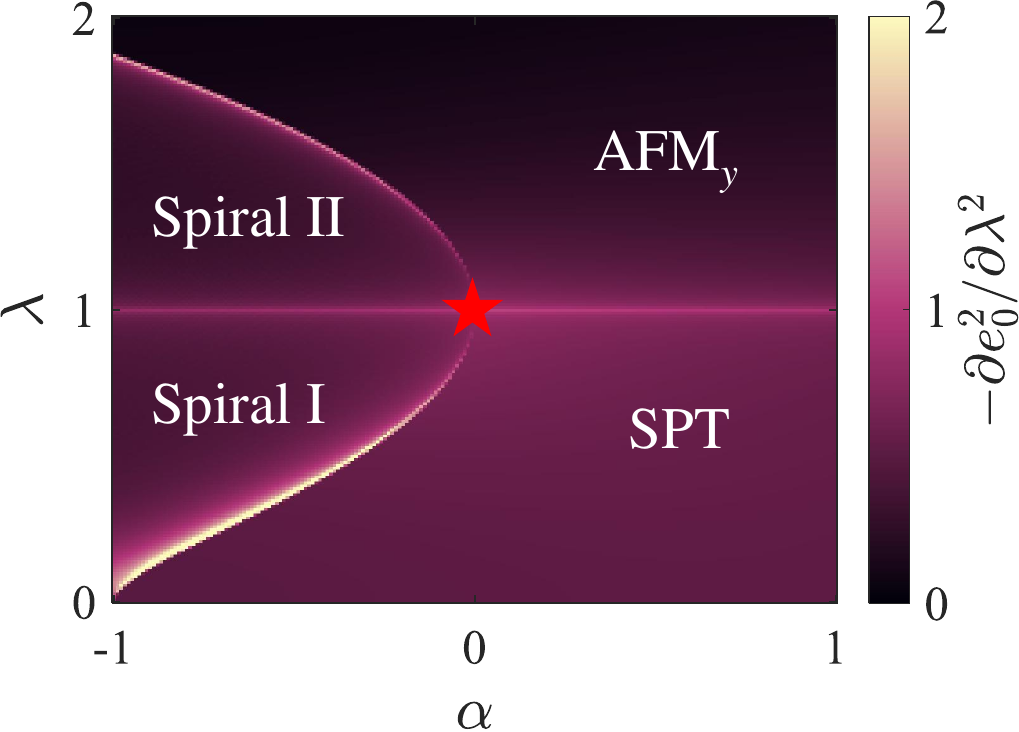}
    \caption{(Color Online) The global phase diagram of the cluster Ising chain with off-diagonal Gamma interactions is shown as a function of the tuning parameters ($\alpha$, $\lambda$), diagnosed by the second derivative of the ground-state energy density, $-\partial^2 e_0/\partial \lambda^{2}$. The diagram comprises four distinct regions: the cluster SPT phase, the $\text{AFM}_y$ phase, and two distinct gapless spiral phases (I and II). The phase transitions from SPT to $\text{AFM}_y$ and from Spiral I to Spiral II are conformal and described by the Ising and three-copy Ising CFTs, respectively. In contrast, the transitions between the gapless spiral phases and the gapped phases are associated with non-conformal Lifshitz criticality. However, the red star denotes a multicritical Lifshitz point featuring emergent conformal symmetry.}
    \label{fig2}
\end{figure}

(i) $\text{AFM}_y$: This gapped phase exhibits antiferromagnetic long-range order along the $y$-direction and spontaneously breaks the  $\mathbb{Z}_2$ spin-flip symmetry. It is characterized by a long-range order in spin-spin correlation function $|G_{yy}(r)|$, while the nonlocal string order parameter $|O_{x}(r)|$ decay exponentially at long distances. The entanglement entropy obeys an area law. 

(ii) SPT: This gapped phase hosts nontrivial symmetry-protected topological edge states. The spin-spin correlation function $|G_{yy}(r)|$ decays exponentially, while the string order parameter $|O_{x}(r)|$ exhibits long-range order at the long-distance limit. The entanglement entropy also follows an area law.

(iii) Spiral I: This gapless phase exhibits oscillatory power-law decay $\sim 1/\sqrt{r}$ in the vector chiral correlation $|O_{xy}(r)|$. The entanglement entropy follows a logarithmic scaling, indicative of a critical phase. In addition, the local spin order decays exponentially, while the string order decays as $\sim 1/\sqrt{r}$ with oscillations.

(iv) Spiral II: This is another gapless phase, also characterized by $\sim 1/\sqrt{r}$ oscillatory decay in the vector chiral correlations and logarithmic entanglement scaling. However, in contrast to Spiral I, the local spin order exhibits oscillatory power-law decay $\sim 1/\sqrt{r}$, while the string order decays exponentially. The behaviors of the order parameters in Spiral II are the dual counterparts of those in Spiral I, and the two phases are related by a duality transformation.

Intriguingly, these four quantum phases are separated by four distinct quantum critical lines:

(i) The conformal critical line between the gapped SPT phase and the $\text{AFM}_{y}$ phase belongs to the (1+1)D Ising universality class, characterized by a correlation length exponent $\nu=1$, anomalous dimension $\eta = 1/4$, and central charge $c=1/2$.

(ii) The transition between the two distinct gapless spiral phases described by three-copy Ising CFT, with critical exponents $\nu=1$, $\eta = 3/4$, and central charge $c=3/2$.

(iii) The transitions between the gapless spiral phases and the gapped phases (SPT or $\text{AFM}_y$) are not described by any known CFT. Instead, they correspond to non-conformal Lifshitz critical points with dynamical critical exponent $z=2$.

(iv) Remarkably, the multicritical point (indicated by the red star in Fig.~\ref{fig2}) at the intersection of all four critical lines represents a novel type of Lifshitz criticality with emergent conformal symmetry, fundamentally different from the non-conformal Lifshitz points described in (iii).

\section{Emergent gapless spiral phases with off-diagonal Gamma interaction}\label{sec:order}
After presenting the global phase diagram, this section provides a detailed characterization of the various quantum phases, diagnosed through the energy gap, entanglement entropy, and, in particular, the order parameters associated with each phase. The definitions and exact expressions of these physical quantities can be found in Appendix~\ref{appA}.

We first map out the global phase diagram by calculating the energy gap of the model for system sizes up to $N = 3000$, as shown in Fig.~\ref{fig8} in Appendix~\ref{appB}. The numerical results reveal three regions in the phase diagram:  a gapless phase (black region) that occupies a large portion of the diagram, predominantly on the $\alpha < 0$ side, and two gapped regions separated by gapless transition lines for $\alpha > 0$, indicating that the two gapped phases are qualitatively distinct. The nature—gapped or gapless—of these phases is further corroborated by the scaling behavior of the entanglement entropy. In the two gapped phases, the entanglement entropy obeys the area law, while in the gapless region, a logarithmic scaling $S_L \sim \frac{1}{3} \ln L$ emerges (see Fig.~\ref{fig4} (d) for $\alpha = -0.5$), indicating that the gapless phase is a critical phase described by a CFT with central charge $c = 1$.

\begin{figure}[tbhp] \centering
\includegraphics[width=8.5cm]{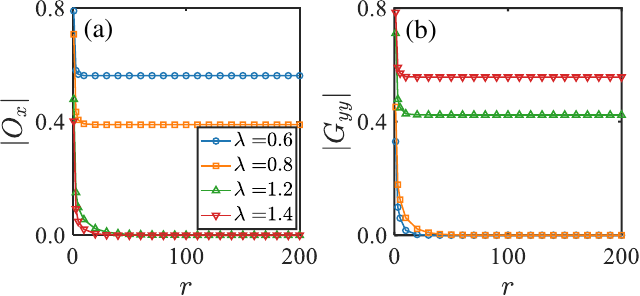}
\caption{(Color online). Long-distance behavior of order parameters in the gapped region for $\alpha=0.5$. The string order parameter $|O_{x}|$ is shown as a function of distance $r$ for various values of $\lambda$ in panel (a), while the spin correlation function $|G_{yy}|$ is plotted against $r$ in panel (b). Notably, when $\lambda<1.0$, the string order parameter displays long-range order, suggesting the presence of a cluster SPT order. Conversely, when $\lambda>1.0$, the spin correlation exhibits long-range order, indicating that the ground state features $\text{AFM}_y$ order. The simulated system size for panels (a)–(b) is $N =3000$.}
\label{fig3}
\end{figure}

\begin{figure}[tbhp] \centering
\includegraphics[width=8.5cm]{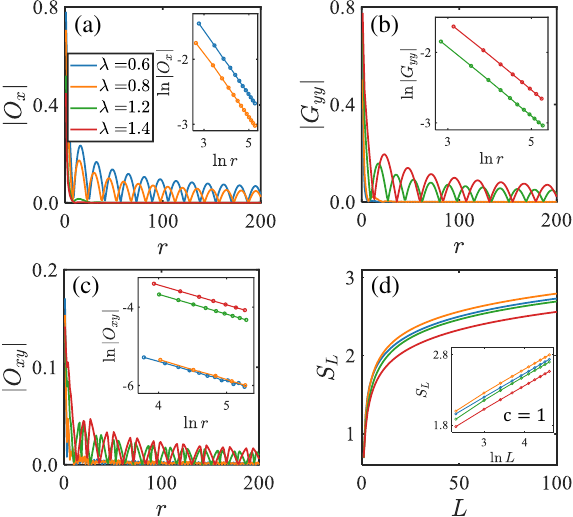}
\caption{(Color online). Long-distance behavior of order parameters and scaling of entanglement entropy in the gapless region for $\alpha = -0.5$. The string order parameter $|O_{x}|$ is plotted as a function of distance $r$ for different values of $\lambda$ in panel (a); the spin correlation function $|G_{yy}|$ is shown as a function of $r$ in panel (b); and the vector chiral order parameter $|O_{xy}|$ is plotted versus $r$ in panel (c). The insets in panels (a)–(c) display the corresponding data on a log-log scale, confirming power-law scaling as $1/\sqrt{r}$. Panel (d) shows the entanglement entropy $S_L$ as a function of subsystem size for various $\lambda$, with the inset showing the same data on a log-log scale, indicating a logarithmic scaling $S_L \sim (1/3)\ln L$. The simulated system size for panels (a)–(d) is $N = 3000$.}
\label{fig4}
\end{figure}

To further clarify the nature of these phases, we analyze the long-distance behavior of various order parameters, which include both local and nonlocal observables: the local spin-spin correlation function $|G_{yy}(r)|$ diagnoses the $\text{AFM}y$ long-range order while the vector-chiral order parameter $|O_{xy}(r)|$ characterizes the gapless spiral phases. The nonlocal string order parameter $|O_x(r)|$ detects topological order in SPT phases. The definitions and exact expressions of these quantities can be found in Appendix~\ref{appA}. Specifically, we first focus on the gapped region at $\alpha = 0.5$ and analytically compute the string order parameter $|O_x(r)|$ and the spin-spin correlation function $|G_{yy}(r)|$ as functions of the lattice distance $r$ for various values of $\lambda$, as shown in Fig.~\ref{fig3} (a) and (b). The results reveal that for $\lambda < 1.0$, $|O_x(r)|$ displays long-range string order, while $|G_{yy}(r)|$ exhibits exponential decay. Conversely, for $\lambda > 1.0$, $|G_{yy}(r)|$ remains finite at large distances, indicating long-range magnetic order, while $|O_x(r)|$ decays exponentially. These observations unambiguously demonstrate that the two gapped phases correspond to the SPT phase for $\lambda < 1.0$ and the $\text{AFM}_y$ phase for $\lambda > 1.0$.

In the gapless region, the physics becomes even more intriguing. We begin by calculating the vector chiral order parameter $|O_{xy}(r)|$ at selected points within the gapless phases for $\alpha = -0.5$. As shown in Fig.~\ref{fig4} (c), this order parameter exhibits a power-law decay with oscillations, scaling as $\sim 1/\sqrt{r}$ for both $\lambda > 1$ and $\lambda < 1$ (see the inset of Fig.~\ref{fig4} (c)). This behavior is consistent with the presence of gapless spiral phases induced by DM interactions. However, more interestingly, the local spin correlation and the nonlocal string order parameters display opposite behaviors depending on the sign of $\lambda-1.0$, as illustrated in Fig.~\ref{fig4} (a) and (b). Specifically, for $\lambda > 1.0$, the spin-spin correlation function exhibits power-law decay with oscillations, sharing the same exponent as the vector chiral order parameter, while the string order parameter decays exponentially. Conversely, for $\lambda < 1.0$, the string order parameter shows power-law decay with oscillations $\sim 1/\sqrt{r}$, while the spin-spin correlation decays exponentially. These opposite behaviors of local and nonlocal order parameters strongly suggest the existence of two distinct gapless phases (labeled Spiral I and Spiral II in the phase diagram), which are related by a duality transformation. In fact, the local spin and nonlocal string order parameters are related through the Kennedy-Tasaki transformation~\cite{Kennedy1992PRB,Oshikawa_1992,Li2025SciPost}, and the two dual gapless spiral phases are separated by a continuous phase transition line, as will be confirmed and discussed in detail in the next section.

Overall, the behaviors of the energy gap $\Delta$, entanglement entropy $S_{L}$, and various order parameters for each quantum phase are summarized in Table~\ref{tab1}.

\begin{ruledtabular}
\label{tab1}
\begin{table*}[hbtp]
\caption{A summary of the scaling behaviors of the energy gap $\Delta$, entanglement entropy $S_{L}$, and various types of order parameters—SPT ($|O_{x}(r)|$), $\text{AFM}{y}$ ($|G_{yy}(r)|$), and Spiral I/II ($|O_{xy}(r)|$)—in each region of the phase diagram.}
\label{tab1}
\begin{tabular}{cccccccc}
 & $\Delta$ & $|S_L|$ & $|O_{x}(r)|$ & $|G_{yy}(r)|$ & $|O_{xy}(r)|$  \\
\hline
 SPT & non-zero & constant & constant & exponential decay & 0 \\
 $\text{AFM}_y$ & non-zero & constant & exponential decay & constant & 0 \\
 Spiral I & 0 & $\sim \frac{1}{3} \ln L$ & oscillating decay as $r^{-1/2}$ & exponential decay & oscillating decay as $r^{-1/2}$ \\
 Spiral II & 0 & $\sim \frac{1}{3} \ln L$ & exponential decay & oscillating decay as $r^{-1/2}$ & oscillating decay as $r^{-1/2}$ \\
\end{tabular}
\end{table*}
\end{ruledtabular}

\section{Phase transitions and critical behaviors}
\label{sec:variation}
After delineating all the quantum phases in the phase diagram, we now turn our attention to the more intriguing QPTs between them. In the previous discussion, we mapped out the global phase diagram using the second derivative of the ground-state energy (Fig.~\ref{fig2}) and identified four distinct continuous phase transition lines separating the four quantum phases. In the following subsection, we will examine the critical behavior of these transitions in more detail.

\subsection{The phase transition lines along $\lambda=1.0$}
\label{sec:variation1}
\begin{figure}[tbhp] \centering
\includegraphics[width=8.5cm]{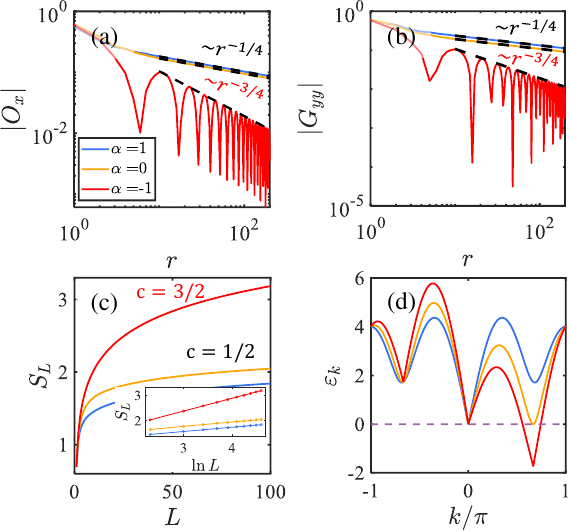}
\caption{(Color online). Long-distance behavior of order parameters, entanglement entropy scaling, and energy spectra along the phase transition line at $\lambda = 1$ for various values of $\alpha$. Panel (a) shows the string order parameter $|O_{x}|$ as a function of distance $r$ on a log-log scale for different $\alpha$ values. Panel (b) displays the spin correlation function $|G_{yy}|$ versus $r$, also on a log-log scale. In panel (c), the entanglement entropy $S_{L}$ is plotted against the subsystem size $L$. The inset highlights the logarithmic scaling behavior: $S_L \sim (1/6)\ln L$ for $\alpha = 1, 0$ and $S_L \sim (1/2)\ln L$ for $\alpha = -1$. Panel (d) shows the energy spectrum $\varepsilon_{k}$ as a function of momentum $k$. The simulated system size for panels (a)–(d) is $N = 3000$.}
\label{fig5}
\end{figure}

We begin by examining the phase transitions between either two gapless phases or two gapped phases, where the critical point is located at $\lambda = 1.0$. For the $\text{AFM}_y$ to SPT transition, we take $\alpha = 1.0$ as an example. At the critical point $\alpha = 1.0$, both the string order parameter $|O_x(r)|$ and the spin-spin correlation function $|G_{yy}(r)|$ exhibit critical power-law decay scaling as $1/r^{1/4}$, corresponding to a scaling dimension of $1/8$, which is consistent with the predictions of the Ising CFT, as shown by the blue lines in Fig.~\ref{fig5} (a) and (b). We also performed a finite-size scaling of the fidelity susceptibility at criticality (see Appendix~\ref{appC}) and extracted the correlation length exponent $\nu = 1.0$. Moreover, the entanglement entropy shows logarithmic scaling at the critical point, following $S_{L} \sim \frac{1}{6} \ln L$ (see the blue line in Fig.~\ref{fig5} (c)), indicating conformal invariance with central charge $c = 1/2$. Together, these numerical results unambiguity demonstrate that the transition between the two gapped phases belongs to the standard (1+1)-dimensional Ising universality class and is described by an Ising CFT with $c = 1/2$.

Conversely, the transition between the two dual gapless spiral phases (Spiral I and II) can be systematically investigated at $\alpha = -1.0$, where the off-diagonal Gamma interactions reduce to DM interactions. As discussed earlier, although both phases exhibit similar behaviors in the vector chiral order parameter, they are not smoothly connected; instead, a phase transition occurs between them, as evidenced by a sharp peak in the fidelity susceptibility near $\lambda = 1.0$ (see Appendix~\ref{appC}). At this critical point, different from the gapped-to-gapped transition, both the string order parameter and the spin-spin correlation function exhibit identical critical power-law decay with oscillations, scaling as $\sim 1/r^{3/4}$ (see the red lines in Fig.~\ref{fig5} (a) and (b)). This corresponds to a scaling dimension three times larger than that of the Ising CFT ($1/8$), suggesting that the critical point belongs to the three-copy Ising universality class. To further substantiate this, we calculate the entanglement entropy at $\lambda = 1.0$, $\alpha = -1.0$, and perform finite-size scaling. The entropy shows logarithmic scaling $S_L \sim \frac{1}{2} \ln L$ (see the red line in Fig.~\ref{fig5} (c)), indicating a conformal critical point with central charge $c = 3/2$, which is again three times that of the standard Ising CFT. Taken together, these numerical observations provide compelling evidence that the transition between the Spiral I and Spiral II phases belongs to the three-copy Ising universality class. The critical behaviors at other representative points along the transition lines are presented in Appendix~\ref{appD}, showing the same features as those at $\alpha = 1.0$ and $\alpha = -1.0$. This indicates that the entire transition line at $\lambda = 1.0$ is characterized by the Ising CFT for $\alpha > 0$ and by the three-copy Ising CFT for $\alpha < 0$.

To gain an intuitive understanding of why the conformal transitions between the $\text{AFM}_y$ and SPT phases, and between the gapless Spiral I and II phases, exhibit different central charges, we compute the single-particle energy spectra at these critical points using the free fermion models under Jordan-Wigner duality. The results are shown as the blue and red lines for $\alpha = 1.0$ and $\alpha = -1.0$, respectively, in Fig.~\ref{fig5} (d). The spectra unambiguity reveals that the number of linear fermionic dispersion points at the Fermi level $\varepsilon_{F} = 0$ is one for $\alpha = 1.0$ and three for $\alpha = -1.0$, each occurring at specific momenta. This indicates that the number of gapless fermionic modes at the Spiral I–II transition is three times greater than at the $\text{AFM}_y$–SPT transition. Since each gapless fermion mode contributes a central charge of $c = 1/2$ in the infrared limit, the total central charges at the respective critical points are $c = 1/2$ and $c = 3/2$, in agreement with the entanglement entropy results discussed earlier.



\subsection{Nonconformal Lishiftz transition between gapped and gapless phases}
\label{sec:variation2}
\begin{figure}[tbhp] \centering
\includegraphics[width=8.5cm]{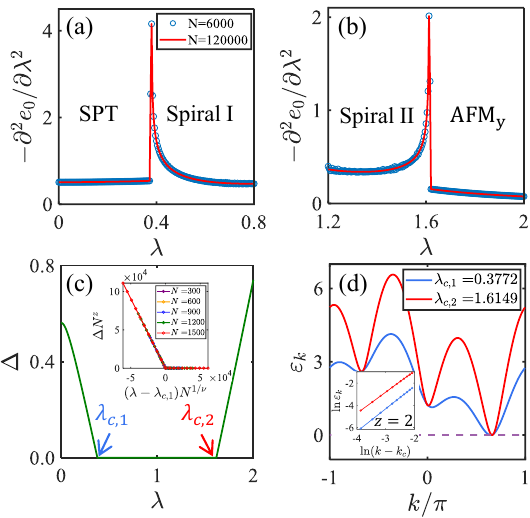}
\caption{(Color online). Critical behavior at the gapless-gapped phase transition points for $\alpha = -0.5$. Panel (a) shows the second derivative of the ground-state energy, $-\partial^{2} e_{0}/\partial \lambda^{2}$, as a function of $\lambda$ near the transition point between the SPT and Spiral I phases. Panel (b) displays the same quantity near the transition between the Spiral II and $\text{AFM}_y$ phases. The energy gap $\Delta$ as a function of $\lambda$ is presented in panel (c), while panel (d) shows the single-particle energy spectrum $\varepsilon_k$ as a function of momentum $k$. The inset in panel (c) presents the data collapse of the energy gap, confirming the critical exponents $\nu = 1/2$ and $z = 2$ at the transition point $\lambda_{c,1}$. The inset in panel (d) reveals a quadratic dispersion relation near the Fermi level, $\varepsilon_k \sim (k - k_c)^2$. The simulated system size for panels (c)–(d) is $N =3000$.}
\label{fig6}
\end{figure}

We now turn to the transitions between gapless and gapped phases, specifically the transitions from Spiral I to the SPT phase and from Spiral II to the $\text{AFM}_y$ phase. As a representative example, we examine the case of $\alpha = -0.5$, as illustrated in the main text (additional data are provided in Appendix~\ref{appB}). As a first step, we accurately locate the critical points between the gapless and gapped phases by analyzing the second derivative of the ground-state energy density, as shown in Fig.~\ref{fig6} (a) and (b). The numerical results reveal two distinct peaks in $-\partial^{2} e_{0}/\partial \lambda^{2}$ at fixed $\alpha = -0.5$, indicating the presence of two critical points, $\lambda_{c,1}$ and $\lambda_{c,2}$, which separate the SPT and Spiral I phases, as well as the $\text{AFM}_y$ and Spiral II phases, respectively. 
To further characterize the nature of these critical points, we compute the energy gap as a function of $\lambda$ at fixed $\alpha = -0.5$, shown in Fig.~\ref{fig6} (c). The results unambiguity show that the energy gap closes exactly at $\lambda_{c,1}$ and $\lambda_{c,2}$. The insets show the data collapse of the energy gap near the critical point $\lambda_{c,1}$ (the corresponding collapse near $\lambda_{c,2}$ is provided in Appendix~\ref{appB}), following $\Delta N^{z} =( \lambda - \lambda_c)N^{1/\nu}$, which confirms that both transitions are nonconformal Lifshitz critical points with a dynamical exponent $z = 2$ and a correlation length exponent $\nu = 1/2$. To provide further evidence, we compute the single-particle energy spectrum at both nonconformal critical points (see Fig.~\ref{fig6} (d)). The results reveal a purely quadratic (parabolic) dispersion near the Fermi level $\varepsilon_{F} = 0$, indicating the absence of relativistic low-energy fermions and confirming that the dynamical exponent at both critical points is indeed $z = 2$.

\subsection{The novel Lifshitz multicritical point with emergent conformal symmetry}
\label{sec:variation3}

\begin{figure*}[tbhp] \centering
\includegraphics[width=16cm]{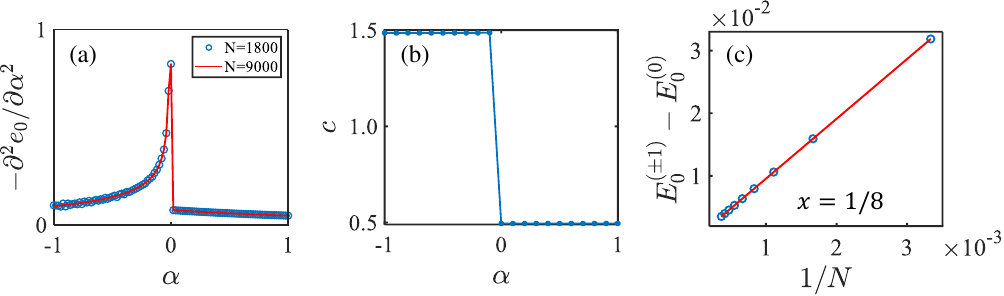}
\caption{(Color online). Confirmation of emergent conformal symmetry at the multicritical point. Panel (a) shows the second derivative of the ground-state energy, $-\frac{\partial^{2} e_{0}}{\partial \alpha^{2}}$, plotted as a function of $\alpha$ along the transition line $\lambda = 1.0$ for different system sizes $N$. Panel (b) presents the central charge $c$ as a function of $\alpha$ along the same transition line. In panel (c), the finite-size scaling of the free energy difference $E_0^{(\pm1)} - E_0^{(0)}$ is shown, which perfectly follows the scaling relation predicted by CFT (see Appendix~\ref{appB} for details). The simulated system size for panels (b) is $N = 3000$.}
\label{fig7}
\end{figure*}
The most striking phenomenon occurs at the multicritical point where the four distinct phase transition lines discussed above intersect. To elaborate on this point, we fix $\lambda = 1.0$ and vary $\alpha$ to locate the position of the multicritical point. Specifically, we compute the second derivative of the ground-state energy density (see Fig.~\ref{fig7} (a)) and observe a sharp peak at $\alpha = 0.0$, indicating the presence of a continuous phase transition. Furthermore, we calculate the central charge as a function of $\alpha$ along the transition line at $\lambda = 1.0$, as shown in Fig.~\ref{fig7} (b). The results reveal a clear jump in the central charge from $c = 3/2$ to $c = 1/2$ near $\alpha = 0.0$, indicating a phase transition of Ising transition characterized by different central charges. At the multicritical point, we also compute various order parameters and the entanglement entropy, shown by the orange curves in Fig.~\ref{fig5} (a)–(c). Both the string order parameter and the spin-spin correlation function exhibit critical power-law decay scaling as $1/r^{1/4}$, corresponding to a scaling dimension consistent with the prediction of the (1+1)D Ising CFT. This is further confirmed by the entanglement entropy, which shows logarithmic scaling with central charge $c = 1/2$ at the same multicritical point (see the orange line in Fig.~\ref{fig5} (c)). In Fig.~\ref{fig5} (d), we also analyze the single-particle energy spectrum at the multicritical point (orange line) and find one linear and one quadratic (parabolic) mode near the Fermi level, located at momenta $k = 0$ and $k = 2\pi/3$, respectively. This suggests that the multicritical point corresponds to a Lifshitz transition with emergent relativistic fermionic modes at low energies, implying the possible emergence of conformal symmetry. To further confirm the presence of conformal symmetry at the multicritical point, we compute the free energy of the system (see the Appendix~\ref{appB} for details). As shown in Fig.~\ref{fig7} (c), the free energy exhibits a perfect scaling relation predicted by CFT~\cite{ginsparg1988applied,francesco2012conformal,Alcaraz2024PRE}, and the extracted scaling dimension is consistent with the Ising CFT. This provides strong evidence for emergent conformal symmetry at the Lifshitz multicritical point and identifies a novel type of Lifshitz transition that is fundamentally different from the conventional (nonconformal) Lifshitz transitions discussed in the previous section.

\section{CONCLUSION AND OUTLOOK}
\label{sec:summary}
In summary, we have comprehensively explored the interplay between SPT physics and off-diagonal interactions, revealing novel quantum phases and phase transitions in an exactly solvable cluster Ising chain with a Gamma exchange interaction. By applying the Jordan-Wigner transformation, we map out a global phase diagram, which is diagnosed using the second derivative of the ground-state energy density and energy gap. The system exhibits distinct gapped and gapless phases depending on the sign of the relative weight of the off-diagonal interaction $\alpha$. For $\alpha > 0$, the system displays a gapped SPT phase for $\lambda < 1$ and a long-range ordered $\text{AFM}_{y}$ phase for $\lambda > 1$, as verified numerically via the scaling behaviors of order parameters and entanglement entropy. These two gapped phases are separated by a continuous phase transition line described by the standard (1+1)D Ising CFT. In contrast, for $\alpha < 0$, tuning $\lambda$ leads to the emergence of two distinct gapless intermediate phases—referred to as Spiral I and Spiral II—between the SPT and $\text{AFM}_{y}$ phases. These spiral phases exhibit power-law decay of correlations with oscillations, $|O_{xy}(r)| \sim 1/\sqrt{r}$, and a logarithmic scaling of entanglement entropy, $S_L \sim \frac{1}{3} \ln L$. Interestingly, the spin correlation function $|G_{yy}(r)|$ and the string order parameter $|O_{x}(r)|$ show opposite scaling behaviors, indicating that these gapless phases are qualitatively distinct and related by the Kennedy-Tasaki duality transformation. These dual-related gapless spiral phases undergo a continuous phase transition that belongs to the three-copy Ising universality class, characterized by three linearly dispersing fermionic modes at low energy, as supported by various numerical simulations. The transitions between the gapped and gapless phases—such as $\text{AFM}_{y}$–Spiral II and SPT–Spiral I—are found to exhibit nonconformal Lifshitz criticality with dynamical exponent $z = 2$, where the low-energy spectra are dominated by a quadratic dispersion fermion. Most remarkably, at the multicritical point $(\alpha = 0.0, \lambda = 1.0)$, where four distinct phase transition lines intersect, we numerically identify a novel type of Lifshitz criticality that exhibits emergent conformal symmetry. At this point, the low-energy spectra consist of one linear mode at momentum $k = 0$ and one quadratic mode at $k = 2\pi/3$, and the system displays a scaling dimension of $1/8$ consistent with the (1+1)D Ising CFT. Looking ahead, an intriguing direction for future research is to explore the topological properties of the gapless regions induced by the off-diagonal $\Gamma$-exchange interactions and to examine the stability of these gapless phases against interactions and disorder. Our work provides valuable insights into studying novel quantum phases and critical phenomena in exactly solvable many-body systems.

\begin{acknowledgments}
X.-J. Yu was supported by the National Natural Science Foundation of China (Grant No.12405034). This work was supported by the National Key Research and Development Program of China (Grant No.~2022YFA1405300). Zhi Li was supported by Open Fund of Key Laboratory of Atomic and Subatomic Structure and Quantum Control (Ministry of Education). \\

\end{acknowledgments}

\appendix

\section{DIAGNOSTIC PHYSICAL QUANTITIES FOR QUANTUM PHASES AND TRANSITIONS}
\label{appA}
\subsection{Order parameters}
The most compelling evidence for identifying various quantum phases is analyzing appropriate order parameters and their long-distance behaviors. The first type of local order parameter is the spin-spin correlation function, which can diagnose antiferromagnetic (AFM) long-range order and is defined as
\begin{equation}
\label{appA_E1}
G_{\alpha\beta}(r) = \left\langle \sigma_{j}^{\alpha} \sigma_{l}^{\beta} \right\rangle,
\end{equation}
where $r = j - l$ and $\alpha = x,~y,~z$. In our model, we identify AFM long-range order along the $y$-direction. In this case, the spin correlation function $|G_{yy}(r)|$ saturates to a constant as $r \to \infty$, indicating long-range order. In contrast, it decays exponentially in symmetric gapped regions without such order.

On the other hand, the gapped cluster SPT phase—characterized by the absence of spontaneous symmetry breaking—cannot be captured by local order parameters. Instead, it can be diagnosed by a non-local string order parameter~\cite{Pollmann2012PRB, Verresen2017PRB}, defined as
\begin{equation}
\label{appA_E2}
O_{x}(r) = \lim_{r \rightarrow \infty} (-1)^{r} \left\langle \sigma_{1}^{x} \sigma_{2}^{y} \left( \prod_{k=3}^{r} \sigma_{k}^{z} \right) \sigma_{r+1}^{y} \sigma_{r+2}^{x} \right\rangle,
\end{equation}
which exhibits long-range order in the nontrivial cluster SPT phase and decays exponentially in the $\text{AFM}_{y}$ phase.

In addition to the gapped phases, the gapless spiral phases induced by off-diagonal Gamma interactions can also be diagnosed by the behavior of a local vector chiral order parameter, defined as
\begin{equation}
\label{appA_E3}
O_{xy}(r) = \left| G_{xy}(r) \right| - \left| G_{yx}(r) \right|.
\end{equation}
This quantity vanishes in the gapped phases but shows a power-law decay with oscillations in the gapless spiral phases.

\subsection{The second derivative of ground state energy density and energy gap}
Thanks to the exact solvability of the spin chain model discussed in the main text, we can analytically obtain the single-particle excitation spectrum $\varepsilon_k$ of the dual free-fermion model, and define the energy gap as $\Delta = \min_{k} \varepsilon_k$. This quantity is diagnostic for distinguishing gapped and gapless quantum phases, and a vanishing gap $\Delta = 0$ indicates a continuous phase transition.

Complementarily, continuous phase transition points can also be identified by analyzing the second derivative of the ground-state energy density, defined as
\begin{equation}
\frac{\partial^2 e_0}{\partial \lambda^2}, \quad \text{where} \quad e_0 = -\frac{1}{N} \sum_k |\varepsilon_k|.
\end{equation}
Specifically, this second derivative exhibits a pronounced peak near a continuous critical point, and the sharpness of the peak increases with system size, providing a reliable numerical signature of continuous phase transitions.

\subsection{Entanglement entropy and central charge}
Quantum entanglement serves as a powerful tool for characterizing quantum phases and phase transitions. A widely used measure of entanglement is the entanglement entropy, defined as
\begin{equation}
\label{appA_E4}
S_{L} = -\mathrm{Tr}\left[\rho_{L} \ln \rho_{L}\right],
\end{equation}
where $\rho_{L}$ is the reduced density matrix of a subsystem consisting of the first $L$ sites, i.e., $\{1, \dots, L\}$.

Since the Hamiltonian in Eq.~(\ref{Hami}) is quadratic and exactly solvable, the entanglement entropy $S_{L}$ can be efficiently computed using the correlation matrix method~\cite{Vidal2003PRL,Peschel_2009}:
\begin{equation}
\label{appA_E5}
S_{L} = -\sum_{j=1}^{L} \left[ v_{j} \ln v_{j} + (1 - v_{j}) \ln (1 - v_{j}) \right],
\end{equation}
where $v_j$ are the eigenvalues of the correlation matrix restricted to the subsystem.

According to conformal field theory (CFT), in (1+1)-dimensional critical systems with periodic boundary conditions (PBC), the entanglement entropy exhibits a universal logarithmic scaling~\cite{ginsparg1988applied,francesco2012conformal}:
\begin{equation}
\label{appA_E6}
S_{L} \sim \frac{c}{3} \ln L,
\end{equation}
where $c$ is the central charge. For a broad class of conformally invariant quantum critical points, the central charge $c$ characterizes the universality class of the phase transition. Notably, this logarithmic scaling of entanglement entropy also holds in certain gapless critical phases described by CFT.

\subsection{Conformal data extracted from free energy scaling}
\begin{figure*}[tbhp] \centering
\includegraphics[width=15cm]{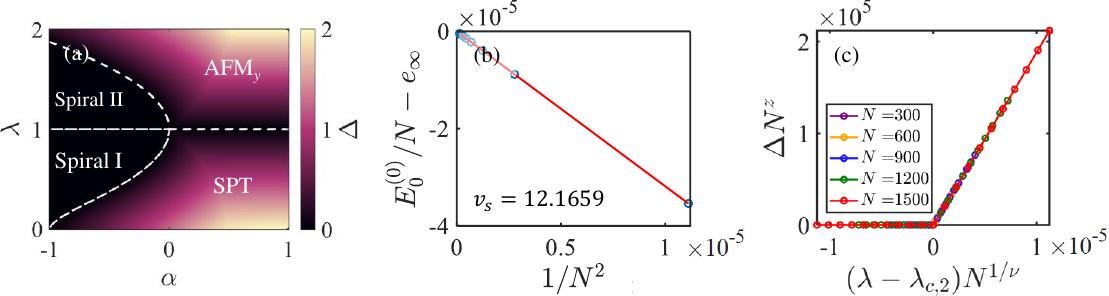}
\caption{(Color online). Panel (a) shows the phase diagram determined by the energy gap $\Delta$. The white dashed lines serve as guides to separate different phases, as identified by other physical observables. Panel (b) presents the finite-size scaling behavior of the ground-state energy density at the multicritical point ($\alpha = 0,~\lambda = 1$). The data is fitted using Eq.~(\ref{appA_E7}), yielding the sound velocity $v_{s} = 12.1659$. Panel (c) shows the data collapse of the energy gap near the critical point $\lambda_{c,2}$ for $\alpha = -0.5$, confirming the critical exponents $\nu = 1/2$ and $z = 2$, which characterize the transition between the Spiral II and $\text{AFM}_y$ phases.}
\label{fig8}
\end{figure*}
\begin{figure*}[!htb] \centering
\includegraphics[width=15cm]{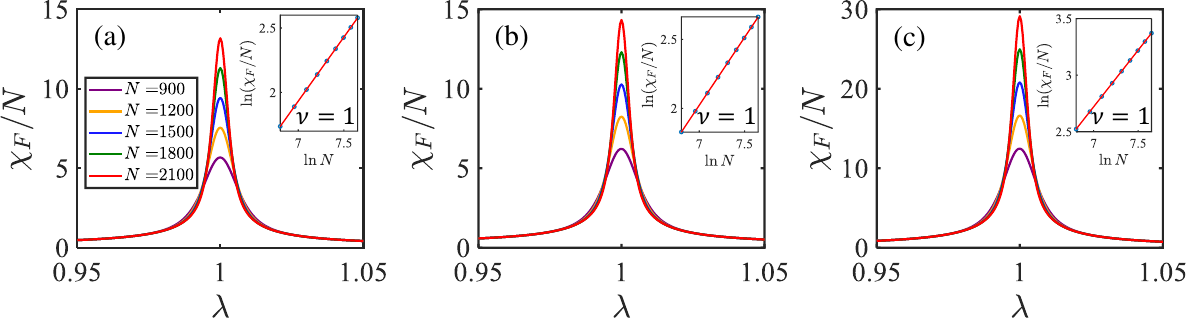}
\caption{(Color online). The fidelity susceptibility per site $\chi_{F}/N$ versus $\lambda$ for (a) $\alpha=1$, (b) $\alpha=0$, (c) $\alpha=-1$. The insets show the log-log plot of the fidelity susceptibility per site $\chi_{F}/N$ against the system sizes $N$ at the critical point $\lambda=1$, and the correlation length critical exponent $\nu=1$ can be inferred from the slope of the fitted straight line.}
\label{fig9}
\end{figure*}

To verify conformal invariance and extract universal data at a conformally invariant quantum critical point, we compute the free energy of the system and fit it using the universal scaling relations predicted by CFT~\cite{ginsparg1988applied,francesco2012conformal,Alcaraz2024PRE}. In our case, since we consider quantum phase transitions at zero temperature, the free energy is equivalent to the eigenenergy of the system.

For a conformally invariant critical system with PBC, the ground-state energy $E^{(0)}_0(N)$ exhibits the following asymptotic finite-size scaling behavior:
\begin{equation}
\label{appA_E7}
\frac{E^{(0)}_0(N)}{N} = e_\infty - \frac{\pi v_s c}{6N^2} + O(N^{-2}),
\end{equation}
where $e_\infty$ is the ground-state energy per site in the thermodynamic limit, $c$ is the central charge, and $v_s$ is the sound velocity, which can be extracted from the energy-momentum dispersion relation.

In practice, we numerically compute the ground-state energy $E^{(0)}_0(N)$ for various system sizes and perform a finite-size scaling analysis to determine $v_s$. With $v_s$ in hand, we can further extract conformal data at criticality, such as the scaling dimensions of primary operators. Specifically, we compute the excitation energy in the sectors with quantum number $Q = \pm 1$, relative to the ground-state sector $Q = 0$, as follows:
\begin{equation}
\label{appA_E8}
E^{(\pm 1)}_0 - E^{(0)}_0 = \frac{2\pi v_s}{N} x + O(N^{-2}),
\end{equation}
where $x$ is the scaling dimension of a conformal operator. The agreement of the numerical data with this scaling form confirms the conformal invariance of the system and enables the identification of the corresponding scaling dimension of the conformal operator.

\section{Complementary numerical results to the main text}
\label{appB}
In this section, we provide complementary numerical results to illustrate further the key physics presented in the main text.

Using the definition of the energy gap provided in Appendix~\ref{appA}, we map out the global phase diagram based on this quantity, as shown in Fig.~\ref{fig8}(a). The results demonstrate that the phase diagram consists of two distinct gapped phases, separated by a gap-closing transition line, and a gapless phase that occupies a large region of the parameter space.

To verify conformal invariance and extract conformal data at the multicritical point, we follow the CFT scaling relation for eigenenergies detailed in Appendix~\ref{appA}. Specifically, we numerically extract the sound velocity $v_s$ by performing finite-size scaling of the ground-state energy density $E^{(0)}_{0}$, as illustrated in Fig.~\ref{fig8}(b). The data exhibits linear scaling behavior consistent with CFT predictions, from which we obtain $v_s = 12.1659$ for our model.

\begin{figure*}[tbhp] \centering
\includegraphics[width=15cm]{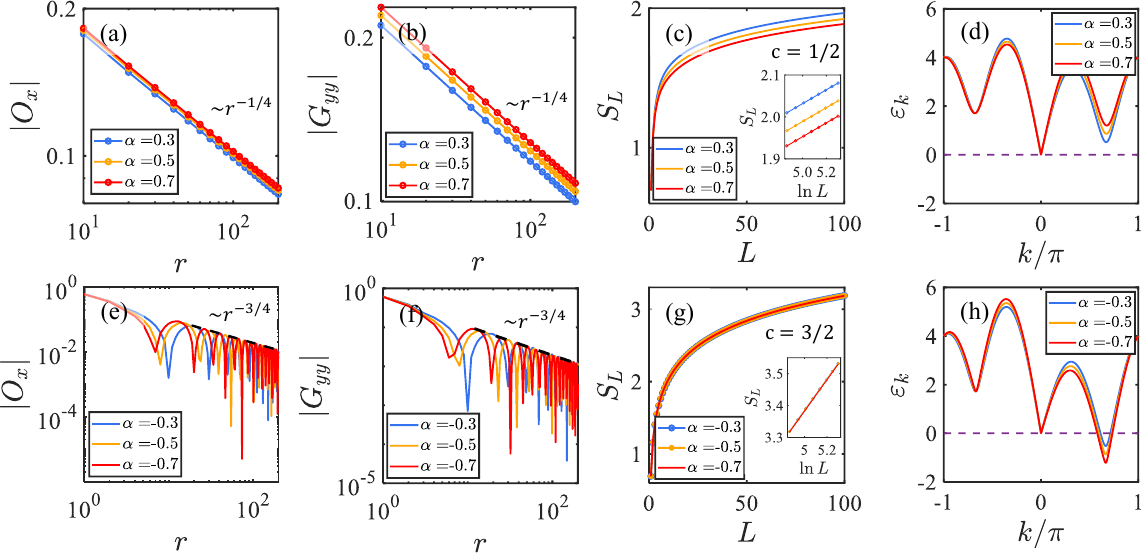}
\caption{(Color online). Long-distance behavior of order parameters, entanglement entropy scaling, and energy spectra along the SPT-$\text{AFM}_y$ transitions ((a)-(d)) and spiral I-spiral II ((e)-(h)) phase transition line at $\lambda = 1.0$ for various values of $\alpha$. Panel (a) shows the string order parameter $|O_{x}|$ as a function of distance $r$ on a log-log scale for different $\alpha$ values. Panel (b) displays the spin correlation function $|G_{yy}|$ versus $r$, also on a log-log scale. In panel (c), the entanglement entropy $S_{L}$ is plotted against the subsystem size $L$. The inset highlights the logarithmic scaling behavior: $S_L \sim (1/6)\ln L$ for $\alpha = 0.3,~0.5,~0.7$. Panel (d) shows the energy spectrum $\varepsilon_{k}$ as a function of momentum $k$. Panels (e)-(h) display the same quantity at the spiral I-spiral II phase transition line with $\alpha=-0.3,~-0.5,~-0.7$. The simulated system size for panels (a)–(h) is $N = 3000$.}
\label{fig10}
\end{figure*}

Lastly, while the main text focuses on the critical behavior at the transition point $\lambda_{c,1}$ between the SPT and Spiral I phases (for $\alpha = -0.5$), here we analyze another gapless-to-gapped phase transition, namely the Spiral II-$\text{AFM}_y$ transition. The data collapse for the energy gap $\Delta$ is presented in Fig.~\ref{fig8}(c), which follows the universal scaling form $\Delta L^z = (\lambda - \lambda_{c,2}) L^{1/\nu}$. From the fitting, we extract the correlation length exponent $\nu = 1/2$ and dynamical critical exponent $z = 2$, indicating that both gapless-to-gapped transitions are characterized by non-conformal Lifshitz criticality.

\section{Finite-size scaling of the fidelity susceptibility along the phase transition line $\lambda=1$}
\label{appC}
In this section, we detect and extract the critical exponent of the correlation length, $\nu$, along the phase transition line $\lambda = 1$, using the finite-size scaling of the fidelity susceptibility. The quantum ground-state fidelity $F(\lambda, \lambda + \delta \lambda)$, which quantifies the similarity between two ground states $\ket{G(\lambda)}$ and $\ket{G(\lambda + \delta \lambda)}$, is defined as
\begin{equation}
\label{appC_E1}
F(\lambda, \lambda + \delta \lambda) = \langle G(\lambda) | G(\lambda + \delta \lambda) \rangle.
\end{equation}
Near the quantum critical point $\lambda_c$, the ground states before and after crossing $\lambda_c$ differ significantly, leading to a substantial drop in fidelity, i.e., $F(\lambda_c, \lambda_c + \delta \lambda) \sim 0$. To probe quantum phase transitions more precisely, we introduce the fidelity susceptibility by expanding the fidelity to the second order:
\begin{equation}
\label{appC_E2}
\chi_F = \lim_{\delta \lambda \to 0} \frac{2[1 - F(\lambda, \lambda + \delta \lambda)]}{(\delta \lambda)^2}.
\end{equation}

For a continuous quantum phase transition in a finite system of size $N$, the fidelity susceptibility $\chi_F(\lambda)$ exhibits a sharp peak at the critical point. Near the criticality, it follows the finite-size scaling form:
\begin{equation}
\label{appC_E3}
\chi_F(\lambda_{c}(N))/N^{d} = N^{2/\nu - d},
\end{equation}
where $\nu$ is the correlation length exponent, $d$ is the spatial dimension of the system, and $\lambda_{c}(N)$ is the pseudocritical point with finite system size $N$.

Along the transition line $\lambda = 1.0$, we compute the fidelity susceptibility and perform finite-size scaling to extract the critical exponent $\nu$, based on the scaling relation in Eq.~(\ref{appC_E3}) for various values of $\alpha = 1,~0,~-1$, as shown in Fig.~\ref{fig9}. The results reveal a pronounced peak at $\lambda_c = 1$, which becomes increasingly sharp with growing system size $N$, indicating a continuous phase transition. The insets of Fig.~\ref{fig9} demonstrate that the correlation length exponent along this transition line is $\nu = 1.0$.

\section{Numerical results for additional selected points along the four distinct transition lines.}
\label{appD}
\subsection{The critical behavior of transition line between SPT-$\text{AFM}_y$ phases}
\label{appD1}
In this section, we present additional numerical results on the critical behavior of $\lambda=1.0$, $0.0<\alpha<1.0$, i.e. transition line between SPT-$\text{AFM}_y$ phases.

\begin{figure*}[tbhp]
\includegraphics[width=16cm]{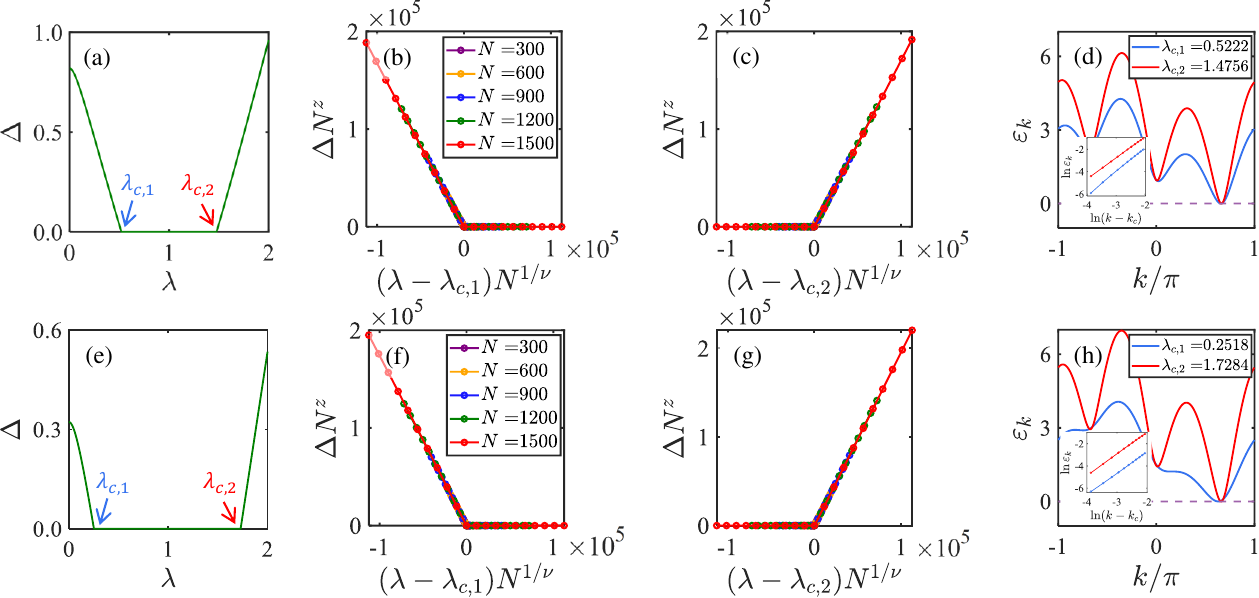}
\caption{(Color online). Critical behavior at the gapless-gapped phase transition points for $\alpha = -0.3$ (a)-(d), $\alpha = -0.7$ (e)-(h). Panel (a) shows the energy gap $\Delta$ as a function of $\lambda$. $\lambda_{c,1}$ corresponds to SPT-Spiral I transition and $\lambda_{c,2}$ corresponds to Spiral II-$\text{AFM}_{y}$ transition, respectively. Panel (b) presents the data collapse of the energy gap, confirming the critical exponents $\nu = 1/2$ and $z = 2$ at the transition point $\lambda_{c,1}$. Panel (c) displays the same quantity near the transition between the Spiral II and $\text{AFM}_y$ phases. Panel (d) shows the single-particle energy spectrum $\varepsilon_k$ as a function of momentum $k$. The inset in panel (d) reveals a quadratic dispersion relation near the Fermi level, $\varepsilon_k \sim (k - k_c)^2$. Panels (e)-(h) display the same quantity near the transition between the gapless and gapped phases with $\alpha = -0.7$. The simulated system size for panels (a)(e)(d)(h) is $N =3000$.}
\label{fig11}
\end{figure*}

At the selected points $\alpha = 0.3$, $0.5$, and $0.7$, both the string order parameter $|O_x(r)|$ and the spin-spin correlation function $|G_{yy}(r)|$ exhibit critical power-law decay, scaling as $1/r^{1/4}$ (see Fig.~\ref{fig10} (a) and (b)), which is consistent with the predicted scaling dimension $1/8$ from the Ising CFT. Moreover, the entanglement entropy at the critical point displays a logarithmic scaling behavior, $S_{L} \sim \frac{1}{6} \ln L$ (see Fig.~\ref{fig10} (c)), indicating conformal invariance with central charge $c = 1/2$. Taken together, these numerical results demonstrate that the transition between the SPT and $\text{AFM}_y$ phases belongs to the standard (1+1)-dimensional Ising universality class and is governed by an Ising CFT with $c = 1/2$. In addition, the energy spectra clearly show that the number of linearly dispersing fermionic modes at the Fermi energy $\varepsilon_{F} = 0$ is one for each of the cases $\alpha = 0.3$, $0.5$ and $0.7$, with the linear crossing occurring at momentum $k = 0$, as shown in Fig.~\ref{fig10} (d).

\subsection{The critical behavior of transition line between Spiral I-Spiral II phases}
\label{appD2}
In this section, we present additional numerical results on the critical behavior along the transition line $\lambda = 1.0$, for $-1.0 < \alpha < 0.0$, corresponding to the phase boundary between the gapless Spiral I and Spiral II phases.

At the selected values $\alpha = -0.3$, $-0.5$ and $-0.7$, both the string order parameter $|O_x(r)|$ and the spin-spin correlation function $|G_{yy}(r)|$ exhibit critical power-law decay with oscillatory behavior, scaling as $1/r^{3/4}$ (see Fig.~\ref{fig10} (e) and (f)). This scaling differs significantly from the $1/r^{1/4}$ decay expected from a single-copy (1+1)-dimensional Ising CFT. Moreover, the entanglement entropy at these critical points follows the logarithmic scaling $S_L \sim \frac{1}{2} \ln L$ (see Fig.~\ref{fig10} (g)),  indicating that the critical points are conformally invariant with a central charge $c = 3/2$. These numerical results demonstrate that the Spiral I–Spiral II transition belongs to the three-copy Ising universality class. In addition, the energy spectra unambiguously show that the number of linearly dispersing fermionic modes at the Fermi level $\varepsilon_{F} = 0$ is three for each case, with the linear crossings shown explicitly in Fig.~\ref{fig10} (h).\\

\subsection{The critical behavior of transition lines between gapped-gapless phases}
\label{appD3}
In this section, we present additional numerical results for more selected points along the transition lin between gapped-gapless phases for $\alpha=-0.3, ~-0.7$, i.e., SPT- Spiral I or $\text{AFM}_{y}$-Spiral II transition. 

For given $\alpha$, we located transition point between gapped and gapless phases through the energy gap evolution with $\lambda$. As shown in Fig.~\ref{fig11} (a) (for $\alpha = -0.3$) and (e) (for $\alpha=-0.7$), which clearly observe two gap-closing point has been emerge, correspondting to SPT-Spiral I ($\lambda_{c,1}$) and $\text{AFM}_{y}$-Spiral II transition ($\lambda_{c,2}$), respectively. Furthermore, we performed data collapse of energy gap at both transition points for each $\alpha$, as shown in Fig.~\ref{fig11} (b), (c) for $\alpha=-0.3$ and (f) and (g) for $\alpha=-0.7$. Both numerical results unambigulity demonstrate the transition line between gapped and gapless phases are characterized by nonconformal Lifshitz criticality with correlation length exponent $\nu=1/2$ and dynamical exponent $z=2$. These nonconformal nature can also be revealed by the quadratic fermionic dispersion at Fermi level $\varepsilon_{F} = 0$ for $\alpha=-0.3$ in Fig.~\ref{fig11} (d) and for $\alpha=-0.7$ in Fig.~\ref{fig11} (h).

\bibliography{main}

\begin{thebibliography}{115}%
\makeatletter
\providecommand \@ifxundefined [1]{%
 \@ifx{#1\undefined}
}%
\providecommand \@ifnum [1]{%
 \ifnum #1\expandafter \@firstoftwo
 \else \expandafter \@secondoftwo
 \fi
}%
\providecommand \@ifx [1]{%
 \ifx #1\expandafter \@firstoftwo
 \else \expandafter \@secondoftwo
 \fi
}%
\providecommand \natexlab [1]{#1}%
\providecommand \enquote  [1]{``#1''}%
\providecommand \bibnamefont  [1]{#1}%
\providecommand \bibfnamefont [1]{#1}%
\providecommand \citenamefont [1]{#1}%
\providecommand \href@noop [0]{\@secondoftwo}%
\providecommand \href [0]{\begingroup \@sanitize@url \@href}%
\providecommand \@href[1]{\@@startlink{#1}\@@href}%
\providecommand \@@href[1]{\endgroup#1\@@endlink}%
\providecommand \@sanitize@url [0]{\catcode `\\12\catcode `\$12\catcode
  `\&12\catcode `\#12\catcode `\^12\catcode `\_12\catcode `\%12\relax}%
\providecommand \@@startlink[1]{}%
\providecommand \@@endlink[0]{}%
\providecommand \url  [0]{\begingroup\@sanitize@url \@url }%
\providecommand \@url [1]{\endgroup\@href {#1}{\urlprefix }}%
\providecommand \urlprefix  [0]{URL }%
\providecommand \Eprint [0]{\href }%
\providecommand \doibase [0]{https://doi.org/}%
\providecommand \selectlanguage [0]{\@gobble}%
\providecommand \bibinfo  [0]{\@secondoftwo}%
\providecommand \bibfield  [0]{\@secondoftwo}%
\providecommand \translation [1]{[#1]}%
\providecommand \BibitemOpen [0]{}%
\providecommand \bibitemStop [0]{}%
\providecommand \bibitemNoStop [0]{.\EOS\space}%
\providecommand \EOS [0]{\spacefactor3000\relax}%
\providecommand \BibitemShut  [1]{\csname bibitem#1\endcsname}%
\let\auto@bib@innerbib\@empty
\bibitem [{\citenamefont {Landau}\ and\ \citenamefont
  {Lifshitz}(2013)}]{landau2013statistical}%
  \BibitemOpen
  \bibfield  {author} {\bibinfo {author} {\bibfnamefont {L.~D.}\ \bibnamefont
  {Landau}}\ and\ \bibinfo {author} {\bibfnamefont {E.~M.}\ \bibnamefont
  {Lifshitz}},\ }\href@noop {} {\emph {\bibinfo {title} {Statistical Physics:
  Volume 5}}},\ Vol.~\bibinfo {volume} {5}\ (\bibinfo  {publisher} {Elsevier},\
  \bibinfo {year} {2013})\BibitemShut {NoStop}%
\bibitem [{\citenamefont {Cardy}(1996)}]{cardy1996scaling}%
  \BibitemOpen
  \bibfield  {author} {\bibinfo {author} {\bibfnamefont {J.}~\bibnamefont
  {Cardy}},\ }\href@noop {} {\emph {\bibinfo {title} {Scaling and
  renormalization in statistical physics}}},\ Vol.~\bibinfo {volume} {5}\
  (\bibinfo  {publisher} {Cambridge university press},\ \bibinfo {year}
  {1996})\BibitemShut {NoStop}%
\bibitem [{\citenamefont {Sachdev}(1999)}]{sachdev1999quantum}%
  \BibitemOpen
  \bibfield  {author} {\bibinfo {author} {\bibfnamefont {S.}~\bibnamefont
  {Sachdev}},\ }\bibfield  {title} {\bibinfo {title} {Quantum phase
  transitions},\ }\href@noop {} {\bibfield  {journal} {\bibinfo  {journal}
  {Physics world}\ }\textbf {\bibinfo {volume} {12}},\ \bibinfo {pages} {33}
  (\bibinfo {year} {1999})}\BibitemShut {NoStop}%
\bibitem [{\citenamefont {Sachdev}(2023)}]{sachdev2023quantum}%
  \BibitemOpen
  \bibfield  {author} {\bibinfo {author} {\bibfnamefont {S.}~\bibnamefont
  {Sachdev}},\ }\href@noop {} {\emph {\bibinfo {title} {Quantum phases of
  matter}}}\ (\bibinfo  {publisher} {Cambridge University Press},\ \bibinfo
  {year} {2023})\BibitemShut {NoStop}%
\bibitem [{\citenamefont {Hasan}\ and\ \citenamefont
  {Kane}(2010)}]{Hasan2010RMP}%
  \BibitemOpen
  \bibfield  {author} {\bibinfo {author} {\bibfnamefont {M.~Z.}\ \bibnamefont
  {Hasan}}\ and\ \bibinfo {author} {\bibfnamefont {C.~L.}\ \bibnamefont
  {Kane}},\ }\bibfield  {title} {\bibinfo {title} {Colloquium: Topological
  insulators},\ }\href {https://doi.org/10.1103/RevModPhys.82.3045} {\bibfield
  {journal} {\bibinfo  {journal} {Rev. Mod. Phys.}\ }\textbf {\bibinfo {volume}
  {82}},\ \bibinfo {pages} {3045} (\bibinfo {year} {2010})}\BibitemShut
  {NoStop}%
\bibitem [{\citenamefont {Qi}\ and\ \citenamefont {Zhang}(2011)}]{Qi2011RMP}%
  \BibitemOpen
  \bibfield  {author} {\bibinfo {author} {\bibfnamefont {X.-L.}\ \bibnamefont
  {Qi}}\ and\ \bibinfo {author} {\bibfnamefont {S.-C.}\ \bibnamefont {Zhang}},\
  }\bibfield  {title} {\bibinfo {title} {Topological insulators and
  superconductors},\ }\href {https://doi.org/10.1103/RevModPhys.83.1057}
  {\bibfield  {journal} {\bibinfo  {journal} {Rev. Mod. Phys.}\ }\textbf
  {\bibinfo {volume} {83}},\ \bibinfo {pages} {1057} (\bibinfo {year}
  {2011})}\BibitemShut {NoStop}%
\bibitem [{\citenamefont {Senthil}(2015)}]{senthil2015symmetry}%
  \BibitemOpen
  \bibfield  {author} {\bibinfo {author} {\bibfnamefont {T.}~\bibnamefont
  {Senthil}},\ }\bibfield  {title} {\bibinfo {title} {Symmetry-protected
  topological phases of quantum matter},\ }\href@noop {} {\bibfield  {journal}
  {\bibinfo  {journal} {Annu. Rev. Condens. Matter Phys.}\ }\textbf {\bibinfo
  {volume} {6}},\ \bibinfo {pages} {299} (\bibinfo {year} {2015})}\BibitemShut
  {NoStop}%
\bibitem [{\citenamefont {Wen}(2017)}]{Wen2017RMP}%
  \BibitemOpen
  \bibfield  {author} {\bibinfo {author} {\bibfnamefont {X.-G.}\ \bibnamefont
  {Wen}},\ }\bibfield  {title} {\bibinfo {title} {Colloquium: Zoo of
  quantum-topological phases of matter},\ }\href
  {https://doi.org/10.1103/RevModPhys.89.041004} {\bibfield  {journal}
  {\bibinfo  {journal} {Rev. Mod. Phys.}\ }\textbf {\bibinfo {volume} {89}},\
  \bibinfo {pages} {041004} (\bibinfo {year} {2017})}\BibitemShut {NoStop}%
\bibitem [{\citenamefont {Kane}\ and\ \citenamefont
  {Mele}(2005{\natexlab{a}})}]{Kane2005PRL}%
  \BibitemOpen
  \bibfield  {author} {\bibinfo {author} {\bibfnamefont {C.~L.}\ \bibnamefont
  {Kane}}\ and\ \bibinfo {author} {\bibfnamefont {E.~J.}\ \bibnamefont
  {Mele}},\ }\bibfield  {title} {\bibinfo {title} {Quantum spin hall effect in
  graphene},\ }\href {https://doi.org/10.1103/PhysRevLett.95.226801} {\bibfield
   {journal} {\bibinfo  {journal} {Phys. Rev. Lett.}\ }\textbf {\bibinfo
  {volume} {95}},\ \bibinfo {pages} {226801} (\bibinfo {year}
  {2005}{\natexlab{a}})}\BibitemShut {NoStop}%
\bibitem [{\citenamefont {Kane}\ and\ \citenamefont
  {Mele}(2005{\natexlab{b}})}]{Kane2005PRL_b}%
  \BibitemOpen
  \bibfield  {author} {\bibinfo {author} {\bibfnamefont {C.~L.}\ \bibnamefont
  {Kane}}\ and\ \bibinfo {author} {\bibfnamefont {E.~J.}\ \bibnamefont
  {Mele}},\ }\bibfield  {title} {\bibinfo {title} {${Z}_{2}$ topological order
  and the quantum spin hall effect},\ }\href
  {https://doi.org/10.1103/PhysRevLett.95.146802} {\bibfield  {journal}
  {\bibinfo  {journal} {Phys. Rev. Lett.}\ }\textbf {\bibinfo {volume} {95}},\
  \bibinfo {pages} {146802} (\bibinfo {year} {2005}{\natexlab{b}})}\BibitemShut
  {NoStop}%
\bibitem [{\citenamefont {Bernevig}\ and\ \citenamefont
  {Zhang}(2006)}]{Bernevig2006PRL}%
  \BibitemOpen
  \bibfield  {author} {\bibinfo {author} {\bibfnamefont {B.~A.}\ \bibnamefont
  {Bernevig}}\ and\ \bibinfo {author} {\bibfnamefont {S.-C.}\ \bibnamefont
  {Zhang}},\ }\bibfield  {title} {\bibinfo {title} {Quantum spin hall effect},\
  }\href {https://doi.org/10.1103/PhysRevLett.96.106802} {\bibfield  {journal}
  {\bibinfo  {journal} {Phys. Rev. Lett.}\ }\textbf {\bibinfo {volume} {96}},\
  \bibinfo {pages} {106802} (\bibinfo {year} {2006})}\BibitemShut {NoStop}%
\bibitem [{\citenamefont {Bernevig}\ \emph {et~al.}(2006)\citenamefont
  {Bernevig}, \citenamefont {Hughes},\ and\ \citenamefont
  {Zhang}}]{Bernevig2006Science}%
  \BibitemOpen
  \bibfield  {author} {\bibinfo {author} {\bibfnamefont {B.~A.}\ \bibnamefont
  {Bernevig}}, \bibinfo {author} {\bibfnamefont {T.~L.}\ \bibnamefont
  {Hughes}},\ and\ \bibinfo {author} {\bibfnamefont {S.-C.}\ \bibnamefont
  {Zhang}},\ }\bibfield  {title} {\bibinfo {title} {Quantum spin hall effect
  and topological phase transition in hgte quantum wells},\ }\href
  {https://doi.org/10.1126/science.1133734} {\bibfield  {journal} {\bibinfo
  {journal} {Science}\ }\textbf {\bibinfo {volume} {314}},\ \bibinfo {pages}
  {1757} (\bibinfo {year} {2006})},\ \Eprint
  {https://arxiv.org/abs/https://www.science.org/doi/pdf/10.1126/science.1133734}
  {https://www.science.org/doi/pdf/10.1126/science.1133734} \BibitemShut
  {NoStop}%
\bibitem [{\citenamefont {Fu}\ \emph {et~al.}(2007)\citenamefont {Fu},
  \citenamefont {Kane},\ and\ \citenamefont {Mele}}]{Fu2007PRL}%
  \BibitemOpen
  \bibfield  {author} {\bibinfo {author} {\bibfnamefont {L.}~\bibnamefont
  {Fu}}, \bibinfo {author} {\bibfnamefont {C.~L.}\ \bibnamefont {Kane}},\ and\
  \bibinfo {author} {\bibfnamefont {E.~J.}\ \bibnamefont {Mele}},\ }\bibfield
  {title} {\bibinfo {title} {Topological insulators in three dimensions},\
  }\href {https://doi.org/10.1103/PhysRevLett.98.106803} {\bibfield  {journal}
  {\bibinfo  {journal} {Phys. Rev. Lett.}\ }\textbf {\bibinfo {volume} {98}},\
  \bibinfo {pages} {106803} (\bibinfo {year} {2007})}\BibitemShut {NoStop}%
\bibitem [{\citenamefont {Gu}\ and\ \citenamefont {Wen}(2009)}]{Gu2009PRB}%
  \BibitemOpen
  \bibfield  {author} {\bibinfo {author} {\bibfnamefont {Z.-C.}\ \bibnamefont
  {Gu}}\ and\ \bibinfo {author} {\bibfnamefont {X.-G.}\ \bibnamefont {Wen}},\
  }\bibfield  {title} {\bibinfo {title} {Tensor-entanglement-filtering
  renormalization approach and symmetry-protected topological order},\ }\href
  {https://doi.org/10.1103/PhysRevB.80.155131} {\bibfield  {journal} {\bibinfo
  {journal} {Phys. Rev. B}\ }\textbf {\bibinfo {volume} {80}},\ \bibinfo
  {pages} {155131} (\bibinfo {year} {2009})}\BibitemShut {NoStop}%
\bibitem [{\citenamefont {Chen}\ \emph {et~al.}(2010)\citenamefont {Chen},
  \citenamefont {Gu},\ and\ \citenamefont {Wen}}]{Chen2010PRB}%
  \BibitemOpen
  \bibfield  {author} {\bibinfo {author} {\bibfnamefont {X.}~\bibnamefont
  {Chen}}, \bibinfo {author} {\bibfnamefont {Z.-C.}\ \bibnamefont {Gu}},\ and\
  \bibinfo {author} {\bibfnamefont {X.-G.}\ \bibnamefont {Wen}},\ }\bibfield
  {title} {\bibinfo {title} {Local unitary transformation, long-range quantum
  entanglement, wave function renormalization, and topological order},\ }\href
  {https://doi.org/10.1103/PhysRevB.82.155138} {\bibfield  {journal} {\bibinfo
  {journal} {Phys. Rev. B}\ }\textbf {\bibinfo {volume} {82}},\ \bibinfo
  {pages} {155138} (\bibinfo {year} {2010})}\BibitemShut {NoStop}%
\bibitem [{\citenamefont {Pollmann}\ and\ \citenamefont
  {Turner}(2012)}]{Pollmann2012PRB}%
  \BibitemOpen
  \bibfield  {author} {\bibinfo {author} {\bibfnamefont {F.}~\bibnamefont
  {Pollmann}}\ and\ \bibinfo {author} {\bibfnamefont {A.~M.}\ \bibnamefont
  {Turner}},\ }\bibfield  {title} {\bibinfo {title} {Detection of
  symmetry-protected topological phases in one dimension},\ }\href
  {https://doi.org/10.1103/PhysRevB.86.125441} {\bibfield  {journal} {\bibinfo
  {journal} {Phys. Rev. B}\ }\textbf {\bibinfo {volume} {86}},\ \bibinfo
  {pages} {125441} (\bibinfo {year} {2012})}\BibitemShut {NoStop}%
\bibitem [{\citenamefont {Chen}\ \emph {et~al.}(2013)\citenamefont {Chen},
  \citenamefont {Gu}, \citenamefont {Liu},\ and\ \citenamefont
  {Wen}}]{Chen2013PRB}%
  \BibitemOpen
  \bibfield  {author} {\bibinfo {author} {\bibfnamefont {X.}~\bibnamefont
  {Chen}}, \bibinfo {author} {\bibfnamefont {Z.-C.}\ \bibnamefont {Gu}},
  \bibinfo {author} {\bibfnamefont {Z.-X.}\ \bibnamefont {Liu}},\ and\ \bibinfo
  {author} {\bibfnamefont {X.-G.}\ \bibnamefont {Wen}},\ }\bibfield  {title}
  {\bibinfo {title} {Symmetry protected topological orders and the group
  cohomology of their symmetry group},\ }\href
  {https://doi.org/10.1103/PhysRevB.87.155114} {\bibfield  {journal} {\bibinfo
  {journal} {Phys. Rev. B}\ }\textbf {\bibinfo {volume} {87}},\ \bibinfo
  {pages} {155114} (\bibinfo {year} {2013})}\BibitemShut {NoStop}%
\bibitem [{\citenamefont {Keselman}\ and\ \citenamefont
  {Berg}(2015)}]{Keselman2015PRB}%
  \BibitemOpen
  \bibfield  {author} {\bibinfo {author} {\bibfnamefont {A.}~\bibnamefont
  {Keselman}}\ and\ \bibinfo {author} {\bibfnamefont {E.}~\bibnamefont
  {Berg}},\ }\bibfield  {title} {\bibinfo {title} {Gapless symmetry-protected
  topological phase of fermions in one dimension},\ }\href
  {https://doi.org/10.1103/PhysRevB.91.235309} {\bibfield  {journal} {\bibinfo
  {journal} {Phys. Rev. B}\ }\textbf {\bibinfo {volume} {91}},\ \bibinfo
  {pages} {235309} (\bibinfo {year} {2015})}\BibitemShut {NoStop}%
\bibitem [{\citenamefont {Scaffidi}\ \emph {et~al.}(2017)\citenamefont
  {Scaffidi}, \citenamefont {Parker},\ and\ \citenamefont
  {Vasseur}}]{Scaffidi2017PRX}%
  \BibitemOpen
  \bibfield  {author} {\bibinfo {author} {\bibfnamefont {T.}~\bibnamefont
  {Scaffidi}}, \bibinfo {author} {\bibfnamefont {D.~E.}\ \bibnamefont
  {Parker}},\ and\ \bibinfo {author} {\bibfnamefont {R.}~\bibnamefont
  {Vasseur}},\ }\bibfield  {title} {\bibinfo {title} {Gapless
  symmetry-protected topological order},\ }\href
  {https://doi.org/10.1103/PhysRevX.7.041048} {\bibfield  {journal} {\bibinfo
  {journal} {Phys. Rev. X}\ }\textbf {\bibinfo {volume} {7}},\ \bibinfo {pages}
  {041048} (\bibinfo {year} {2017})}\BibitemShut {NoStop}%
\bibitem [{\citenamefont {Verresen}\ \emph {et~al.}(2018)\citenamefont
  {Verresen}, \citenamefont {Jones},\ and\ \citenamefont
  {Pollmann}}]{Verresen2018PRL}%
  \BibitemOpen
  \bibfield  {author} {\bibinfo {author} {\bibfnamefont {R.}~\bibnamefont
  {Verresen}}, \bibinfo {author} {\bibfnamefont {N.~G.}\ \bibnamefont
  {Jones}},\ and\ \bibinfo {author} {\bibfnamefont {F.}~\bibnamefont
  {Pollmann}},\ }\bibfield  {title} {\bibinfo {title} {Topology and edge modes
  in quantum critical chains},\ }\href
  {https://doi.org/10.1103/PhysRevLett.120.057001} {\bibfield  {journal}
  {\bibinfo  {journal} {Phys. Rev. Lett.}\ }\textbf {\bibinfo {volume} {120}},\
  \bibinfo {pages} {057001} (\bibinfo {year} {2018})}\BibitemShut {NoStop}%
\bibitem [{\citenamefont {Verresen}\ \emph {et~al.}(2021)\citenamefont
  {Verresen}, \citenamefont {Thorngren}, \citenamefont {Jones},\ and\
  \citenamefont {Pollmann}}]{Verresen2021PRX}%
  \BibitemOpen
  \bibfield  {author} {\bibinfo {author} {\bibfnamefont {R.}~\bibnamefont
  {Verresen}}, \bibinfo {author} {\bibfnamefont {R.}~\bibnamefont {Thorngren}},
  \bibinfo {author} {\bibfnamefont {N.~G.}\ \bibnamefont {Jones}},\ and\
  \bibinfo {author} {\bibfnamefont {F.}~\bibnamefont {Pollmann}},\ }\bibfield
  {title} {\bibinfo {title} {Gapless topological phases and symmetry-enriched
  quantum criticality},\ }\href {https://doi.org/10.1103/PhysRevX.11.041059}
  {\bibfield  {journal} {\bibinfo  {journal} {Phys. Rev. X}\ }\textbf {\bibinfo
  {volume} {11}},\ \bibinfo {pages} {041059} (\bibinfo {year}
  {2021})}\BibitemShut {NoStop}%
\bibitem [{\citenamefont {Yu}\ \emph {et~al.}(2022)\citenamefont {Yu},
  \citenamefont {Huang}, \citenamefont {Song}, \citenamefont {Xu},
  \citenamefont {Ding},\ and\ \citenamefont {Zhang}}]{Yu2022PRL}%
  \BibitemOpen
  \bibfield  {author} {\bibinfo {author} {\bibfnamefont {X.-J.}\ \bibnamefont
  {Yu}}, \bibinfo {author} {\bibfnamefont {R.-Z.}\ \bibnamefont {Huang}},
  \bibinfo {author} {\bibfnamefont {H.-H.}\ \bibnamefont {Song}}, \bibinfo
  {author} {\bibfnamefont {L.}~\bibnamefont {Xu}}, \bibinfo {author}
  {\bibfnamefont {C.}~\bibnamefont {Ding}},\ and\ \bibinfo {author}
  {\bibfnamefont {L.}~\bibnamefont {Zhang}},\ }\bibfield  {title} {\bibinfo
  {title} {Conformal boundary conditions of symmetry-enriched quantum critical
  spin chains},\ }\href {https://doi.org/10.1103/PhysRevLett.129.210601}
  {\bibfield  {journal} {\bibinfo  {journal} {Phys. Rev. Lett.}\ }\textbf
  {\bibinfo {volume} {129}},\ \bibinfo {pages} {210601} (\bibinfo {year}
  {2022})}\BibitemShut {NoStop}%
\bibitem [{\citenamefont {Yu}\ \emph {et~al.}(2024)\citenamefont {Yu},
  \citenamefont {Yang}, \citenamefont {Lin},\ and\ \citenamefont
  {Jian}}]{Yu2024PRL}%
  \BibitemOpen
  \bibfield  {author} {\bibinfo {author} {\bibfnamefont {X.-J.}\ \bibnamefont
  {Yu}}, \bibinfo {author} {\bibfnamefont {S.}~\bibnamefont {Yang}}, \bibinfo
  {author} {\bibfnamefont {H.-Q.}\ \bibnamefont {Lin}},\ and\ \bibinfo {author}
  {\bibfnamefont {S.-K.}\ \bibnamefont {Jian}},\ }\bibfield  {title} {\bibinfo
  {title} {Universal entanglement spectrum in one-dimensional gapless symmetry
  protected topological states},\ }\href
  {https://doi.org/10.1103/PhysRevLett.133.026601} {\bibfield  {journal}
  {\bibinfo  {journal} {Phys. Rev. Lett.}\ }\textbf {\bibinfo {volume} {133}},\
  \bibinfo {pages} {026601} (\bibinfo {year} {2024})}\BibitemShut {NoStop}%
\bibitem [{\citenamefont {Parker}\ \emph {et~al.}(2018)\citenamefont {Parker},
  \citenamefont {Scaffidi},\ and\ \citenamefont {Vasseur}}]{Parker2018PRB}%
  \BibitemOpen
  \bibfield  {author} {\bibinfo {author} {\bibfnamefont {D.~E.}\ \bibnamefont
  {Parker}}, \bibinfo {author} {\bibfnamefont {T.}~\bibnamefont {Scaffidi}},\
  and\ \bibinfo {author} {\bibfnamefont {R.}~\bibnamefont {Vasseur}},\
  }\bibfield  {title} {\bibinfo {title} {Topological luttinger liquids from
  decorated domain walls},\ }\href {https://doi.org/10.1103/PhysRevB.97.165114}
  {\bibfield  {journal} {\bibinfo  {journal} {Phys. Rev. B}\ }\textbf {\bibinfo
  {volume} {97}},\ \bibinfo {pages} {165114} (\bibinfo {year}
  {2018})}\BibitemShut {NoStop}%
\bibitem [{\citenamefont {Jiang}\ \emph {et~al.}(2018)\citenamefont {Jiang},
  \citenamefont {Li}, \citenamefont {Seidel},\ and\ \citenamefont
  {Lee}}]{JIANG2018753}%
  \BibitemOpen
  \bibfield  {author} {\bibinfo {author} {\bibfnamefont {H.-C.}\ \bibnamefont
  {Jiang}}, \bibinfo {author} {\bibfnamefont {Z.-X.}\ \bibnamefont {Li}},
  \bibinfo {author} {\bibfnamefont {A.}~\bibnamefont {Seidel}},\ and\ \bibinfo
  {author} {\bibfnamefont {D.-H.}\ \bibnamefont {Lee}},\ }\bibfield  {title}
  {\bibinfo {title} {Symmetry protected topological luttinger liquids and the
  phase transition between them},\ }\href
  {https://doi.org/https://doi.org/10.1016/j.scib.2018.05.010} {\bibfield
  {journal} {\bibinfo  {journal} {Science Bulletin}\ }\textbf {\bibinfo
  {volume} {63}},\ \bibinfo {pages} {753} (\bibinfo {year} {2018})}\BibitemShut
  {NoStop}%
\bibitem [{\citenamefont {Jones}\ and\ \citenamefont
  {Verresen}(2019)}]{Jones2019JSP}%
  \BibitemOpen
  \bibfield  {author} {\bibinfo {author} {\bibfnamefont {N.~G.}\ \bibnamefont
  {Jones}}\ and\ \bibinfo {author} {\bibfnamefont {R.}~\bibnamefont
  {Verresen}},\ }\bibfield  {title} {\bibinfo {title} {Asymptotic correlations
  in gapped and critical topological phases of 1d quantum systems},\ }\href
  {https://doi.org/10.1007/s10955-019-02257-9} {\bibfield  {journal} {\bibinfo
  {journal} {Journal of Statistical Physics}\ }\textbf {\bibinfo {volume}
  {175}},\ \bibinfo {pages} {1164} (\bibinfo {year} {2019})}\BibitemShut
  {NoStop}%
\bibitem [{\citenamefont
  {Verresen}(2020)}]{verresen2020topologyedgestatessurvive}%
  \BibitemOpen
  \bibfield  {author} {\bibinfo {author} {\bibfnamefont {R.}~\bibnamefont
  {Verresen}},\ }\href {https://arxiv.org/abs/2003.05453} {\bibinfo {title}
  {Topology and edge states survive quantum criticality between topological
  insulators}} (\bibinfo {year} {2020}),\ \Eprint
  {https://arxiv.org/abs/2003.05453} {arXiv:2003.05453 [cond-mat.str-el]}
  \BibitemShut {NoStop}%
\bibitem [{\citenamefont {Thorngren}\ \emph {et~al.}(2021)\citenamefont
  {Thorngren}, \citenamefont {Vishwanath},\ and\ \citenamefont
  {Verresen}}]{Thorngren2021PRB}%
  \BibitemOpen
  \bibfield  {author} {\bibinfo {author} {\bibfnamefont {R.}~\bibnamefont
  {Thorngren}}, \bibinfo {author} {\bibfnamefont {A.}~\bibnamefont
  {Vishwanath}},\ and\ \bibinfo {author} {\bibfnamefont {R.}~\bibnamefont
  {Verresen}},\ }\bibfield  {title} {\bibinfo {title} {Intrinsically gapless
  topological phases},\ }\href {https://doi.org/10.1103/PhysRevB.104.075132}
  {\bibfield  {journal} {\bibinfo  {journal} {Phys. Rev. B}\ }\textbf {\bibinfo
  {volume} {104}},\ \bibinfo {pages} {075132} (\bibinfo {year}
  {2021})}\BibitemShut {NoStop}%
\bibitem [{\citenamefont {Duque}\ \emph {et~al.}(2021)\citenamefont {Duque},
  \citenamefont {Hu}, \citenamefont {You}, \citenamefont {Khemani},
  \citenamefont {Verresen},\ and\ \citenamefont {Vasseur}}]{Duque2021PRB}%
  \BibitemOpen
  \bibfield  {author} {\bibinfo {author} {\bibfnamefont {C.~M.}\ \bibnamefont
  {Duque}}, \bibinfo {author} {\bibfnamefont {H.-Y.}\ \bibnamefont {Hu}},
  \bibinfo {author} {\bibfnamefont {Y.-Z.}\ \bibnamefont {You}}, \bibinfo
  {author} {\bibfnamefont {V.}~\bibnamefont {Khemani}}, \bibinfo {author}
  {\bibfnamefont {R.}~\bibnamefont {Verresen}},\ and\ \bibinfo {author}
  {\bibfnamefont {R.}~\bibnamefont {Vasseur}},\ }\bibfield  {title} {\bibinfo
  {title} {Topological and symmetry-enriched random quantum critical points},\
  }\href {https://doi.org/10.1103/PhysRevB.103.L100207} {\bibfield  {journal}
  {\bibinfo  {journal} {Phys. Rev. B}\ }\textbf {\bibinfo {volume} {103}},\
  \bibinfo {pages} {L100207} (\bibinfo {year} {2021})}\BibitemShut {NoStop}%
\bibitem [{\citenamefont {Jones}\ \emph {et~al.}(2023)\citenamefont {Jones},
  \citenamefont {Thorngren},\ and\ \citenamefont {Verresen}}]{Jones2023PRL}%
  \BibitemOpen
  \bibfield  {author} {\bibinfo {author} {\bibfnamefont {N.~G.}\ \bibnamefont
  {Jones}}, \bibinfo {author} {\bibfnamefont {R.}~\bibnamefont {Thorngren}},\
  and\ \bibinfo {author} {\bibfnamefont {R.}~\bibnamefont {Verresen}},\
  }\bibfield  {title} {\bibinfo {title} {Bulk-boundary correspondence and
  singularity-filling in long-range free-fermion chains},\ }\href
  {https://doi.org/10.1103/PhysRevLett.130.246601} {\bibfield  {journal}
  {\bibinfo  {journal} {Phys. Rev. Lett.}\ }\textbf {\bibinfo {volume} {130}},\
  \bibinfo {pages} {246601} (\bibinfo {year} {2023})}\BibitemShut {NoStop}%
\bibitem [{\citenamefont {Wen}\ and\ \citenamefont
  {Potter}(2023)}]{Wen2023PRB}%
  \BibitemOpen
  \bibfield  {author} {\bibinfo {author} {\bibfnamefont {R.}~\bibnamefont
  {Wen}}\ and\ \bibinfo {author} {\bibfnamefont {A.~C.}\ \bibnamefont
  {Potter}},\ }\bibfield  {title} {\bibinfo {title} {Bulk-boundary
  correspondence for intrinsically gapless symmetry-protected topological
  phases from group cohomology},\ }\href
  {https://doi.org/10.1103/PhysRevB.107.245127} {\bibfield  {journal} {\bibinfo
   {journal} {Phys. Rev. B}\ }\textbf {\bibinfo {volume} {107}},\ \bibinfo
  {pages} {245127} (\bibinfo {year} {2023})}\BibitemShut {NoStop}%
\bibitem [{\citenamefont {Huang}\ and\ \citenamefont
  {Cheng}(2025)}]{huang2025topologicalholographyquantumcriticality}%
  \BibitemOpen
  \bibfield  {author} {\bibinfo {author} {\bibfnamefont {S.-J.}\ \bibnamefont
  {Huang}}\ and\ \bibinfo {author} {\bibfnamefont {M.}~\bibnamefont {Cheng}},\
  }\href {https://arxiv.org/abs/2310.16878} {\bibinfo {title} {Topological
  holography, quantum criticality, and boundary states}} (\bibinfo {year}
  {2025}),\ \Eprint {https://arxiv.org/abs/2310.16878} {arXiv:2310.16878
  [cond-mat.str-el]} \BibitemShut {NoStop}%
\bibitem [{\citenamefont {Mondal}\ \emph {et~al.}(2023)\citenamefont {Mondal},
  \citenamefont {Agarwala}, \citenamefont {Mishra},\ and\ \citenamefont
  {Prakash}}]{Mondal2023PRB}%
  \BibitemOpen
  \bibfield  {author} {\bibinfo {author} {\bibfnamefont {S.}~\bibnamefont
  {Mondal}}, \bibinfo {author} {\bibfnamefont {A.}~\bibnamefont {Agarwala}},
  \bibinfo {author} {\bibfnamefont {T.}~\bibnamefont {Mishra}},\ and\ \bibinfo
  {author} {\bibfnamefont {A.}~\bibnamefont {Prakash}},\ }\bibfield  {title}
  {\bibinfo {title} {Symmetry-enriched criticality in a coupled spin ladder},\
  }\href {https://doi.org/10.1103/PhysRevB.108.245135} {\bibfield  {journal}
  {\bibinfo  {journal} {Phys. Rev. B}\ }\textbf {\bibinfo {volume} {108}},\
  \bibinfo {pages} {245135} (\bibinfo {year} {2023})}\BibitemShut {NoStop}%
\bibitem [{\citenamefont {Yu}\ and\ \citenamefont {Li}(2024)}]{Yu2024PRB}%
  \BibitemOpen
  \bibfield  {author} {\bibinfo {author} {\bibfnamefont {X.-J.}\ \bibnamefont
  {Yu}}\ and\ \bibinfo {author} {\bibfnamefont {W.-L.}\ \bibnamefont {Li}},\
  }\bibfield  {title} {\bibinfo {title} {Fidelity susceptibility at the
  lifshitz transition between the noninteracting topologically distinct quantum
  critical points},\ }\href {https://doi.org/10.1103/PhysRevB.110.045119}
  {\bibfield  {journal} {\bibinfo  {journal} {Phys. Rev. B}\ }\textbf {\bibinfo
  {volume} {110}},\ \bibinfo {pages} {045119} (\bibinfo {year}
  {2024})}\BibitemShut {NoStop}%
\bibitem [{\citenamefont {Prembabu}\ \emph {et~al.}(2024)\citenamefont
  {Prembabu}, \citenamefont {Thorngren},\ and\ \citenamefont
  {Verresen}}]{Prembabu2024PRB}%
  \BibitemOpen
  \bibfield  {author} {\bibinfo {author} {\bibfnamefont {S.}~\bibnamefont
  {Prembabu}}, \bibinfo {author} {\bibfnamefont {R.}~\bibnamefont
  {Thorngren}},\ and\ \bibinfo {author} {\bibfnamefont {R.}~\bibnamefont
  {Verresen}},\ }\bibfield  {title} {\bibinfo {title} {Boundary-deconfined
  quantum criticality at transitions between symmetry-protected topological
  chains},\ }\href {https://doi.org/10.1103/PhysRevB.109.L201112} {\bibfield
  {journal} {\bibinfo  {journal} {Phys. Rev. B}\ }\textbf {\bibinfo {volume}
  {109}},\ \bibinfo {pages} {L201112} (\bibinfo {year} {2024})}\BibitemShut
  {NoStop}%
\bibitem [{\citenamefont {Su}\ and\ \citenamefont {Zeng}(2024)}]{Su2024PRB}%
  \BibitemOpen
  \bibfield  {author} {\bibinfo {author} {\bibfnamefont {L.}~\bibnamefont
  {Su}}\ and\ \bibinfo {author} {\bibfnamefont {M.}~\bibnamefont {Zeng}},\
  }\bibfield  {title} {\bibinfo {title} {Gapless symmetry-protected topological
  phases and generalized deconfined critical points from gauging a finite
  subgroup},\ }\href {https://doi.org/10.1103/PhysRevB.109.245108} {\bibfield
  {journal} {\bibinfo  {journal} {Phys. Rev. B}\ }\textbf {\bibinfo {volume}
  {109}},\ \bibinfo {pages} {245108} (\bibinfo {year} {2024})}\BibitemShut
  {NoStop}%
\bibitem [{\citenamefont {Li}\ \emph {et~al.}(2024{\natexlab{a}})\citenamefont
  {Li}, \citenamefont {Oshikawa},\ and\ \citenamefont {Zheng}}]{Li2024SciPost}%
  \BibitemOpen
  \bibfield  {author} {\bibinfo {author} {\bibfnamefont {L.}~\bibnamefont
  {Li}}, \bibinfo {author} {\bibfnamefont {M.}~\bibnamefont {Oshikawa}},\ and\
  \bibinfo {author} {\bibfnamefont {Y.}~\bibnamefont {Zheng}},\ }\bibfield
  {title} {\bibinfo {title} {{Decorated defect construction of gapless-SPT
  states}},\ }\href {https://doi.org/10.21468/SciPostPhys.17.1.013} {\bibfield
  {journal} {\bibinfo  {journal} {SciPost Phys.}\ }\textbf {\bibinfo {volume}
  {17}},\ \bibinfo {pages} {013} (\bibinfo {year}
  {2024}{\natexlab{a}})}\BibitemShut {NoStop}%
\bibitem [{\citenamefont {Zhong}\ \emph {et~al.}(2024)\citenamefont {Zhong},
  \citenamefont {Li}, \citenamefont {Chen},\ and\ \citenamefont
  {Yu}}]{Zhong2024PRA}%
  \BibitemOpen
  \bibfield  {author} {\bibinfo {author} {\bibfnamefont {W.-H.}\ \bibnamefont
  {Zhong}}, \bibinfo {author} {\bibfnamefont {W.-L.}\ \bibnamefont {Li}},
  \bibinfo {author} {\bibfnamefont {Y.-C.}\ \bibnamefont {Chen}},\ and\
  \bibinfo {author} {\bibfnamefont {X.-J.}\ \bibnamefont {Yu}},\ }\bibfield
  {title} {\bibinfo {title} {Topological edge modes and phase transitions in a
  critical fermionic chain with long-range interactions},\ }\href
  {https://doi.org/10.1103/PhysRevA.110.022212} {\bibfield  {journal} {\bibinfo
   {journal} {Phys. Rev. A}\ }\textbf {\bibinfo {volume} {110}},\ \bibinfo
  {pages} {022212} (\bibinfo {year} {2024})}\BibitemShut {NoStop}%
\bibitem [{\citenamefont
  {Ando}(2024)}]{ando2024gauginglatticegappedgaplesstopological}%
  \BibitemOpen
  \bibfield  {author} {\bibinfo {author} {\bibfnamefont {T.}~\bibnamefont
  {Ando}},\ }\href {https://arxiv.org/abs/2402.03566} {\bibinfo {title}
  {Gauging on the lattice and gapped/gapless topological phases}} (\bibinfo
  {year} {2024}),\ \Eprint {https://arxiv.org/abs/2402.03566} {arXiv:2402.03566
  [cond-mat.str-el]} \BibitemShut {NoStop}%
\bibitem [{\citenamefont {Zhang}\ \emph {et~al.}(2024)\citenamefont {Zhang},
  \citenamefont {Li}, \citenamefont {Yang},\ and\ \citenamefont
  {Yu}}]{Zhang2024PRA}%
  \BibitemOpen
  \bibfield  {author} {\bibinfo {author} {\bibfnamefont {H.-L.}\ \bibnamefont
  {Zhang}}, \bibinfo {author} {\bibfnamefont {H.-Z.}\ \bibnamefont {Li}},
  \bibinfo {author} {\bibfnamefont {S.}~\bibnamefont {Yang}},\ and\ \bibinfo
  {author} {\bibfnamefont {X.-J.}\ \bibnamefont {Yu}},\ }\bibfield  {title}
  {\bibinfo {title} {Quantum phase transition and critical behavior between the
  gapless topological phases},\ }\href
  {https://doi.org/10.1103/PhysRevA.109.062226} {\bibfield  {journal} {\bibinfo
   {journal} {Phys. Rev. A}\ }\textbf {\bibinfo {volume} {109}},\ \bibinfo
  {pages} {062226} (\bibinfo {year} {2024})}\BibitemShut {NoStop}%
\bibitem [{\citenamefont {Yang}\ \emph
  {et~al.}(2025{\natexlab{a}})\citenamefont {Yang}, \citenamefont {Lin},\ and\
  \citenamefont {Yu}}]{Yang2025CP}%
  \BibitemOpen
  \bibfield  {author} {\bibinfo {author} {\bibfnamefont {S.}~\bibnamefont
  {Yang}}, \bibinfo {author} {\bibfnamefont {H.-Q.}\ \bibnamefont {Lin}},\ and\
  \bibinfo {author} {\bibfnamefont {X.-J.}\ \bibnamefont {Yu}},\ }\bibfield
  {title} {\bibinfo {title} {Gapless topological behaviors in a long-range
  quantum spin chain},\ }\href {https://doi.org/10.1038/s42005-025-01947-z}
  {\bibfield  {journal} {\bibinfo  {journal} {Communications Physics}\ }\textbf
  {\bibinfo {volume} {8}},\ \bibinfo {pages} {27} (\bibinfo {year}
  {2025}{\natexlab{a}})}\BibitemShut {NoStop}%
\bibitem [{\citenamefont {Zhou}\ \emph {et~al.}(2025)\citenamefont {Zhou},
  \citenamefont {Gong},\ and\ \citenamefont {Yu}}]{Zhou2025CP}%
  \BibitemOpen
  \bibfield  {author} {\bibinfo {author} {\bibfnamefont {L.}~\bibnamefont
  {Zhou}}, \bibinfo {author} {\bibfnamefont {J.}~\bibnamefont {Gong}},\ and\
  \bibinfo {author} {\bibfnamefont {X.-J.}\ \bibnamefont {Yu}},\ }\bibfield
  {title} {\bibinfo {title} {Topological edge states at floquet quantum
  criticality},\ }\href {https://doi.org/10.1038/s42005-025-02137-7} {\bibfield
   {journal} {\bibinfo  {journal} {Communications Physics}\ }\textbf {\bibinfo
  {volume} {8}},\ \bibinfo {pages} {214} (\bibinfo {year} {2025})}\BibitemShut
  {NoStop}%
\bibitem [{\citenamefont {Cardoso}\ \emph {et~al.}(2025)\citenamefont
  {Cardoso}, \citenamefont {Yeh}, \citenamefont {Korneev}, \citenamefont
  {Abanov},\ and\ \citenamefont {Mitra}}]{Cardoso2025PRB}%
  \BibitemOpen
  \bibfield  {author} {\bibinfo {author} {\bibfnamefont {G.}~\bibnamefont
  {Cardoso}}, \bibinfo {author} {\bibfnamefont {H.-C.}\ \bibnamefont {Yeh}},
  \bibinfo {author} {\bibfnamefont {L.}~\bibnamefont {Korneev}}, \bibinfo
  {author} {\bibfnamefont {A.~G.}\ \bibnamefont {Abanov}},\ and\ \bibinfo
  {author} {\bibfnamefont {A.}~\bibnamefont {Mitra}},\ }\bibfield  {title}
  {\bibinfo {title} {Gapless floquet topology},\ }\href
  {https://doi.org/10.1103/PhysRevB.111.125162} {\bibfield  {journal} {\bibinfo
   {journal} {Phys. Rev. B}\ }\textbf {\bibinfo {volume} {111}},\ \bibinfo
  {pages} {125162} (\bibinfo {year} {2025})}\BibitemShut {NoStop}%
\bibitem [{\citenamefont {Flores-Calder\'on}\ \emph {et~al.}(2025)\citenamefont
  {Flores-Calder\'on}, \citenamefont {K\"onig},\ and\ \citenamefont
  {Cook}}]{Flores2025PRL}%
  \BibitemOpen
  \bibfield  {author} {\bibinfo {author} {\bibfnamefont {R.}~\bibnamefont
  {Flores-Calder\'on}}, \bibinfo {author} {\bibfnamefont {E.~J.}\ \bibnamefont
  {K\"onig}},\ and\ \bibinfo {author} {\bibfnamefont {A.~M.}\ \bibnamefont
  {Cook}},\ }\bibfield  {title} {\bibinfo {title} {Topological quantum
  criticality from multiplicative topological phases},\ }\href
  {https://doi.org/10.1103/PhysRevLett.134.116602} {\bibfield  {journal}
  {\bibinfo  {journal} {Phys. Rev. Lett.}\ }\textbf {\bibinfo {volume} {134}},\
  \bibinfo {pages} {116602} (\bibinfo {year} {2025})}\BibitemShut {NoStop}%
\bibitem [{\citenamefont {Wen}\ and\ \citenamefont
  {Potter}(2025)}]{Wen2025PRB}%
  \BibitemOpen
  \bibfield  {author} {\bibinfo {author} {\bibfnamefont {R.}~\bibnamefont
  {Wen}}\ and\ \bibinfo {author} {\bibfnamefont {A.~C.}\ \bibnamefont
  {Potter}},\ }\bibfield  {title} {\bibinfo {title} {Classification of
  $1+1\mathrm{D}$ gapless symmetry protected phases via topological
  holography},\ }\href {https://doi.org/10.1103/PhysRevB.111.115161} {\bibfield
   {journal} {\bibinfo  {journal} {Phys. Rev. B}\ }\textbf {\bibinfo {volume}
  {111}},\ \bibinfo {pages} {115161} (\bibinfo {year} {2025})}\BibitemShut
  {NoStop}%
\bibitem [{\citenamefont {Li}\ \emph {et~al.}(2025{\natexlab{a}})\citenamefont
  {Li}, \citenamefont {Oshikawa},\ and\ \citenamefont {Zheng}}]{Li2025SciPost}%
  \BibitemOpen
  \bibfield  {author} {\bibinfo {author} {\bibfnamefont {L.}~\bibnamefont
  {Li}}, \bibinfo {author} {\bibfnamefont {M.}~\bibnamefont {Oshikawa}},\ and\
  \bibinfo {author} {\bibfnamefont {Y.}~\bibnamefont {Zheng}},\ }\bibfield
  {title} {\bibinfo {title} {{Intrinsically/purely gapless-SPT from
  non-invertible duality transformations}},\ }\href
  {https://doi.org/10.21468/SciPostPhys.18.5.153} {\bibfield  {journal}
  {\bibinfo  {journal} {SciPost Phys.}\ }\textbf {\bibinfo {volume} {18}},\
  \bibinfo {pages} {153} (\bibinfo {year} {2025}{\natexlab{a}})}\BibitemShut
  {NoStop}%
\bibitem [{\citenamefont {Yu}\ \emph {et~al.}(2025)\citenamefont {Yu},
  \citenamefont {Yang}, \citenamefont {Liu}, \citenamefont {Lin},\ and\
  \citenamefont {Jian}}]{yu2025gaplesssymmetryprotectedtopologicalstates}%
  \BibitemOpen
  \bibfield  {author} {\bibinfo {author} {\bibfnamefont {X.-J.}\ \bibnamefont
  {Yu}}, \bibinfo {author} {\bibfnamefont {S.}~\bibnamefont {Yang}}, \bibinfo
  {author} {\bibfnamefont {S.}~\bibnamefont {Liu}}, \bibinfo {author}
  {\bibfnamefont {H.-Q.}\ \bibnamefont {Lin}},\ and\ \bibinfo {author}
  {\bibfnamefont {S.-K.}\ \bibnamefont {Jian}},\ }\href
  {https://arxiv.org/abs/2501.03851} {\bibinfo {title} {Gapless
  symmetry-protected topological states in measurement-only circuits}}
  (\bibinfo {year} {2025}),\ \Eprint {https://arxiv.org/abs/2501.03851}
  {arXiv:2501.03851 [cond-mat.str-el]} \BibitemShut {NoStop}%
\bibitem [{\citenamefont {Tan}\ \emph {et~al.}(2025)\citenamefont {Tan},
  \citenamefont {Wang}, \citenamefont {Yang}, \citenamefont {Shen},
  \citenamefont {Jin}, \citenamefont {Zhu}, \citenamefont {Ji}, \citenamefont
  {Xu}, \citenamefont {Chen}, \citenamefont {Wu}, \citenamefont {Zhang},
  \citenamefont {Gao}, \citenamefont {Wang}, \citenamefont {Zou}, \citenamefont
  {Zhang}, \citenamefont {Li}, \citenamefont {Bao}, \citenamefont {Zhu},
  \citenamefont {Zhong}, \citenamefont {Cui}, \citenamefont {Han},
  \citenamefont {He}, \citenamefont {Wang}, \citenamefont {Yang}, \citenamefont
  {Wang}, \citenamefont {Shen}, \citenamefont {Liu}, \citenamefont {Song},
  \citenamefont {Deng}, \citenamefont {Dong}, \citenamefont {Zhang},
  \citenamefont {Jian}, \citenamefont {Li}, \citenamefont {Wang}, \citenamefont
  {Guo}, \citenamefont {Song}, \citenamefont {Yu}, \citenamefont {Wang},
  \citenamefont {Lin},\ and\ \citenamefont
  {Wu}}]{tan2025exploringnontrivialtopologyquantum}%
  \BibitemOpen
  \bibfield  {author} {\bibinfo {author} {\bibfnamefont {Z.}~\bibnamefont
  {Tan}}, \bibinfo {author} {\bibfnamefont {K.}~\bibnamefont {Wang}}, \bibinfo
  {author} {\bibfnamefont {S.}~\bibnamefont {Yang}}, \bibinfo {author}
  {\bibfnamefont {F.}~\bibnamefont {Shen}}, \bibinfo {author} {\bibfnamefont
  {F.}~\bibnamefont {Jin}}, \bibinfo {author} {\bibfnamefont {X.}~\bibnamefont
  {Zhu}}, \bibinfo {author} {\bibfnamefont {Y.}~\bibnamefont {Ji}}, \bibinfo
  {author} {\bibfnamefont {S.}~\bibnamefont {Xu}}, \bibinfo {author}
  {\bibfnamefont {J.}~\bibnamefont {Chen}}, \bibinfo {author} {\bibfnamefont
  {Y.}~\bibnamefont {Wu}}, \bibinfo {author} {\bibfnamefont {C.}~\bibnamefont
  {Zhang}}, \bibinfo {author} {\bibfnamefont {Y.}~\bibnamefont {Gao}}, \bibinfo
  {author} {\bibfnamefont {N.}~\bibnamefont {Wang}}, \bibinfo {author}
  {\bibfnamefont {Y.}~\bibnamefont {Zou}}, \bibinfo {author} {\bibfnamefont
  {A.}~\bibnamefont {Zhang}}, \bibinfo {author} {\bibfnamefont
  {T.}~\bibnamefont {Li}}, \bibinfo {author} {\bibfnamefont {Z.}~\bibnamefont
  {Bao}}, \bibinfo {author} {\bibfnamefont {Z.}~\bibnamefont {Zhu}}, \bibinfo
  {author} {\bibfnamefont {J.}~\bibnamefont {Zhong}}, \bibinfo {author}
  {\bibfnamefont {Z.}~\bibnamefont {Cui}}, \bibinfo {author} {\bibfnamefont
  {Y.}~\bibnamefont {Han}}, \bibinfo {author} {\bibfnamefont {Y.}~\bibnamefont
  {He}}, \bibinfo {author} {\bibfnamefont {H.}~\bibnamefont {Wang}}, \bibinfo
  {author} {\bibfnamefont {J.}~\bibnamefont {Yang}}, \bibinfo {author}
  {\bibfnamefont {Y.}~\bibnamefont {Wang}}, \bibinfo {author} {\bibfnamefont
  {J.}~\bibnamefont {Shen}}, \bibinfo {author} {\bibfnamefont {G.}~\bibnamefont
  {Liu}}, \bibinfo {author} {\bibfnamefont {Z.}~\bibnamefont {Song}}, \bibinfo
  {author} {\bibfnamefont {J.}~\bibnamefont {Deng}}, \bibinfo {author}
  {\bibfnamefont {H.}~\bibnamefont {Dong}}, \bibinfo {author} {\bibfnamefont
  {P.}~\bibnamefont {Zhang}}, \bibinfo {author} {\bibfnamefont {S.-K.}\
  \bibnamefont {Jian}}, \bibinfo {author} {\bibfnamefont {H.}~\bibnamefont
  {Li}}, \bibinfo {author} {\bibfnamefont {Z.}~\bibnamefont {Wang}}, \bibinfo
  {author} {\bibfnamefont {Q.}~\bibnamefont {Guo}}, \bibinfo {author}
  {\bibfnamefont {C.}~\bibnamefont {Song}}, \bibinfo {author} {\bibfnamefont
  {X.-J.}\ \bibnamefont {Yu}}, \bibinfo {author} {\bibfnamefont
  {H.}~\bibnamefont {Wang}}, \bibinfo {author} {\bibfnamefont {H.-Q.}\
  \bibnamefont {Lin}},\ and\ \bibinfo {author} {\bibfnamefont {F.}~\bibnamefont
  {Wu}},\ }\href {https://arxiv.org/abs/2501.04679} {\bibinfo {title}
  {Exploring nontrivial topology at quantum criticality in a superconducting
  processor}} (\bibinfo {year} {2025}),\ \Eprint
  {https://arxiv.org/abs/2501.04679} {arXiv:2501.04679 [quant-ph]} \BibitemShut
  {NoStop}%
\bibitem [{\citenamefont {Zhong}\ \emph {et~al.}(2025)\citenamefont {Zhong},
  \citenamefont {Lin},\ and\ \citenamefont
  {Yu}}]{zhong2025quantumentanglementfermionicgapless}%
  \BibitemOpen
  \bibfield  {author} {\bibinfo {author} {\bibfnamefont {W.-H.}\ \bibnamefont
  {Zhong}}, \bibinfo {author} {\bibfnamefont {H.-Q.}\ \bibnamefont {Lin}},\
  and\ \bibinfo {author} {\bibfnamefont {X.-J.}\ \bibnamefont {Yu}},\ }\href
  {https://arxiv.org/abs/2502.18178} {\bibinfo {title} {Quantum entanglement of
  fermionic gapless symmetry protected topological phases in one dimension}}
  (\bibinfo {year} {2025}),\ \Eprint {https://arxiv.org/abs/2502.18178}
  {arXiv:2502.18178 [cond-mat.str-el]} \BibitemShut {NoStop}%
\bibitem [{\citenamefont {Yang}\ \emph
  {et~al.}(2025{\natexlab{b}})\citenamefont {Yang}, \citenamefont {Xu},
  \citenamefont {Lu}, \citenamefont {You}, \citenamefont {Lin},\ and\
  \citenamefont {Yu}}]{yang2025deconfinedcriticalityintrinsicallygapless}%
  \BibitemOpen
  \bibfield  {author} {\bibinfo {author} {\bibfnamefont {S.}~\bibnamefont
  {Yang}}, \bibinfo {author} {\bibfnamefont {F.}~\bibnamefont {Xu}}, \bibinfo
  {author} {\bibfnamefont {D.-C.}\ \bibnamefont {Lu}}, \bibinfo {author}
  {\bibfnamefont {Y.-Z.}\ \bibnamefont {You}}, \bibinfo {author} {\bibfnamefont
  {H.-Q.}\ \bibnamefont {Lin}},\ and\ \bibinfo {author} {\bibfnamefont {X.-J.}\
  \bibnamefont {Yu}},\ }\href {https://arxiv.org/abs/2503.01198} {\bibinfo
  {title} {Deconfined criticality as intrinsically gapless topological state in
  one dimension}} (\bibinfo {year} {2025}{\natexlab{b}}),\ \Eprint
  {https://arxiv.org/abs/2503.01198} {arXiv:2503.01198 [cond-mat.str-el]}
  \BibitemShut {NoStop}%
\bibitem [{\citenamefont {Li}\ \emph {et~al.}(2024{\natexlab{b}})\citenamefont
  {Li}, \citenamefont {Huang},\ and\ \citenamefont
  {Cao}}]{li2024noninvertiblesymmetryenrichedquantumcritical}%
  \BibitemOpen
  \bibfield  {author} {\bibinfo {author} {\bibfnamefont {L.}~\bibnamefont
  {Li}}, \bibinfo {author} {\bibfnamefont {R.-Z.}\ \bibnamefont {Huang}},\ and\
  \bibinfo {author} {\bibfnamefont {W.}~\bibnamefont {Cao}},\ }\href
  {https://arxiv.org/abs/2411.19034} {\bibinfo {title} {Noninvertible
  symmetry-enriched quantum critical point}} (\bibinfo {year}
  {2024}{\natexlab{b}}),\ \Eprint {https://arxiv.org/abs/2411.19034}
  {arXiv:2411.19034 [cond-mat.str-el]} \BibitemShut {NoStop}%
\bibitem [{\citenamefont
  {Wen}(2025{\natexlab{a}})}]{wen2025topologicalholography21dgapped}%
  \BibitemOpen
  \bibfield  {author} {\bibinfo {author} {\bibfnamefont {R.}~\bibnamefont
  {Wen}},\ }\href {https://arxiv.org/abs/2503.13685} {\bibinfo {title}
  {Topological holography for 2+1-d gapped and gapless phases with generalized
  symmetries}} (\bibinfo {year} {2025}{\natexlab{a}}),\ \Eprint
  {https://arxiv.org/abs/2503.13685} {arXiv:2503.13685 [hep-th]} \BibitemShut
  {NoStop}%
\bibitem [{\citenamefont
  {Wen}(2025{\natexlab{b}})}]{wen2025stringcondensationtopologicalholography}%
  \BibitemOpen
  \bibfield  {author} {\bibinfo {author} {\bibfnamefont {R.}~\bibnamefont
  {Wen}},\ }\href {https://arxiv.org/abs/2408.05801} {\bibinfo {title} {String
  condensation and topological holography for 2+1d gapless spt}} (\bibinfo
  {year} {2025}{\natexlab{b}}),\ \Eprint {https://arxiv.org/abs/2408.05801}
  {arXiv:2408.05801 [cond-mat.str-el]} \BibitemShut {NoStop}%
\bibitem [{\citenamefont {Son}\ \emph {et~al.}(2011)\citenamefont {Son},
  \citenamefont {Amico}, \citenamefont {Fazio}, \citenamefont {Hamma},
  \citenamefont {Pascazio},\ and\ \citenamefont {Vedral}}]{Son_2011}%
  \BibitemOpen
  \bibfield  {author} {\bibinfo {author} {\bibfnamefont {W.}~\bibnamefont
  {Son}}, \bibinfo {author} {\bibfnamefont {L.}~\bibnamefont {Amico}}, \bibinfo
  {author} {\bibfnamefont {R.}~\bibnamefont {Fazio}}, \bibinfo {author}
  {\bibfnamefont {A.}~\bibnamefont {Hamma}}, \bibinfo {author} {\bibfnamefont
  {S.}~\bibnamefont {Pascazio}},\ and\ \bibinfo {author} {\bibfnamefont
  {V.}~\bibnamefont {Vedral}},\ }\bibfield  {title} {\bibinfo {title} {Quantum
  phase transition between cluster and antiferromagnetic states},\ }\href
  {https://doi.org/10.1209/0295-5075/95/50001} {\bibfield  {journal} {\bibinfo
  {journal} {Europhysics Letters}\ }\textbf {\bibinfo {volume} {95}},\ \bibinfo
  {pages} {50001} (\bibinfo {year} {2011})}\BibitemShut {NoStop}%
\bibitem [{\citenamefont {Smacchia}\ \emph {et~al.}(2011)\citenamefont
  {Smacchia}, \citenamefont {Amico}, \citenamefont {Facchi}, \citenamefont
  {Fazio}, \citenamefont {Florio}, \citenamefont {Pascazio},\ and\
  \citenamefont {Vedral}}]{Smacchia2011PRA}%
  \BibitemOpen
  \bibfield  {author} {\bibinfo {author} {\bibfnamefont {P.}~\bibnamefont
  {Smacchia}}, \bibinfo {author} {\bibfnamefont {L.}~\bibnamefont {Amico}},
  \bibinfo {author} {\bibfnamefont {P.}~\bibnamefont {Facchi}}, \bibinfo
  {author} {\bibfnamefont {R.}~\bibnamefont {Fazio}}, \bibinfo {author}
  {\bibfnamefont {G.}~\bibnamefont {Florio}}, \bibinfo {author} {\bibfnamefont
  {S.}~\bibnamefont {Pascazio}},\ and\ \bibinfo {author} {\bibfnamefont
  {V.}~\bibnamefont {Vedral}},\ }\bibfield  {title} {\bibinfo {title}
  {Statistical mechanics of the cluster ising model},\ }\href
  {https://doi.org/10.1103/PhysRevA.84.022304} {\bibfield  {journal} {\bibinfo
  {journal} {Phys. Rev. A}\ }\textbf {\bibinfo {volume} {84}},\ \bibinfo
  {pages} {022304} (\bibinfo {year} {2011})}\BibitemShut {NoStop}%
\bibitem [{\citenamefont {Montes}\ and\ \citenamefont
  {Hamma}(2012)}]{Montes2012PRE}%
  \BibitemOpen
  \bibfield  {author} {\bibinfo {author} {\bibfnamefont {S.}~\bibnamefont
  {Montes}}\ and\ \bibinfo {author} {\bibfnamefont {A.}~\bibnamefont {Hamma}},\
  }\bibfield  {title} {\bibinfo {title} {Phase diagram and quench dynamics of
  the cluster-$xy$ spin chain},\ }\href
  {https://doi.org/10.1103/PhysRevE.86.021101} {\bibfield  {journal} {\bibinfo
  {journal} {Phys. Rev. E}\ }\textbf {\bibinfo {volume} {86}},\ \bibinfo
  {pages} {021101} (\bibinfo {year} {2012})}\BibitemShut {NoStop}%
\bibitem [{\citenamefont {Lahtinen}\ and\ \citenamefont
  {Ardonne}(2015)}]{Lahtinen2015PRL}%
  \BibitemOpen
  \bibfield  {author} {\bibinfo {author} {\bibfnamefont {V.}~\bibnamefont
  {Lahtinen}}\ and\ \bibinfo {author} {\bibfnamefont {E.}~\bibnamefont
  {Ardonne}},\ }\bibfield  {title} {\bibinfo {title} {Realizing all
  $so(n{)}_{1}$ quantum criticalities in symmetry protected cluster models},\
  }\href {https://doi.org/10.1103/PhysRevLett.115.237203} {\bibfield  {journal}
  {\bibinfo  {journal} {Phys. Rev. Lett.}\ }\textbf {\bibinfo {volume} {115}},\
  \bibinfo {pages} {237203} (\bibinfo {year} {2015})}\BibitemShut {NoStop}%
\bibitem [{\citenamefont {Giampaolo}\ and\ \citenamefont
  {Hiesmayr}(2015)}]{Giampaolo2015PRA}%
  \BibitemOpen
  \bibfield  {author} {\bibinfo {author} {\bibfnamefont {S.~M.}\ \bibnamefont
  {Giampaolo}}\ and\ \bibinfo {author} {\bibfnamefont {B.~C.}\ \bibnamefont
  {Hiesmayr}},\ }\bibfield  {title} {\bibinfo {title} {Topological and nematic
  ordered phases in many-body cluster-ising models},\ }\href
  {https://doi.org/10.1103/PhysRevA.92.012306} {\bibfield  {journal} {\bibinfo
  {journal} {Phys. Rev. A}\ }\textbf {\bibinfo {volume} {92}},\ \bibinfo
  {pages} {012306} (\bibinfo {year} {2015})}\BibitemShut {NoStop}%
\bibitem [{\citenamefont {Ohta}\ \emph {et~al.}(2016)\citenamefont {Ohta},
  \citenamefont {Tanaka}, \citenamefont {Danshita},\ and\ \citenamefont
  {Totsuka}}]{Ohta2016PRB}%
  \BibitemOpen
  \bibfield  {author} {\bibinfo {author} {\bibfnamefont {T.}~\bibnamefont
  {Ohta}}, \bibinfo {author} {\bibfnamefont {S.}~\bibnamefont {Tanaka}},
  \bibinfo {author} {\bibfnamefont {I.}~\bibnamefont {Danshita}},\ and\
  \bibinfo {author} {\bibfnamefont {K.}~\bibnamefont {Totsuka}},\ }\bibfield
  {title} {\bibinfo {title} {Topological and dynamical properties of a
  generalized cluster model in one dimension},\ }\href
  {https://doi.org/10.1103/PhysRevB.93.165423} {\bibfield  {journal} {\bibinfo
  {journal} {Phys. Rev. B}\ }\textbf {\bibinfo {volume} {93}},\ \bibinfo
  {pages} {165423} (\bibinfo {year} {2016})}\BibitemShut {NoStop}%
\bibitem [{\citenamefont {Verresen}\ \emph {et~al.}(2017)\citenamefont
  {Verresen}, \citenamefont {Moessner},\ and\ \citenamefont
  {Pollmann}}]{Verresen2017PRB}%
  \BibitemOpen
  \bibfield  {author} {\bibinfo {author} {\bibfnamefont {R.}~\bibnamefont
  {Verresen}}, \bibinfo {author} {\bibfnamefont {R.}~\bibnamefont {Moessner}},\
  and\ \bibinfo {author} {\bibfnamefont {F.}~\bibnamefont {Pollmann}},\
  }\bibfield  {title} {\bibinfo {title} {One-dimensional symmetry protected
  topological phases and their transitions},\ }\href
  {https://doi.org/10.1103/PhysRevB.96.165124} {\bibfield  {journal} {\bibinfo
  {journal} {Phys. Rev. B}\ }\textbf {\bibinfo {volume} {96}},\ \bibinfo
  {pages} {165124} (\bibinfo {year} {2017})}\BibitemShut {NoStop}%
\bibitem [{\citenamefont {Nie}\ \emph {et~al.}(2017)\citenamefont {Nie},
  \citenamefont {Mei}, \citenamefont {Amico},\ and\ \citenamefont
  {Kwek}}]{Nie2017PRE}%
  \BibitemOpen
  \bibfield  {author} {\bibinfo {author} {\bibfnamefont {W.}~\bibnamefont
  {Nie}}, \bibinfo {author} {\bibfnamefont {F.}~\bibnamefont {Mei}}, \bibinfo
  {author} {\bibfnamefont {L.}~\bibnamefont {Amico}},\ and\ \bibinfo {author}
  {\bibfnamefont {L.~C.}\ \bibnamefont {Kwek}},\ }\bibfield  {title} {\bibinfo
  {title} {Scaling of geometric phase versus band structure in cluster-ising
  models},\ }\href {https://doi.org/10.1103/PhysRevE.96.020106} {\bibfield
  {journal} {\bibinfo  {journal} {Phys. Rev. E}\ }\textbf {\bibinfo {volume}
  {96}},\ \bibinfo {pages} {020106} (\bibinfo {year} {2017})}\BibitemShut
  {NoStop}%
\bibitem [{\citenamefont {Ding}(2019)}]{Ding2019PRE}%
  \BibitemOpen
  \bibfield  {author} {\bibinfo {author} {\bibfnamefont {C.}~\bibnamefont
  {Ding}},\ }\bibfield  {title} {\bibinfo {title} {Phase transitions of a
  cluster ising model},\ }\href {https://doi.org/10.1103/PhysRevE.100.042131}
  {\bibfield  {journal} {\bibinfo  {journal} {Phys. Rev. E}\ }\textbf {\bibinfo
  {volume} {100}},\ \bibinfo {pages} {042131} (\bibinfo {year}
  {2019})}\BibitemShut {NoStop}%
\bibitem [{\citenamefont {Jones}\ \emph {et~al.}(2021)\citenamefont {Jones},
  \citenamefont {Bibo}, \citenamefont {Jobst}, \citenamefont {Pollmann},
  \citenamefont {Smith},\ and\ \citenamefont {Verresen}}]{Jones2021PRR}%
  \BibitemOpen
  \bibfield  {author} {\bibinfo {author} {\bibfnamefont {N.~G.}\ \bibnamefont
  {Jones}}, \bibinfo {author} {\bibfnamefont {J.}~\bibnamefont {Bibo}},
  \bibinfo {author} {\bibfnamefont {B.}~\bibnamefont {Jobst}}, \bibinfo
  {author} {\bibfnamefont {F.}~\bibnamefont {Pollmann}}, \bibinfo {author}
  {\bibfnamefont {A.}~\bibnamefont {Smith}},\ and\ \bibinfo {author}
  {\bibfnamefont {R.}~\bibnamefont {Verresen}},\ }\bibfield  {title} {\bibinfo
  {title} {Skeleton of matrix-product-state-solvable models connecting
  topological phases of matter},\ }\href
  {https://doi.org/10.1103/PhysRevResearch.3.033265} {\bibfield  {journal}
  {\bibinfo  {journal} {Phys. Rev. Res.}\ }\textbf {\bibinfo {volume} {3}},\
  \bibinfo {pages} {033265} (\bibinfo {year} {2021})}\BibitemShut {NoStop}%
\bibitem [{\citenamefont {Guo}\ \emph {et~al.}(2022)\citenamefont {Guo},
  \citenamefont {Yu}, \citenamefont {Hu},\ and\ \citenamefont
  {Li}}]{Guo2022PRA}%
  \BibitemOpen
  \bibfield  {author} {\bibinfo {author} {\bibfnamefont {Z.-X.}\ \bibnamefont
  {Guo}}, \bibinfo {author} {\bibfnamefont {X.-J.}\ \bibnamefont {Yu}},
  \bibinfo {author} {\bibfnamefont {X.-D.}\ \bibnamefont {Hu}},\ and\ \bibinfo
  {author} {\bibfnamefont {Z.}~\bibnamefont {Li}},\ }\bibfield  {title}
  {\bibinfo {title} {Emergent phase transitions in a cluster ising model with
  dissipation},\ }\href {https://doi.org/10.1103/PhysRevA.105.053311}
  {\bibfield  {journal} {\bibinfo  {journal} {Phys. Rev. A}\ }\textbf {\bibinfo
  {volume} {105}},\ \bibinfo {pages} {053311} (\bibinfo {year}
  {2022})}\BibitemShut {NoStop}%
\bibitem [{\citenamefont {Kuno}\ \emph {et~al.}(2022)\citenamefont {Kuno},
  \citenamefont {Orito},\ and\ \citenamefont {Ichinose}}]{Kuno_2022}%
  \BibitemOpen
  \bibfield  {author} {\bibinfo {author} {\bibfnamefont {Y.}~\bibnamefont
  {Kuno}}, \bibinfo {author} {\bibfnamefont {T.}~\bibnamefont {Orito}},\ and\
  \bibinfo {author} {\bibfnamefont {I.}~\bibnamefont {Ichinose}},\ }\bibfield
  {title} {\bibinfo {title} {Localization and slow-thermalization in a cluster
  spin model},\ }\href {https://doi.org/10.1088/1367-2630/ac7d01} {\bibfield
  {journal} {\bibinfo  {journal} {New Journal of Physics}\ }\textbf {\bibinfo
  {volume} {24}},\ \bibinfo {pages} {073019} (\bibinfo {year}
  {2022})}\BibitemShut {NoStop}%
\bibitem [{\citenamefont {Verga}(2023)}]{Verga2023PRB}%
  \BibitemOpen
  \bibfield  {author} {\bibinfo {author} {\bibfnamefont {A.~D.}\ \bibnamefont
  {Verga}},\ }\bibfield  {title} {\bibinfo {title} {Entanglement dynamics and
  phase transitions of the floquet cluster spin chain},\ }\href
  {https://doi.org/10.1103/PhysRevB.107.085116} {\bibfield  {journal} {\bibinfo
   {journal} {Phys. Rev. B}\ }\textbf {\bibinfo {volume} {107}},\ \bibinfo
  {pages} {085116} (\bibinfo {year} {2023})}\BibitemShut {NoStop}%
\bibitem [{\citenamefont {Alcaraz}\ and\ \citenamefont
  {Ramos}(2024)}]{Alcaraz2024PRE}%
  \BibitemOpen
  \bibfield  {author} {\bibinfo {author} {\bibfnamefont {F.~C.}\ \bibnamefont
  {Alcaraz}}\ and\ \bibinfo {author} {\bibfnamefont {L.~M.}\ \bibnamefont
  {Ramos}},\ }\bibfield  {title} {\bibinfo {title} {Conformally invariant
  free-parafermionic quantum chains with multispin interactions},\ }\href
  {https://doi.org/10.1103/PhysRevE.109.044138} {\bibfield  {journal} {\bibinfo
   {journal} {Phys. Rev. E}\ }\textbf {\bibinfo {volume} {109}},\ \bibinfo
  {pages} {044138} (\bibinfo {year} {2024})}\BibitemShut {NoStop}%
\bibitem [{\citenamefont {Li}\ \emph {et~al.}(2025{\natexlab{b}})\citenamefont
  {Li}, \citenamefont {Chen}, \citenamefont {Guo}, \citenamefont {Yu},\ and\
  \citenamefont {Li}}]{Li2025PRA}%
  \BibitemOpen
  \bibfield  {author} {\bibinfo {author} {\bibfnamefont {W.-L.}\ \bibnamefont
  {Li}}, \bibinfo {author} {\bibfnamefont {Y.-A.}\ \bibnamefont {Chen}},
  \bibinfo {author} {\bibfnamefont {Z.-X.}\ \bibnamefont {Guo}}, \bibinfo
  {author} {\bibfnamefont {X.-J.}\ \bibnamefont {Yu}},\ and\ \bibinfo {author}
  {\bibfnamefont {Z.}~\bibnamefont {Li}},\ }\bibfield  {title} {\bibinfo
  {title} {Global phase diagram of the cluster-$xy$ spin chain with
  dissipation},\ }\href {https://doi.org/10.1103/PhysRevA.111.013316}
  {\bibfield  {journal} {\bibinfo  {journal} {Phys. Rev. A}\ }\textbf {\bibinfo
  {volume} {111}},\ \bibinfo {pages} {013316} (\bibinfo {year}
  {2025}{\natexlab{b}})}\BibitemShut {NoStop}%
\bibitem [{\citenamefont {Becker}\ \emph {et~al.}(2010)\citenamefont {Becker},
  \citenamefont {Soltan-Panahi}, \citenamefont {Kronjäger}, \citenamefont
  {Dörscher}, \citenamefont {Bongs},\ and\ \citenamefont
  {Sengstock}}]{Becker_2010}%
  \BibitemOpen
  \bibfield  {author} {\bibinfo {author} {\bibfnamefont {C.}~\bibnamefont
  {Becker}}, \bibinfo {author} {\bibfnamefont {P.}~\bibnamefont
  {Soltan-Panahi}}, \bibinfo {author} {\bibfnamefont {J.}~\bibnamefont
  {Kronjäger}}, \bibinfo {author} {\bibfnamefont {S.}~\bibnamefont
  {Dörscher}}, \bibinfo {author} {\bibfnamefont {K.}~\bibnamefont {Bongs}},\
  and\ \bibinfo {author} {\bibfnamefont {K.}~\bibnamefont {Sengstock}},\
  }\bibfield  {title} {\bibinfo {title} {Ultracold quantum gases in triangular
  optical lattices},\ }\href {https://doi.org/10.1088/1367-2630/12/6/065025}
  {\bibfield  {journal} {\bibinfo  {journal} {New Journal of Physics}\ }\textbf
  {\bibinfo {volume} {12}},\ \bibinfo {pages} {065025} (\bibinfo {year}
  {2010})}\BibitemShut {NoStop}%
\bibitem [{\citenamefont {Petiziol}\ \emph {et~al.}(2021)\citenamefont
  {Petiziol}, \citenamefont {Sameti}, \citenamefont {Carretta}, \citenamefont
  {Wimberger},\ and\ \citenamefont {Mintert}}]{Petiziol2021PRL}%
  \BibitemOpen
  \bibfield  {author} {\bibinfo {author} {\bibfnamefont {F.}~\bibnamefont
  {Petiziol}}, \bibinfo {author} {\bibfnamefont {M.}~\bibnamefont {Sameti}},
  \bibinfo {author} {\bibfnamefont {S.}~\bibnamefont {Carretta}}, \bibinfo
  {author} {\bibfnamefont {S.}~\bibnamefont {Wimberger}},\ and\ \bibinfo
  {author} {\bibfnamefont {F.}~\bibnamefont {Mintert}},\ }\bibfield  {title}
  {\bibinfo {title} {Quantum simulation of three-body interactions in weakly
  driven quantum systems},\ }\href
  {https://doi.org/10.1103/PhysRevLett.126.250504} {\bibfield  {journal}
  {\bibinfo  {journal} {Phys. Rev. Lett.}\ }\textbf {\bibinfo {volume} {126}},\
  \bibinfo {pages} {250504} (\bibinfo {year} {2021})}\BibitemShut {NoStop}%
\bibitem [{\citenamefont {Smith}\ \emph {et~al.}(2022)\citenamefont {Smith},
  \citenamefont {Jobst}, \citenamefont {Green},\ and\ \citenamefont
  {Pollmann}}]{Smith2022PRR}%
  \BibitemOpen
  \bibfield  {author} {\bibinfo {author} {\bibfnamefont {A.}~\bibnamefont
  {Smith}}, \bibinfo {author} {\bibfnamefont {B.}~\bibnamefont {Jobst}},
  \bibinfo {author} {\bibfnamefont {A.~G.}\ \bibnamefont {Green}},\ and\
  \bibinfo {author} {\bibfnamefont {F.}~\bibnamefont {Pollmann}},\ }\bibfield
  {title} {\bibinfo {title} {Crossing a topological phase transition with a
  quantum computer},\ }\href
  {https://doi.org/10.1103/PhysRevResearch.4.L022020} {\bibfield  {journal}
  {\bibinfo  {journal} {Phys. Rev. Res.}\ }\textbf {\bibinfo {volume} {4}},\
  \bibinfo {pages} {L022020} (\bibinfo {year} {2022})}\BibitemShut {NoStop}%
\bibitem [{\citenamefont {Shen}\ \emph {et~al.}(2025)\citenamefont {Shen},
  \citenamefont {Chen}, \citenamefont {Yang}, \citenamefont {Zhong},\ and\
  \citenamefont {Lee}}]{shen2025robustsimulationsmanybodysymmetryprotected}%
  \BibitemOpen
  \bibfield  {author} {\bibinfo {author} {\bibfnamefont {R.}~\bibnamefont
  {Shen}}, \bibinfo {author} {\bibfnamefont {T.}~\bibnamefont {Chen}}, \bibinfo
  {author} {\bibfnamefont {B.}~\bibnamefont {Yang}}, \bibinfo {author}
  {\bibfnamefont {Y.}~\bibnamefont {Zhong}},\ and\ \bibinfo {author}
  {\bibfnamefont {C.~H.}\ \bibnamefont {Lee}},\ }\href
  {https://arxiv.org/abs/2503.08776} {\bibinfo {title} {Robust simulations of
  many-body symmetry-protected topological phase transitions on a quantum
  processor}} (\bibinfo {year} {2025}),\ \Eprint
  {https://arxiv.org/abs/2503.08776} {arXiv:2503.08776 [quant-ph]} \BibitemShut
  {NoStop}%
\bibitem [{\citenamefont {S\o{}rensen}\ \emph {et~al.}(2021)\citenamefont
  {S\o{}rensen}, \citenamefont {Catuneanu}, \citenamefont {Gordon},\ and\
  \citenamefont {Kee}}]{Sorensen2021PRX}%
  \BibitemOpen
  \bibfield  {author} {\bibinfo {author} {\bibfnamefont {E.~S.}\ \bibnamefont
  {S\o{}rensen}}, \bibinfo {author} {\bibfnamefont {A.}~\bibnamefont
  {Catuneanu}}, \bibinfo {author} {\bibfnamefont {J.~S.}\ \bibnamefont
  {Gordon}},\ and\ \bibinfo {author} {\bibfnamefont {H.-Y.}\ \bibnamefont
  {Kee}},\ }\bibfield  {title} {\bibinfo {title} {Heart of entanglement:
  Chiral, nematic, and incommensurate phases in the kitaev-gamma ladder in a
  field},\ }\href {https://doi.org/10.1103/PhysRevX.11.011013} {\bibfield
  {journal} {\bibinfo  {journal} {Phys. Rev. X}\ }\textbf {\bibinfo {volume}
  {11}},\ \bibinfo {pages} {011013} (\bibinfo {year} {2021})}\BibitemShut
  {NoStop}%
\bibitem [{\citenamefont {Yang}\ \emph {et~al.}(2020)\citenamefont {Yang},
  \citenamefont {Nocera}, \citenamefont {Tummuru}, \citenamefont {Kee},\ and\
  \citenamefont {Affleck}}]{Yang2020PRL}%
  \BibitemOpen
  \bibfield  {author} {\bibinfo {author} {\bibfnamefont {W.}~\bibnamefont
  {Yang}}, \bibinfo {author} {\bibfnamefont {A.}~\bibnamefont {Nocera}},
  \bibinfo {author} {\bibfnamefont {T.}~\bibnamefont {Tummuru}}, \bibinfo
  {author} {\bibfnamefont {H.-Y.}\ \bibnamefont {Kee}},\ and\ \bibinfo {author}
  {\bibfnamefont {I.}~\bibnamefont {Affleck}},\ }\bibfield  {title} {\bibinfo
  {title} {Phase diagram of the spin-$1/2$ kitaev-gamma chain and emergent
  su(2) symmetry},\ }\href {https://doi.org/10.1103/PhysRevLett.124.147205}
  {\bibfield  {journal} {\bibinfo  {journal} {Phys. Rev. Lett.}\ }\textbf
  {\bibinfo {volume} {124}},\ \bibinfo {pages} {147205} (\bibinfo {year}
  {2020})}\BibitemShut {NoStop}%
\bibitem [{\citenamefont {Liu}\ \emph {et~al.}(2020)\citenamefont {Liu},
  \citenamefont {Yi}, \citenamefont {Sun}, \citenamefont {Dong},\ and\
  \citenamefont {You}}]{Liu2020PRE}%
  \BibitemOpen
  \bibfield  {author} {\bibinfo {author} {\bibfnamefont {Z.-A.}\ \bibnamefont
  {Liu}}, \bibinfo {author} {\bibfnamefont {T.-C.}\ \bibnamefont {Yi}},
  \bibinfo {author} {\bibfnamefont {J.-H.}\ \bibnamefont {Sun}}, \bibinfo
  {author} {\bibfnamefont {Y.-L.}\ \bibnamefont {Dong}},\ and\ \bibinfo
  {author} {\bibfnamefont {W.-L.}\ \bibnamefont {You}},\ }\bibfield  {title}
  {\bibinfo {title} {Lifshitz phase transitions in a one-dimensional gamma
  model},\ }\href {https://doi.org/10.1103/PhysRevE.102.032127} {\bibfield
  {journal} {\bibinfo  {journal} {Phys. Rev. E}\ }\textbf {\bibinfo {volume}
  {102}},\ \bibinfo {pages} {032127} (\bibinfo {year} {2020})}\BibitemShut
  {NoStop}%
\bibitem [{\citenamefont {Luo}\ \emph {et~al.}(2021{\natexlab{a}})\citenamefont
  {Luo}, \citenamefont {Zhao}, \citenamefont {Wang},\ and\ \citenamefont
  {Kee}}]{Luo2021PRB}%
  \BibitemOpen
  \bibfield  {author} {\bibinfo {author} {\bibfnamefont {Q.}~\bibnamefont
  {Luo}}, \bibinfo {author} {\bibfnamefont {J.}~\bibnamefont {Zhao}}, \bibinfo
  {author} {\bibfnamefont {X.}~\bibnamefont {Wang}},\ and\ \bibinfo {author}
  {\bibfnamefont {H.-Y.}\ \bibnamefont {Kee}},\ }\bibfield  {title} {\bibinfo
  {title} {Unveiling the phase diagram of a bond-alternating spin-$\frac{1}{2}$
  $k\text{\ensuremath{-}}\mathrm{\ensuremath{\Gamma}}$ chain},\ }\href
  {https://doi.org/10.1103/PhysRevB.103.144423} {\bibfield  {journal} {\bibinfo
   {journal} {Phys. Rev. B}\ }\textbf {\bibinfo {volume} {103}},\ \bibinfo
  {pages} {144423} (\bibinfo {year} {2021}{\natexlab{a}})}\BibitemShut
  {NoStop}%
\bibitem [{\citenamefont {Zhao}\ \emph {et~al.}(2022)\citenamefont {Zhao},
  \citenamefont {Yi}, \citenamefont {Xue},\ and\ \citenamefont
  {You}}]{Zhao2022PRA}%
  \BibitemOpen
  \bibfield  {author} {\bibinfo {author} {\bibfnamefont {Z.}~\bibnamefont
  {Zhao}}, \bibinfo {author} {\bibfnamefont {T.-C.}\ \bibnamefont {Yi}},
  \bibinfo {author} {\bibfnamefont {M.}~\bibnamefont {Xue}},\ and\ \bibinfo
  {author} {\bibfnamefont {W.-L.}\ \bibnamefont {You}},\ }\bibfield  {title}
  {\bibinfo {title} {Characterizing quantum criticality and steered coherence
  in the $xy$-gamma chain},\ }\href
  {https://doi.org/10.1103/PhysRevA.105.063306} {\bibfield  {journal} {\bibinfo
   {journal} {Phys. Rev. A}\ }\textbf {\bibinfo {volume} {105}},\ \bibinfo
  {pages} {063306} (\bibinfo {year} {2022})}\BibitemShut {NoStop}%
\bibitem [{\citenamefont {Liu}\ \emph {et~al.}(2021)\citenamefont {Liu},
  \citenamefont {Dong}, \citenamefont {Wu}, \citenamefont {Wang},\ and\
  \citenamefont {You}}]{LIU2021126122}%
  \BibitemOpen
  \bibfield  {author} {\bibinfo {author} {\bibfnamefont {Z.-A.}\ \bibnamefont
  {Liu}}, \bibinfo {author} {\bibfnamefont {Y.-L.}\ \bibnamefont {Dong}},
  \bibinfo {author} {\bibfnamefont {N.}~\bibnamefont {Wu}}, \bibinfo {author}
  {\bibfnamefont {Y.}~\bibnamefont {Wang}},\ and\ \bibinfo {author}
  {\bibfnamefont {W.-L.}\ \bibnamefont {You}},\ }\bibfield  {title} {\bibinfo
  {title} {Quantum criticality and correlations in the ising-gamma chain},\
  }\href {https://doi.org/https://doi.org/10.1016/j.physa.2021.126122}
  {\bibfield  {journal} {\bibinfo  {journal} {Physica A: Statistical Mechanics
  and its Applications}\ }\textbf {\bibinfo {volume} {579}},\ \bibinfo {pages}
  {126122} (\bibinfo {year} {2021})}\BibitemShut {NoStop}%
\bibitem [{\citenamefont {Yang}\ \emph
  {et~al.}(2025{\natexlab{c}})\citenamefont {Yang}, \citenamefont {Nocera},
  \citenamefont {Xu}, \citenamefont {Adhikary},\ and\ \citenamefont
  {Affleck}}]{yang2025emergentsu21conformalsymmetry}%
  \BibitemOpen
  \bibfield  {author} {\bibinfo {author} {\bibfnamefont {W.}~\bibnamefont
  {Yang}}, \bibinfo {author} {\bibfnamefont {A.}~\bibnamefont {Nocera}},
  \bibinfo {author} {\bibfnamefont {C.}~\bibnamefont {Xu}}, \bibinfo {author}
  {\bibfnamefont {A.}~\bibnamefont {Adhikary}},\ and\ \bibinfo {author}
  {\bibfnamefont {I.}~\bibnamefont {Affleck}},\ }\href
  {https://arxiv.org/abs/2204.13810} {\bibinfo {title} {Emergent su(2)$_1$
  conformal symmetry in the spin-1/2 kitaev-gamma chain with a
  dzyaloshinskii-moriya interaction}} (\bibinfo {year} {2025}{\natexlab{c}}),\
  \Eprint {https://arxiv.org/abs/2204.13810} {arXiv:2204.13810
  [cond-mat.str-el]} \BibitemShut {NoStop}%
\bibitem [{\citenamefont {Kheiri}\ \emph {et~al.}(2024)\citenamefont {Kheiri},
  \citenamefont {Cheraghi}, \citenamefont {Mahdavifar},\ and\ \citenamefont
  {Sedlmayr}}]{Kheiri2024PRB}%
  \BibitemOpen
  \bibfield  {author} {\bibinfo {author} {\bibfnamefont {S.}~\bibnamefont
  {Kheiri}}, \bibinfo {author} {\bibfnamefont {H.}~\bibnamefont {Cheraghi}},
  \bibinfo {author} {\bibfnamefont {S.}~\bibnamefont {Mahdavifar}},\ and\
  \bibinfo {author} {\bibfnamefont {N.}~\bibnamefont {Sedlmayr}},\ }\bibfield
  {title} {\bibinfo {title} {Information propagation in one-dimensional
  $xy\text{\ensuremath{-}}\mathrm{\ensuremath{\Gamma}}$ chains},\ }\href
  {https://doi.org/10.1103/PhysRevB.109.134303} {\bibfield  {journal} {\bibinfo
   {journal} {Phys. Rev. B}\ }\textbf {\bibinfo {volume} {109}},\ \bibinfo
  {pages} {134303} (\bibinfo {year} {2024})}\BibitemShut {NoStop}%
\bibitem [{\citenamefont {Abbasi}\ \emph {et~al.}(2025)\citenamefont {Abbasi},
  \citenamefont {Mahdavifar},\ and\ \citenamefont
  {Motamedifar}}]{Abbasi2025SciPostCore}%
  \BibitemOpen
  \bibfield  {author} {\bibinfo {author} {\bibfnamefont {M.}~\bibnamefont
  {Abbasi}}, \bibinfo {author} {\bibfnamefont {S.}~\bibnamefont {Mahdavifar}},\
  and\ \bibinfo {author} {\bibfnamefont {M.}~\bibnamefont {Motamedifar}},\
  }\bibfield  {title} {\bibinfo {title} {{Spin-1/2 XX chains with modulated
  gamma interaction}},\ }\href
  {https://doi.org/10.21468/SciPostPhysCore.8.1.001} {\bibfield  {journal}
  {\bibinfo  {journal} {SciPost Phys. Core}\ }\textbf {\bibinfo {volume} {8}},\
  \bibinfo {pages} {001} (\bibinfo {year} {2025})}\BibitemShut {NoStop}%
\bibitem [{\citenamefont {Saito}\ and\ \citenamefont
  {Hotta}(2024)}]{Saito_2024}%
  \BibitemOpen
  \bibfield  {author} {\bibinfo {author} {\bibfnamefont {H.}~\bibnamefont
  {Saito}}\ and\ \bibinfo {author} {\bibfnamefont {C.}~\bibnamefont {Hotta}},\
  }\bibfield  {title} {\bibinfo {title} {Phase diagram of the quantum
  spin-$\frac{1}{2}$ heisenberg-$\mathrm{\ensuremath{\Gamma}}$ model on a
  frustrated zigzag chain},\ }\href
  {https://doi.org/10.1103/PhysRevB.110.024409} {\bibfield  {journal} {\bibinfo
   {journal} {Phys. Rev. B}\ }\textbf {\bibinfo {volume} {110}},\ \bibinfo
  {pages} {024409} (\bibinfo {year} {2024})}\BibitemShut {NoStop}%
\bibitem [{\citenamefont {Mahdavifar}\ and\ \citenamefont
  {Liu}(2024)}]{Mahdavifar2024}%
  \BibitemOpen
  \bibfield  {author} {\bibinfo {author} {\bibfnamefont {S.}~\bibnamefont
  {Mahdavifar}}\ and\ \bibinfo {author} {\bibfnamefont {D.~C.}\ \bibnamefont
  {Liu}},\ }\bibfield  {title} {\bibinfo {title} {Anisotropic spin-1/2 xxz
  chains with uniform gamma interaction},\ }\href
  {https://doi.org/10.1038/s41598-024-81404-z} {\bibfield  {journal} {\bibinfo
  {journal} {Scientific Reports}\ }\textbf {\bibinfo {volume} {14}},\ \bibinfo
  {pages} {30024} (\bibinfo {year} {2024})}\BibitemShut {NoStop}%
\bibitem [{\citenamefont {Jackeli}\ and\ \citenamefont
  {Khaliullin}(2009)}]{Jackeli2009PRL}%
  \BibitemOpen
  \bibfield  {author} {\bibinfo {author} {\bibfnamefont {G.}~\bibnamefont
  {Jackeli}}\ and\ \bibinfo {author} {\bibfnamefont {G.}~\bibnamefont
  {Khaliullin}},\ }\bibfield  {title} {\bibinfo {title} {Mott insulators in the
  strong spin-orbit coupling limit: From heisenberg to a quantum compass and
  kitaev models},\ }\href {https://doi.org/10.1103/PhysRevLett.102.017205}
  {\bibfield  {journal} {\bibinfo  {journal} {Phys. Rev. Lett.}\ }\textbf
  {\bibinfo {volume} {102}},\ \bibinfo {pages} {017205} (\bibinfo {year}
  {2009})}\BibitemShut {NoStop}%
\bibitem [{\citenamefont {Hermanns}\ \emph {et~al.}(2018)\citenamefont
  {Hermanns}, \citenamefont {Kimchi},\ and\ \citenamefont
  {Knolle}}]{hermanns2018physics}%
  \BibitemOpen
  \bibfield  {author} {\bibinfo {author} {\bibfnamefont {M.}~\bibnamefont
  {Hermanns}}, \bibinfo {author} {\bibfnamefont {I.}~\bibnamefont {Kimchi}},\
  and\ \bibinfo {author} {\bibfnamefont {J.}~\bibnamefont {Knolle}},\
  }\bibfield  {title} {\bibinfo {title} {Physics of the kitaev model:
  Fractionalization, dynamic correlations, and material connections},\
  }\href@noop {} {\bibfield  {journal} {\bibinfo  {journal} {Annual Review of
  Condensed Matter Physics}\ }\textbf {\bibinfo {volume} {9}},\ \bibinfo
  {pages} {17} (\bibinfo {year} {2018})}\BibitemShut {NoStop}%
\bibitem [{\citenamefont {Luo}\ \emph {et~al.}(2021{\natexlab{b}})\citenamefont
  {Luo}, \citenamefont {Zhao}, \citenamefont {Kee},\ and\ \citenamefont
  {Wang}}]{Luo2021npj}%
  \BibitemOpen
  \bibfield  {author} {\bibinfo {author} {\bibfnamefont {Q.}~\bibnamefont
  {Luo}}, \bibinfo {author} {\bibfnamefont {J.}~\bibnamefont {Zhao}}, \bibinfo
  {author} {\bibfnamefont {H.-Y.}\ \bibnamefont {Kee}},\ and\ \bibinfo {author}
  {\bibfnamefont {X.}~\bibnamefont {Wang}},\ }\bibfield  {title} {\bibinfo
  {title} {Gapless quantum spin liquid in a honeycomb $\gamma$ magnet},\ }\href
  {https://doi.org/10.1038/s41535-021-00356-z} {\bibfield  {journal} {\bibinfo
  {journal} {npj Quantum Materials}\ }\textbf {\bibinfo {volume} {6}},\
  \bibinfo {pages} {57} (\bibinfo {year} {2021}{\natexlab{b}})}\BibitemShut
  {NoStop}%
\bibitem [{\citenamefont {Takikawa}\ and\ \citenamefont
  {Fujimoto}(2019)}]{Takikawa2019PRB}%
  \BibitemOpen
  \bibfield  {author} {\bibinfo {author} {\bibfnamefont {D.}~\bibnamefont
  {Takikawa}}\ and\ \bibinfo {author} {\bibfnamefont {S.}~\bibnamefont
  {Fujimoto}},\ }\bibfield  {title} {\bibinfo {title} {Impact of off-diagonal
  exchange interactions on the kitaev spin-liquid state of
  $\ensuremath{\alpha}\text{\ensuremath{-}}{\mathrm{rucl}}_{3}$},\ }\href
  {https://doi.org/10.1103/PhysRevB.99.224409} {\bibfield  {journal} {\bibinfo
  {journal} {Phys. Rev. B}\ }\textbf {\bibinfo {volume} {99}},\ \bibinfo
  {pages} {224409} (\bibinfo {year} {2019})}\BibitemShut {NoStop}%
\bibitem [{\citenamefont {Wang}\ \emph {et~al.}(2021)\citenamefont {Wang},
  \citenamefont {Qi}, \citenamefont {Xi}, \citenamefont {Wang}, \citenamefont
  {Yu},\ and\ \citenamefont {Li}}]{WangPRB2021}%
  \BibitemOpen
  \bibfield  {author} {\bibinfo {author} {\bibfnamefont {S.}~\bibnamefont
  {Wang}}, \bibinfo {author} {\bibfnamefont {Z.}~\bibnamefont {Qi}}, \bibinfo
  {author} {\bibfnamefont {B.}~\bibnamefont {Xi}}, \bibinfo {author}
  {\bibfnamefont {W.}~\bibnamefont {Wang}}, \bibinfo {author} {\bibfnamefont
  {S.-L.}\ \bibnamefont {Yu}},\ and\ \bibinfo {author} {\bibfnamefont {J.-X.}\
  \bibnamefont {Li}},\ }\bibfield  {title} {\bibinfo {title} {Comprehensive
  study of the global phase diagram of the
  $j\ensuremath{-}k\ensuremath{-}\mathrm{\ensuremath{\Gamma}}$ model on a
  triangular lattice},\ }\href {https://doi.org/10.1103/PhysRevB.103.054410}
  {\bibfield  {journal} {\bibinfo  {journal} {Phys. Rev. B}\ }\textbf {\bibinfo
  {volume} {103}},\ \bibinfo {pages} {054410} (\bibinfo {year}
  {2021})}\BibitemShut {NoStop}%
\bibitem [{\citenamefont {R{\"o}{\ss}ler}\ \emph {et~al.}(2006)\citenamefont
  {R{\"o}{\ss}ler}, \citenamefont {Bogdanov},\ and\ \citenamefont
  {Pfleiderer}}]{Rossler2006Nature}%
  \BibitemOpen
  \bibfield  {author} {\bibinfo {author} {\bibfnamefont {U.~K.}\ \bibnamefont
  {R{\"o}{\ss}ler}}, \bibinfo {author} {\bibfnamefont {A.~N.}\ \bibnamefont
  {Bogdanov}},\ and\ \bibinfo {author} {\bibfnamefont {C.}~\bibnamefont
  {Pfleiderer}},\ }\bibfield  {title} {\bibinfo {title} {Spontaneous skyrmion
  ground states in magnetic metals},\ }\href
  {https://doi.org/10.1038/nature05056} {\bibfield  {journal} {\bibinfo
  {journal} {Nature}\ }\textbf {\bibinfo {volume} {442}},\ \bibinfo {pages}
  {797} (\bibinfo {year} {2006})}\BibitemShut {NoStop}%
\bibitem [{\citenamefont {Heinze}\ \emph {et~al.}(2011)\citenamefont {Heinze},
  \citenamefont {von Bergmann}, \citenamefont {Menzel}, \citenamefont {Brede},
  \citenamefont {Kubetzka}, \citenamefont {Wiesendanger}, \citenamefont
  {Bihlmayer},\ and\ \citenamefont {Bl{\"u}gel}}]{Heinze2011NP}%
  \BibitemOpen
  \bibfield  {author} {\bibinfo {author} {\bibfnamefont {S.}~\bibnamefont
  {Heinze}}, \bibinfo {author} {\bibfnamefont {K.}~\bibnamefont {von
  Bergmann}}, \bibinfo {author} {\bibfnamefont {M.}~\bibnamefont {Menzel}},
  \bibinfo {author} {\bibfnamefont {J.}~\bibnamefont {Brede}}, \bibinfo
  {author} {\bibfnamefont {A.}~\bibnamefont {Kubetzka}}, \bibinfo {author}
  {\bibfnamefont {R.}~\bibnamefont {Wiesendanger}}, \bibinfo {author}
  {\bibfnamefont {G.}~\bibnamefont {Bihlmayer}},\ and\ \bibinfo {author}
  {\bibfnamefont {S.}~\bibnamefont {Bl{\"u}gel}},\ }\bibfield  {title}
  {\bibinfo {title} {Spontaneous atomic-scale magnetic skyrmion lattice in two
  dimensions},\ }\href {https://doi.org/10.1038/nphys2045} {\bibfield
  {journal} {\bibinfo  {journal} {Nature Physics}\ }\textbf {\bibinfo {volume}
  {7}},\ \bibinfo {pages} {713} (\bibinfo {year} {2011})}\BibitemShut {NoStop}%
\bibitem [{\citenamefont {Furukawa}\ \emph {et~al.}(2012)\citenamefont
  {Furukawa}, \citenamefont {Sato}, \citenamefont {Onoda},\ and\ \citenamefont
  {Furusaki}}]{Furukawa2012PRB}%
  \BibitemOpen
  \bibfield  {author} {\bibinfo {author} {\bibfnamefont {S.}~\bibnamefont
  {Furukawa}}, \bibinfo {author} {\bibfnamefont {M.}~\bibnamefont {Sato}},
  \bibinfo {author} {\bibfnamefont {S.}~\bibnamefont {Onoda}},\ and\ \bibinfo
  {author} {\bibfnamefont {A.}~\bibnamefont {Furusaki}},\ }\bibfield  {title}
  {\bibinfo {title} {Ground-state phase diagram of a spin-$\frac{1}{2}$
  frustrated ferromagnetic xxz chain: Haldane dimer phase and gapped/gapless
  chiral phases},\ }\href {https://doi.org/10.1103/PhysRevB.86.094417}
  {\bibfield  {journal} {\bibinfo  {journal} {Phys. Rev. B}\ }\textbf {\bibinfo
  {volume} {86}},\ \bibinfo {pages} {094417} (\bibinfo {year}
  {2012})}\BibitemShut {NoStop}%
\bibitem [{\citenamefont {Luo}(2022)}]{Luo2022PRB}%
  \BibitemOpen
  \bibfield  {author} {\bibinfo {author} {\bibfnamefont {Q.}~\bibnamefont
  {Luo}},\ }\bibfield  {title} {\bibinfo {title} {Analytical results for the
  unusual gr\"uneisen ratio in the quantum ising model with
  dzyaloshinskii-moriya interaction},\ }\href
  {https://doi.org/10.1103/PhysRevB.105.L060401} {\bibfield  {journal}
  {\bibinfo  {journal} {Phys. Rev. B}\ }\textbf {\bibinfo {volume} {105}},\
  \bibinfo {pages} {L060401} (\bibinfo {year} {2022})}\BibitemShut {NoStop}%
\bibitem [{\citenamefont {Ginsparg}(1988)}]{ginsparg1988applied}%
  \BibitemOpen
  \bibfield  {author} {\bibinfo {author} {\bibfnamefont {P.}~\bibnamefont
  {Ginsparg}},\ }\bibfield  {title} {\bibinfo {title} {Applied conformal field
  theory},\ }\href@noop {} {\bibfield  {journal} {\bibinfo  {journal} {arXiv
  preprint hep-th/9108028}\ } (\bibinfo {year} {1988})}\BibitemShut {NoStop}%
\bibitem [{\citenamefont {Francesco}\ \emph {et~al.}(2012)\citenamefont
  {Francesco}, \citenamefont {Mathieu},\ and\ \citenamefont
  {S{\'e}n{\'e}chal}}]{francesco2012conformal}%
  \BibitemOpen
  \bibfield  {author} {\bibinfo {author} {\bibfnamefont {P.}~\bibnamefont
  {Francesco}}, \bibinfo {author} {\bibfnamefont {P.}~\bibnamefont {Mathieu}},\
  and\ \bibinfo {author} {\bibfnamefont {D.}~\bibnamefont {S{\'e}n{\'e}chal}},\
  }\href@noop {} {\emph {\bibinfo {title} {Conformal field theory}}}\ (\bibinfo
   {publisher} {Springer Science \& Business Media},\ \bibinfo {year}
  {2012})\BibitemShut {NoStop}%
\bibitem [{\citenamefont {Hornreich}(1980)}]{hornreich1980lifshitz}%
  \BibitemOpen
  \bibfield  {author} {\bibinfo {author} {\bibfnamefont {R.}~\bibnamefont
  {Hornreich}},\ }\bibfield  {title} {\bibinfo {title} {The lifshitz point:
  Phase diagrams and critical behavior},\ }\href@noop {} {\bibfield  {journal}
  {\bibinfo  {journal} {Journal of Magnetism and Magnetic Materials}\ }\textbf
  {\bibinfo {volume} {15}},\ \bibinfo {pages} {387} (\bibinfo {year}
  {1980})}\BibitemShut {NoStop}%
\bibitem [{\citenamefont {Chepiga}\ and\ \citenamefont
  {Mila}(2021)}]{Chepiga2021PRR}%
  \BibitemOpen
  \bibfield  {author} {\bibinfo {author} {\bibfnamefont {N.}~\bibnamefont
  {Chepiga}}\ and\ \bibinfo {author} {\bibfnamefont {F.}~\bibnamefont {Mila}},\
  }\bibfield  {title} {\bibinfo {title} {Lifshitz point at commensurate melting
  of chains of rydberg atoms},\ }\href
  {https://doi.org/10.1103/PhysRevResearch.3.023049} {\bibfield  {journal}
  {\bibinfo  {journal} {Phys. Rev. Res.}\ }\textbf {\bibinfo {volume} {3}},\
  \bibinfo {pages} {023049} (\bibinfo {year} {2021})}\BibitemShut {NoStop}%
\bibitem [{\citenamefont {Wang}\ and\ \citenamefont
  {Sedrakyan}(2022)}]{Wang2022SciPost}%
  \BibitemOpen
  \bibfield  {author} {\bibinfo {author} {\bibfnamefont {K.}~\bibnamefont
  {Wang}}\ and\ \bibinfo {author} {\bibfnamefont {T.~A.}\ \bibnamefont
  {Sedrakyan}},\ }\bibfield  {title} {\bibinfo {title} {{Universal finite-size
  amplitude and anomalous entanglement entropy of $z=2$ quantum Lifshitz
  criticalities in topological chains}},\ }\href
  {https://doi.org/10.21468/SciPostPhys.12.4.134} {\bibfield  {journal}
  {\bibinfo  {journal} {SciPost Phys.}\ }\textbf {\bibinfo {volume} {12}},\
  \bibinfo {pages} {134} (\bibinfo {year} {2022})}\BibitemShut {NoStop}%
\bibitem [{\citenamefont {Kunimi}\ \emph {et~al.}(2024)\citenamefont {Kunimi},
  \citenamefont {Tomita}, \citenamefont {Katsura},\ and\ \citenamefont
  {Kato}}]{Kunimi2024PRA}%
  \BibitemOpen
  \bibfield  {author} {\bibinfo {author} {\bibfnamefont {M.}~\bibnamefont
  {Kunimi}}, \bibinfo {author} {\bibfnamefont {T.}~\bibnamefont {Tomita}},
  \bibinfo {author} {\bibfnamefont {H.}~\bibnamefont {Katsura}},\ and\ \bibinfo
  {author} {\bibfnamefont {Y.}~\bibnamefont {Kato}},\ }\bibfield  {title}
  {\bibinfo {title} {Proposal for simulating quantum spin models with the
  dzyaloshinskii-moriya interaction using rydberg atoms and the construction of
  asymptotic quantum many-body scar states},\ }\href
  {https://doi.org/10.1103/PhysRevA.110.043312} {\bibfield  {journal} {\bibinfo
   {journal} {Phys. Rev. A}\ }\textbf {\bibinfo {volume} {110}},\ \bibinfo
  {pages} {043312} (\bibinfo {year} {2024})}\BibitemShut {NoStop}%
\bibitem [{\citenamefont {Pitsios}\ \emph {et~al.}(2017)\citenamefont
  {Pitsios}, \citenamefont {Banchi}, \citenamefont {Rab}, \citenamefont
  {Bentivegna}, \citenamefont {Caprara}, \citenamefont {Crespi}, \citenamefont
  {Spagnolo}, \citenamefont {Bose}, \citenamefont {Mataloni}, \citenamefont
  {Osellame},\ and\ \citenamefont {Sciarrino}}]{Pitsios2017NC}%
  \BibitemOpen
  \bibfield  {author} {\bibinfo {author} {\bibfnamefont {I.}~\bibnamefont
  {Pitsios}}, \bibinfo {author} {\bibfnamefont {L.}~\bibnamefont {Banchi}},
  \bibinfo {author} {\bibfnamefont {A.~S.}\ \bibnamefont {Rab}}, \bibinfo
  {author} {\bibfnamefont {M.}~\bibnamefont {Bentivegna}}, \bibinfo {author}
  {\bibfnamefont {D.}~\bibnamefont {Caprara}}, \bibinfo {author} {\bibfnamefont
  {A.}~\bibnamefont {Crespi}}, \bibinfo {author} {\bibfnamefont
  {N.}~\bibnamefont {Spagnolo}}, \bibinfo {author} {\bibfnamefont
  {S.}~\bibnamefont {Bose}}, \bibinfo {author} {\bibfnamefont {P.}~\bibnamefont
  {Mataloni}}, \bibinfo {author} {\bibfnamefont {R.}~\bibnamefont {Osellame}},\
  and\ \bibinfo {author} {\bibfnamefont {F.}~\bibnamefont {Sciarrino}},\
  }\bibfield  {title} {\bibinfo {title} {Photonic simulation of entanglement
  growth and engineering after a spin chain quench},\ }\href
  {https://doi.org/10.1038/s41467-017-01589-y} {\bibfield  {journal} {\bibinfo
  {journal} {Nature Communications}\ }\textbf {\bibinfo {volume} {8}},\
  \bibinfo {pages} {1569} (\bibinfo {year} {2017})}\BibitemShut {NoStop}%
\bibitem [{\citenamefont {Douglas}\ \emph {et~al.}(2015)\citenamefont
  {Douglas}, \citenamefont {Habibian}, \citenamefont {Hung}, \citenamefont
  {Gorshkov}, \citenamefont {Kimble},\ and\ \citenamefont
  {Chang}}]{Douglas2015}%
  \BibitemOpen
  \bibfield  {author} {\bibinfo {author} {\bibfnamefont {J.~S.}\ \bibnamefont
  {Douglas}}, \bibinfo {author} {\bibfnamefont {H.}~\bibnamefont {Habibian}},
  \bibinfo {author} {\bibfnamefont {C.-L.}\ \bibnamefont {Hung}}, \bibinfo
  {author} {\bibfnamefont {A.~V.}\ \bibnamefont {Gorshkov}}, \bibinfo {author}
  {\bibfnamefont {H.~J.}\ \bibnamefont {Kimble}},\ and\ \bibinfo {author}
  {\bibfnamefont {D.~E.}\ \bibnamefont {Chang}},\ }\bibfield  {title} {\bibinfo
  {title} {Quantum many-body models with cold atoms coupled to photonic
  crystals},\ }\href {https://doi.org/10.1038/nphoton.2015.57} {\bibfield
  {journal} {\bibinfo  {journal} {Nature Photonics}\ }\textbf {\bibinfo
  {volume} {9}},\ \bibinfo {pages} {326} (\bibinfo {year} {2015})}\BibitemShut
  {NoStop}%
\bibitem [{\citenamefont {Gonz{\'a}lez-Tudela}\ \emph
  {et~al.}(2015)\citenamefont {Gonz{\'a}lez-Tudela}, \citenamefont {Hung},
  \citenamefont {Chang}, \citenamefont {Cirac},\ and\ \citenamefont
  {Kimble}}]{GT2015NP}%
  \BibitemOpen
  \bibfield  {author} {\bibinfo {author} {\bibfnamefont {A.}~\bibnamefont
  {Gonz{\'a}lez-Tudela}}, \bibinfo {author} {\bibfnamefont {C.-L.}\
  \bibnamefont {Hung}}, \bibinfo {author} {\bibfnamefont {D.~E.}\ \bibnamefont
  {Chang}}, \bibinfo {author} {\bibfnamefont {J.~I.}\ \bibnamefont {Cirac}},\
  and\ \bibinfo {author} {\bibfnamefont {H.~J.}\ \bibnamefont {Kimble}},\
  }\bibfield  {title} {\bibinfo {title} {Subwavelength vacuum lattices and
  atom--atom interactions in two-dimensional photonic crystals},\ }\href
  {https://doi.org/10.1038/nphoton.2015.54} {\bibfield  {journal} {\bibinfo
  {journal} {Nature Photonics}\ }\textbf {\bibinfo {volume} {9}},\ \bibinfo
  {pages} {320} (\bibinfo {year} {2015})}\BibitemShut {NoStop}%
\bibitem [{\citenamefont {Kuepferling}\ \emph {et~al.}(2023)\citenamefont
  {Kuepferling}, \citenamefont {Casiraghi}, \citenamefont {Soares},
  \citenamefont {Durin}, \citenamefont {Garcia-Sanchez}, \citenamefont {Chen},
  \citenamefont {Back}, \citenamefont {Marrows}, \citenamefont {Tacchi},\ and\
  \citenamefont {Carlotti}}]{Kuepferling2023RMP}%
  \BibitemOpen
  \bibfield  {author} {\bibinfo {author} {\bibfnamefont {M.}~\bibnamefont
  {Kuepferling}}, \bibinfo {author} {\bibfnamefont {A.}~\bibnamefont
  {Casiraghi}}, \bibinfo {author} {\bibfnamefont {G.}~\bibnamefont {Soares}},
  \bibinfo {author} {\bibfnamefont {G.}~\bibnamefont {Durin}}, \bibinfo
  {author} {\bibfnamefont {F.}~\bibnamefont {Garcia-Sanchez}}, \bibinfo
  {author} {\bibfnamefont {L.}~\bibnamefont {Chen}}, \bibinfo {author}
  {\bibfnamefont {C.~H.}\ \bibnamefont {Back}}, \bibinfo {author}
  {\bibfnamefont {C.~H.}\ \bibnamefont {Marrows}}, \bibinfo {author}
  {\bibfnamefont {S.}~\bibnamefont {Tacchi}},\ and\ \bibinfo {author}
  {\bibfnamefont {G.}~\bibnamefont {Carlotti}},\ }\bibfield  {title} {\bibinfo
  {title} {Measuring interfacial dzyaloshinskii-moriya interaction in ultrathin
  magnetic films},\ }\href {https://doi.org/10.1103/RevModPhys.95.015003}
  {\bibfield  {journal} {\bibinfo  {journal} {Rev. Mod. Phys.}\ }\textbf
  {\bibinfo {volume} {95}},\ \bibinfo {pages} {015003} (\bibinfo {year}
  {2023})}\BibitemShut {NoStop}%
\bibitem [{\citenamefont {Yang}\ \emph {et~al.}(2023)\citenamefont {Yang},
  \citenamefont {Liang},\ and\ \citenamefont {Cui}}]{Yang2023NRP}%
  \BibitemOpen
  \bibfield  {author} {\bibinfo {author} {\bibfnamefont {H.}~\bibnamefont
  {Yang}}, \bibinfo {author} {\bibfnamefont {J.}~\bibnamefont {Liang}},\ and\
  \bibinfo {author} {\bibfnamefont {Q.}~\bibnamefont {Cui}},\ }\bibfield
  {title} {\bibinfo {title} {First-principles calculations for
  dzyaloshinskii--moriya interaction},\ }\href
  {https://doi.org/10.1038/s42254-022-00529-0} {\bibfield  {journal} {\bibinfo
  {journal} {Nature Reviews Physics}\ }\textbf {\bibinfo {volume} {5}},\
  \bibinfo {pages} {43} (\bibinfo {year} {2023})}\BibitemShut {NoStop}%
\bibitem [{\citenamefont {Dmitrienko}\ \emph {et~al.}(2014)\citenamefont
  {Dmitrienko}, \citenamefont {Ovchinnikova}, \citenamefont {Collins},
  \citenamefont {Nisbet}, \citenamefont {Beutier}, \citenamefont {Kvashnin},
  \citenamefont {Mazurenko}, \citenamefont {Lichtenstein},\ and\ \citenamefont
  {Katsnelson}}]{Dmitrienko2014NP}%
  \BibitemOpen
  \bibfield  {author} {\bibinfo {author} {\bibfnamefont {V.~E.}\ \bibnamefont
  {Dmitrienko}}, \bibinfo {author} {\bibfnamefont {E.~N.}\ \bibnamefont
  {Ovchinnikova}}, \bibinfo {author} {\bibfnamefont {S.~P.}\ \bibnamefont
  {Collins}}, \bibinfo {author} {\bibfnamefont {G.}~\bibnamefont {Nisbet}},
  \bibinfo {author} {\bibfnamefont {G.}~\bibnamefont {Beutier}}, \bibinfo
  {author} {\bibfnamefont {Y.~O.}\ \bibnamefont {Kvashnin}}, \bibinfo {author}
  {\bibfnamefont {V.~V.}\ \bibnamefont {Mazurenko}}, \bibinfo {author}
  {\bibfnamefont {A.~I.}\ \bibnamefont {Lichtenstein}},\ and\ \bibinfo {author}
  {\bibfnamefont {M.~I.}\ \bibnamefont {Katsnelson}},\ }\bibfield  {title}
  {\bibinfo {title} {Measuring the dzyaloshinskii--moriya interaction in a weak
  ferromagnet},\ }\href {https://doi.org/10.1038/nphys2859} {\bibfield
  {journal} {\bibinfo  {journal} {Nature Physics}\ }\textbf {\bibinfo {volume}
  {10}},\ \bibinfo {pages} {202} (\bibinfo {year} {2014})}\BibitemShut
  {NoStop}%
\bibitem [{\citenamefont {Bitko}\ \emph {et~al.}(1996)\citenamefont {Bitko},
  \citenamefont {Rosenbaum},\ and\ \citenamefont {Aeppli}}]{BD1996PRL}%
  \BibitemOpen
  \bibfield  {author} {\bibinfo {author} {\bibfnamefont {D.}~\bibnamefont
  {Bitko}}, \bibinfo {author} {\bibfnamefont {T.~F.}\ \bibnamefont
  {Rosenbaum}},\ and\ \bibinfo {author} {\bibfnamefont {G.}~\bibnamefont
  {Aeppli}},\ }\bibfield  {title} {\bibinfo {title} {Quantum critical behavior
  for a model magnet},\ }\href {https://doi.org/10.1103/PhysRevLett.77.940}
  {\bibfield  {journal} {\bibinfo  {journal} {Phys. Rev. Lett.}\ }\textbf
  {\bibinfo {volume} {77}},\ \bibinfo {pages} {940} (\bibinfo {year}
  {1996})}\BibitemShut {NoStop}%
\bibitem [{\citenamefont {Coldea}\ \emph {et~al.}(2010)\citenamefont {Coldea},
  \citenamefont {Tennant}, \citenamefont {Wheeler}, \citenamefont {Wawrzynska},
  \citenamefont {Prabhakaran}, \citenamefont {Telling}, \citenamefont
  {Habicht}, \citenamefont {Smeibidl},\ and\ \citenamefont
  {Kiefer}}]{RC2010Sci}%
  \BibitemOpen
  \bibfield  {author} {\bibinfo {author} {\bibfnamefont {R.}~\bibnamefont
  {Coldea}}, \bibinfo {author} {\bibfnamefont {D.~A.}\ \bibnamefont {Tennant}},
  \bibinfo {author} {\bibfnamefont {E.~M.}\ \bibnamefont {Wheeler}}, \bibinfo
  {author} {\bibfnamefont {E.}~\bibnamefont {Wawrzynska}}, \bibinfo {author}
  {\bibfnamefont {D.}~\bibnamefont {Prabhakaran}}, \bibinfo {author}
  {\bibfnamefont {M.}~\bibnamefont {Telling}}, \bibinfo {author} {\bibfnamefont
  {K.}~\bibnamefont {Habicht}}, \bibinfo {author} {\bibfnamefont
  {P.}~\bibnamefont {Smeibidl}},\ and\ \bibinfo {author} {\bibfnamefont
  {K.}~\bibnamefont {Kiefer}},\ }\bibfield  {title} {\bibinfo {title} {Quantum
  criticality in an ising chain: Experimental evidence for emergent
  e<sub>8</sub> symmetry},\ }\href {https://doi.org/10.1126/science.1180085}
  {\bibfield  {journal} {\bibinfo  {journal} {Science}\ }\textbf {\bibinfo
  {volume} {327}},\ \bibinfo {pages} {177} (\bibinfo {year} {2010})},\ \Eprint
  {https://arxiv.org/abs/https://www.science.org/doi/pdf/10.1126/science.1180085}
  {https://www.science.org/doi/pdf/10.1126/science.1180085} \BibitemShut
  {NoStop}%
\bibitem [{\citenamefont {Kenzelmann}\ \emph {et~al.}(2002)\citenamefont
  {Kenzelmann}, \citenamefont {Coldea}, \citenamefont {Tennant}, \citenamefont
  {Visser}, \citenamefont {Hofmann}, \citenamefont {Smeibidl},\ and\
  \citenamefont {Tylczynski}}]{KM2002PRB}%
  \BibitemOpen
  \bibfield  {author} {\bibinfo {author} {\bibfnamefont {M.}~\bibnamefont
  {Kenzelmann}}, \bibinfo {author} {\bibfnamefont {R.}~\bibnamefont {Coldea}},
  \bibinfo {author} {\bibfnamefont {D.~A.}\ \bibnamefont {Tennant}}, \bibinfo
  {author} {\bibfnamefont {D.}~\bibnamefont {Visser}}, \bibinfo {author}
  {\bibfnamefont {M.}~\bibnamefont {Hofmann}}, \bibinfo {author} {\bibfnamefont
  {P.}~\bibnamefont {Smeibidl}},\ and\ \bibinfo {author} {\bibfnamefont
  {Z.}~\bibnamefont {Tylczynski}},\ }\bibfield  {title} {\bibinfo {title}
  {Order-to-disorder transition in the $\mathrm{XY}$-like quantum magnet
  ${\mathrm{cs}}_{2}{\mathrm{cocl}}_{4}$ induced by noncommuting applied
  fields},\ }\href {https://doi.org/10.1103/PhysRevB.65.144432} {\bibfield
  {journal} {\bibinfo  {journal} {Phys. Rev. B}\ }\textbf {\bibinfo {volume}
  {65}},\ \bibinfo {pages} {144432} (\bibinfo {year} {2002})}\BibitemShut
  {NoStop}%
\bibitem [{\citenamefont {Breunig}\ \emph {et~al.}(2013)\citenamefont
  {Breunig}, \citenamefont {Garst}, \citenamefont {Sela}, \citenamefont
  {Buldmann}, \citenamefont {Becker}, \citenamefont {Bohat\'y}, \citenamefont
  {M\"uller},\ and\ \citenamefont {Lorenz}}]{BO2013PRL}%
  \BibitemOpen
  \bibfield  {author} {\bibinfo {author} {\bibfnamefont {O.}~\bibnamefont
  {Breunig}}, \bibinfo {author} {\bibfnamefont {M.}~\bibnamefont {Garst}},
  \bibinfo {author} {\bibfnamefont {E.}~\bibnamefont {Sela}}, \bibinfo {author}
  {\bibfnamefont {B.}~\bibnamefont {Buldmann}}, \bibinfo {author}
  {\bibfnamefont {P.}~\bibnamefont {Becker}}, \bibinfo {author} {\bibfnamefont
  {L.}~\bibnamefont {Bohat\'y}}, \bibinfo {author} {\bibfnamefont
  {R.}~\bibnamefont {M\"uller}},\ and\ \bibinfo {author} {\bibfnamefont
  {T.}~\bibnamefont {Lorenz}},\ }\bibfield  {title} {\bibinfo {title}
  {Spin-$\frac{1}{2}$ $xxz$ chain system ${\mathrm{cs}}_{2}{\mathrm{cocl}}_{4}$
  in a transverse magnetic field},\ }\href
  {https://doi.org/10.1103/PhysRevLett.111.187202} {\bibfield  {journal}
  {\bibinfo  {journal} {Phys. Rev. Lett.}\ }\textbf {\bibinfo {volume} {111}},\
  \bibinfo {pages} {187202} (\bibinfo {year} {2013})}\BibitemShut {NoStop}%
\bibitem [{\citenamefont {Jafari}\ \emph {et~al.}(2008)\citenamefont {Jafari},
  \citenamefont {Kargarian}, \citenamefont {Langari},\ and\ \citenamefont
  {Siahatgar}}]{Jafari2008PRB}%
  \BibitemOpen
  \bibfield  {author} {\bibinfo {author} {\bibfnamefont {R.}~\bibnamefont
  {Jafari}}, \bibinfo {author} {\bibfnamefont {M.}~\bibnamefont {Kargarian}},
  \bibinfo {author} {\bibfnamefont {A.}~\bibnamefont {Langari}},\ and\ \bibinfo
  {author} {\bibfnamefont {M.}~\bibnamefont {Siahatgar}},\ }\bibfield  {title}
  {\bibinfo {title} {Phase diagram and entanglement of the ising model with
  dzyaloshinskii-moriya interaction},\ }\href
  {https://doi.org/10.1103/PhysRevB.78.214414} {\bibfield  {journal} {\bibinfo
  {journal} {Phys. Rev. B}\ }\textbf {\bibinfo {volume} {78}},\ \bibinfo
  {pages} {214414} (\bibinfo {year} {2008})}\BibitemShut {NoStop}%
\bibitem [{\citenamefont {Kargarian}\ \emph {et~al.}(2009)\citenamefont
  {Kargarian}, \citenamefont {Jafari},\ and\ \citenamefont
  {Langari}}]{Kargarian2009PRA}%
  \BibitemOpen
  \bibfield  {author} {\bibinfo {author} {\bibfnamefont {M.}~\bibnamefont
  {Kargarian}}, \bibinfo {author} {\bibfnamefont {R.}~\bibnamefont {Jafari}},\
  and\ \bibinfo {author} {\bibfnamefont {A.}~\bibnamefont {Langari}},\
  }\bibfield  {title} {\bibinfo {title} {Dzyaloshinskii-moriya interaction and
  anisotropy effects on the entanglement of the heisenberg model},\ }\href
  {https://doi.org/10.1103/PhysRevA.79.042319} {\bibfield  {journal} {\bibinfo
  {journal} {Phys. Rev. A}\ }\textbf {\bibinfo {volume} {79}},\ \bibinfo
  {pages} {042319} (\bibinfo {year} {2009})}\BibitemShut {NoStop}%
\bibitem [{\citenamefont {Ma}\ \emph {et~al.}(2011)\citenamefont {Ma},
  \citenamefont {Liu},\ and\ \citenamefont {Kong}}]{Ma2011PRA}%
  \BibitemOpen
  \bibfield  {author} {\bibinfo {author} {\bibfnamefont {F.-W.}\ \bibnamefont
  {Ma}}, \bibinfo {author} {\bibfnamefont {S.-X.}\ \bibnamefont {Liu}},\ and\
  \bibinfo {author} {\bibfnamefont {X.-M.}\ \bibnamefont {Kong}},\ }\bibfield
  {title} {\bibinfo {title} {Quantum entanglement and quantum phase transition
  in the $xy$ model with staggered dzyaloshinskii-moriya interaction},\ }\href
  {https://doi.org/10.1103/PhysRevA.84.042302} {\bibfield  {journal} {\bibinfo
  {journal} {Phys. Rev. A}\ }\textbf {\bibinfo {volume} {84}},\ \bibinfo
  {pages} {042302} (\bibinfo {year} {2011})}\BibitemShut {NoStop}%
\bibitem [{\citenamefont {Kennedy}\ and\ \citenamefont
  {Tasaki}(1992)}]{Kennedy1992PRB}%
  \BibitemOpen
  \bibfield  {author} {\bibinfo {author} {\bibfnamefont {T.}~\bibnamefont
  {Kennedy}}\ and\ \bibinfo {author} {\bibfnamefont {H.}~\bibnamefont
  {Tasaki}},\ }\bibfield  {title} {\bibinfo {title} {Hidden
  ${\mathrm{z}}_{2}$\ifmmode\times\else\texttimes\fi{}${\mathrm{z}}_{2}$
  symmetry breaking in haldane-gap antiferromagnets},\ }\href
  {https://doi.org/10.1103/PhysRevB.45.304} {\bibfield  {journal} {\bibinfo
  {journal} {Phys. Rev. B}\ }\textbf {\bibinfo {volume} {45}},\ \bibinfo
  {pages} {304} (\bibinfo {year} {1992})}\BibitemShut {NoStop}%
\bibitem [{\citenamefont {Oshikawa}(1992)}]{Oshikawa_1992}%
  \BibitemOpen
  \bibfield  {author} {\bibinfo {author} {\bibfnamefont {M.}~\bibnamefont
  {Oshikawa}},\ }\bibfield  {title} {\bibinfo {title} {Hidden z2*z2 symmetry in
  quantum spin chains with arbitrary integer spin},\ }\href
  {https://doi.org/10.1088/0953-8984/4/36/019} {\bibfield  {journal} {\bibinfo
  {journal} {Journal of Physics: Condensed Matter}\ }\textbf {\bibinfo {volume}
  {4}},\ \bibinfo {pages} {7469} (\bibinfo {year} {1992})}\BibitemShut
  {NoStop}%
\bibitem [{\citenamefont {Vidal}\ \emph {et~al.}(2003)\citenamefont {Vidal},
  \citenamefont {Latorre}, \citenamefont {Rico},\ and\ \citenamefont
  {Kitaev}}]{Vidal2003PRL}%
  \BibitemOpen
  \bibfield  {author} {\bibinfo {author} {\bibfnamefont {G.}~\bibnamefont
  {Vidal}}, \bibinfo {author} {\bibfnamefont {J.~I.}\ \bibnamefont {Latorre}},
  \bibinfo {author} {\bibfnamefont {E.}~\bibnamefont {Rico}},\ and\ \bibinfo
  {author} {\bibfnamefont {A.}~\bibnamefont {Kitaev}},\ }\bibfield  {title}
  {\bibinfo {title} {Entanglement in quantum critical phenomena},\ }\href
  {https://doi.org/10.1103/PhysRevLett.90.227902} {\bibfield  {journal}
  {\bibinfo  {journal} {Phys. Rev. Lett.}\ }\textbf {\bibinfo {volume} {90}},\
  \bibinfo {pages} {227902} (\bibinfo {year} {2003})}\BibitemShut {NoStop}%
\bibitem [{\citenamefont {Peschel}\ and\ \citenamefont
  {Eisler}(2009)}]{Peschel_2009}%
  \BibitemOpen
  \bibfield  {author} {\bibinfo {author} {\bibfnamefont {I.}~\bibnamefont
  {Peschel}}\ and\ \bibinfo {author} {\bibfnamefont {V.}~\bibnamefont
  {Eisler}},\ }\bibfield  {title} {\bibinfo {title} {Reduced density matrices
  and entanglement entropy in free lattice models},\ }\href
  {https://doi.org/10.1088/1751-8113/42/50/504003} {\bibfield  {journal}
  {\bibinfo  {journal} {Journal of Physics A: Mathematical and Theoretical}\
  }\textbf {\bibinfo {volume} {42}},\ \bibinfo {pages} {504003} (\bibinfo
  {year} {2009})}\BibitemShut {NoStop}%
\end{thebibliography}%

\end{document}